\RequirePackage{ifpdf}
\documentclass[11pt,hyper,letterpaper]{JHEP3}
\usepackage{graphicx,caption,subcaption}
\usepackage{dcolumn}
\usepackage{bm}
\usepackage{cite}
\usepackage{amsmath,amssymb,amsfonts,cases}
\usepackage{bbm}
\usepackage{slashed}
\usepackage{wrapfig}

\newcommand{\beq}{\begin{equation}}
\newcommand{\eeq}{\end{equation}}
\newcommand{\bea}{\begin{eqnarray}}
\newcommand{\eea}{\end{eqnarray}}
\newcommand{\bi}{\begin{itemize}}
\newcommand{\ei}{\end{itemize}}
\newcommand{\ben}{\begin{enumerate}}
\newcommand{\een}{\end{enumerate}}

\newcommand{\tent}{T_{\textrm{ent}}}
\newcommand{\dcrit}{d_{\textrm{crit}}}

\newcommand{\scft}{S_{\textrm{CFT}}}
\newcommand{\sbulk}{S_{\textrm{bulk}}}

\newcommand{\Ttt}{\langle T_{tt}\rangle}
\newcommand{\Op}{\mathcal{O}}
\newcommand{\Opv}{\langle \mathcal{O} \rangle}

\newcommand{\auv}{\a_{\textrm{UV}}}
\newcommand{\air}{\a_{\textrm{IR}}}
\newcommand{\amin}{\mathcal{A}_{\textrm{min}}}
\newcommand{\detg}{\textrm{det}\,g_{MN}}

\renewcommand{\a}{\alpha}

\newcommand{\G}{\Gamma}
\newcommand{\g}{\gamma}

\newcommand{\s}{\sigma}

\newcommand{\z}{\zeta}
\newcommand{\p}{\partial}

\newcommand{\nn}{\nonumber}

\def\a{\alpha}

\def\f{\phi}

\def\th{\theta}                   
\def\s{\sigma}                                   

\def\z{\zeta}
\def\D{\Delta}
\def\F{\Phi}
\def\G{\Gamma}

\title{ \LARGE On Holographic Entanglement Density}
\author{Nikola I. Gushterov,\!$^1$\footnotemark[1]\, Andy O'Bannon,\!$^2$\footnotemark[2]\, and Ronnie Rodgers\!$^2$\footnotemark[3]
\\
$^1$Rudolf Peierls Centre for Theoretical Physics, University of Oxford \\ 1 Keble Road, Oxford OX1 3NP, United Kingdom
\\
$^2$STAG Research Centre, Physics and Astronomy, University of Southampton \\ Highfield, Southampton SO17 1BJ, United Kingdom}

\footnotetext[1]{E-mail address: \email{nikola.gushterov@physics.ox.ac.uk}}
\footnotetext[2]{E-mail address: \email{a.obannon@soton.ac.uk}}
\footnotetext[3]{E-mail address: \email{r.j.rodgers@soton.ac.uk}}

\abstract{We use holographic duality to study the entanglement entropy (EE) of Conformal Field Theories (CFTs) in various spacetime dimensions $d$, in the presence of various deformations: a relevant Lorentz scalar operator with constant source, a temperature $T$, a chemical potential $\mu$, a marginal Lorentz scalar operator with source linear in a spatial coordinate, and a circle-compactified spatial direction. We consider EE between a strip or sphere sub-region and the rest of the system, and define the ``entanglement density'' (ED) as the change in EE due to the deformation, divided by the sub-region's volume. Using the deformed CFTs above, we show how the ED's dependence on the strip width or sphere radius, $L$, is useful for characterizing states of matter. For example, the ED's small-$L$ behavior is determined either by the dimension of the perturbing operator or by the first law of EE. For Lorentz-invariant renormalization group (RG) flows between CFTs, the ``area theorem'' states that the coefficient of the EE's area law term must be larger in the UV than in the IR. In these cases the ED must therefore approach zero from below as $L \to \infty$. However, when Lorentz symmetry is broken and the IR fixed point has different scaling from the UV, we find that the ED often approaches the thermal entropy density from above, indicating area theorem violation.}

\preprint{OUTP-17-11P}

\begin{document}

\section{Introduction, Summary, and Outlook}
\label{intro}

\subsection{Introduction and Motivation}

A central goal of physics is to characterize and classify states of matter. At temperatures $T$ low enough that quantum effects determine the properties of matter, the goal is to characterize and classify patterns of quantum entanglement. A growing body of evidence suggests that entanglement entropy (EE) between a sub-region and the rest of the system, and specifically EE's dependence on the sub-region's size $L$ (the radius of a sphere, for example), can play a central role in reaching that goal. For example, EE receives characteristic contributions $\propto \ln L^{d-1}$ from a Goldstone boson~\cite{Metlitski:2011pr}, $\propto L^{d-1} \ln L$ from a Fermi surface~\cite{Wolf:2006zzb,2006PhRvL..96j0503G,Swingle:2009bf,Swingle:2010yi}, or independent of $L$ from topologically-ordered degrees of freedom~\cite{2006PhRvL..96k0404K,2006PhRvL..96k0405L,2008PhRvB..78o5120C,2011PhRvB..84s5120G}.

In this paper, we study how EE may characterize Conformal Field Theories (CFTs) deformed by: RG flows to infra-red (IR) CFTs (section~\ref{rg}), temperature $T$ (sec.~\ref{adssc}), chemical potential $\mu$ that leads to either a $(0+1)$-dimensional IR fixed point (sec.~\ref{adsrn}) or hyperscaling-violating (HV) fixed point (sec.~\ref{hyper}), a marginal scalar operator with source linear in a spatial coordinate, $x$ (sec.~\ref{translation}), and compactification of $x$ (sec.~\ref{soliton}). Each of these has one or more illustrative features, distinct from all others: all have gapless IR degrees of freedom, except the compactification, all are translationally invariant, except the source linear in $x$, etc.

In this paper we define an ``entanglement density'' (ED)\footnote{Our ED should not be confused with the entanglement density of refs.~\cite{Nozaki:2013wia,Bhattacharya:2014vja}, defined as a second variation of EE under infinitesimal changes to the sub-region's boundary.}, and explore the extent to which it characterizes the deformed CFTs mentioned above. Specifically, given the EE of the deformed CFT, $S$, the EE of the undeformed CFT's vacuum state, $S_{\textrm{CFT}}$, and the volume of the entangling region, $V$, we defined the ED as
\beq
\label{eq:eddef}
\sigma\equiv\frac{S-S_{\textrm{CFT}}}{V}.
\eeq
In continuum quantum field theories (QFTs), $S$ generically has short-distance divergences from large correlations across the entangling surface (the sub-region's boundary). We regulate these with an ultra-violet (UV) cutoff, $\varepsilon$. For our deformed CFTs, these divergences are identical to those of the parent CFT, hence the subtraction $S-S_{\textrm{CFT}}$ renders $\sigma$ finite and cutoff-independent, and therefore physically meaningful. (Actually, regulating $\s$ when $x$ is compactified is slightly more subtle, as we discuss in sec.~\ref{soliton}.)

Of course, we could remove the divergences in other ways, for instance by adding counterterms~\cite{Taylor:2016aoi,Taylor:2016kic,Taylor:2017zzo}, and we could divide by other quantities intrinsic to the entangling surface besides $V$, such as surface area, $A$. However, our definition of ED is motivated by the so-called ``entanglement temperature,'' $\tent$~\cite{Bhattacharya:2012mi,Blanco:2013joa}, defined as follows. For two states infinitesimally close in a QFT's Hilbert space, positivity of their relative entropy implies a ``first law'' of EE (FLEE), namely the difference in EE is equivalent to the change of the expectation value of the modular Hamiltonian~\cite{Blanco:2013joa}. For a spherical sub-region in a CFT, the latter is simply the change of energy inside the sphere, or more precisely the change in the expectation value $\Ttt$, where $T_{\mu\nu}$ is the stress-energy tensor and $t$ is time, divided by the quantity $(d+1)/(2\pi L)$. Similarly, for CFTs holographically dual to Einstein gravity theories in $(d+1)$-dimensional Anti-de Sitter space, $AdS_{d+1}$~\cite{Maldacena:1997re,Witten:1998qj}, and for a strip sub-region, defined as two parallel planes separated by a distance $L$, the change of EE is also equivalent to the change of energy inside the strip, divided by a quantity $\propto 1/L$~\cite{Bhattacharya:2012mi}. In these cases, $\tent$ is defined as the change in energy divided by the change in EE. In other words, $\tent$ is precisely the quantity $\propto 1/L$ in each case, which depends on $d$, but not on any other details of the CFT or its states.

For states with constant $\Ttt$, and for sufficiently small $L$, our $\sigma= \Ttt\,\tent^{-1} \propto \Ttt L$. However, our $\sigma$ generalizes $\tent^{-1}$ to any change of energy, including zero change. For example, our $\sigma$ is well-defined in states with constant $\Ttt$ for any $L$, not just for small $L$, and also for Lorentz-invariant states, which have $\Ttt=0$. Moreover, $\sigma$ is well-defined not only for a change of the state, but also for some changes of the Hamiltonian, as occurs for example in certain RG flows. In short, while $\tent^{-1}$ is the change in EE per unit energy for a change of the state, $\sigma$ is the change in EE per unit volume for a change of the state or Hamiltonian.

Our goal is to characterize the deformed CFTs above using $\sigma$'s dependence on $L$. We will consider only holographic QFTs because holography is currently the easiest way to compute $S$ in interacting QFTs. Typically holographic QFTs are non-Abelian gauge theories in the 't Hooft large-$N$ limit with large 't Hooft coupling~\cite{Aharony:1999ti}. As we review in sec.~\ref{general}, in the holographically dual geometry $S$ is given by the area of the minimal surface that approaches the entangling surface at the asymptotic $AdS_{d+1}$ boundary~\cite{Ryu:2006bv,Ryu:2006ef,Lewkowycz:2013nqa}. We will consider only $d\geq 3$ and only strip or sphere sub-regions. In sec.~\ref{general} we consider an asymptotically $AdS_{d+1}$ metric of a general form that encompasses all our later examples, except that of sec.~\ref{soliton}. In particular, we derive equations for $S$ for the strip or sphere in terms of metric components. In the subsequent sections we then numerically solve for $S$ and hence $\s$ case-by-case.

For deformations of the state but not the Hamiltonian, such as $T$ or $\mu$,\footnote{Crucially, in thermal equilibrium $\mu$ can be introduced either as a deformation of the Hamiltonian, \textit{i.e.} a source for the charge operator, or as a deformation of the state, with no change to the Hamiltonian, \textit{i.e.} restrict the path integral such that a bosonic or fermionic field of charge $q$ acquires a factor $\pm e^{q \mu/T}$ around the Euclidean time circle, respectively. We have the latter approach in mind.} the FLEE requires $\s \propto \Ttt L$ at small $L$, as mentioned above. For deformations of the Hamiltonian, in general $\s$'s small-$L$ behavior is determined by the dimension $\Delta$ of the perturbing operator and whether the operator's source depends on $x$, as we discuss in secs.~\ref{rg}, \ref{hyper}, and~\ref{translation}.

As $L \to \infty$ relative to any other scale, the leading behavior of the EE is
\beq
\label{eq:EElargeL}
S = s \, V + \a \, A + \ldots,
\eeq
where $s$ is the thermodynamic entropy density ($s=0$ in some of our examples), $\a$ is a dimensionful constant, and $\ldots$ represents terms sub-leading in $1/L$ relative to those shown.

The leading ``volume law'' term $\propto V$ in eq.~\eqref{eq:EElargeL} is expected for excited states, such as thermal states. In such cases, intuitively when $L \to \infty$ the sub-region becomes the entire system, and the sub-region's reduced density matrix becomes the total density matrix, which for a thermal state implies $S \to sV$. In holography, the volume law appears in thermal states because the minimal area surface lies along a horizon~\cite{Hubeny:2012ry,Liu:2013una} with Bekenstein-Hawking entropy density $s$~\cite{Witten:1998zw}. For a sphere $V \propto L^{d-1}$ while for the strip $V \propto \textrm{Vol}\left(\mathbb{R}^{d-2}\right) L$, where $\textrm{Vol}\left(\mathbb{R}^{d-2}\right)$ is the (infinite) area of the ``wall'' of the strip.

The sub-leading contribution $\propto A$ in eq.~\eqref{eq:EElargeL} is the well-known ``area law'' term~\cite{Bombelli:1986rw,Srednicki:1993im,2007JSMTE..08...24H,Eisert:2008ur}. For a sphere, $A \propto L^{d-2}$, and in the vacuum of a CFT, the only other scale is the UV cutoff, $\varepsilon$, so that $\a \propto 1/\varepsilon^{d-2}$ by dimensional analysis. If the CFT is then deformed, then in general $\a$ is a sum of terms, including the term $\propto 1/\varepsilon^{d-2}$ plus terms set by whatever scales are available, such as $T$, $\mu$, mass scales, etc. For a strip, $A \propto2\textrm{Vol}\left(\mathbb{R}^{d-2}\right)$, and in the vacuum of a CFT, two other scales are available, $\varepsilon$ and $L$. Indeed, in that case $\a$ is a sum of two terms, one $\propto 1/\varepsilon^{d-2}$ and the other $\propto 1/L^{d-2}$~\cite{Ryu:2006bv,Ryu:2006ef}. If the CFT is then deformed, then in general $\a$ is a sum of terms, including the terms $\propto 1/\varepsilon^{d-2}$ and $\propto 1/L^{d-2}$, and other terms set by whatever scales are available. Some deformations can also produce in $S$ a term $\propto A \ln A$, such as $\mu$ in a free fermion CFT, producing a Fermi surface~\cite{Wolf:2006zzb,2006PhRvL..96j0503G,Swingle:2009bf,Swingle:2010yi}, as mentioned above. For discussions about the conditions under which such ``area law violation'' can occur, see for example ref.~\cite{Swingle:2011np}.

Crucially, for Lorentz-invariant RG flows to a $d$-dimensional CFT in the infra-red (IR), $\a$ obeys a kind of (weak) $c$-theorem, called the ``area theorem''~\cite{Casini:2012ei,Casini:2016udt}: the value of $\a$ in the UV CFT, $\auv$, must be greater than or equal to that of the IR CFT, $\air$. Of course, as mentioned above both $\auv$ and $\air$ include a term $\propto 1/\varepsilon^{d-2}$, and hence diverge as $\varepsilon\to0$. However, these terms $\propto 1/\varepsilon^{d-2}$ cancel in the difference $\Delta \a \equiv \auv - \air$, so the meaningful statement of the area theorem is $\Delta \a\geq0$. To be precise, the area theorem has been proven for a sphere in $d=3$ using strong sub-additivity~\cite{Casini:2012ei} and for a sphere in $d \geq 3$ using positivity of relative entropy~\cite{Casini:2016udt}.\footnote{For the strip, a similar, but distinct, theorem for the coefficient of an area term appears in refs.~\cite{Ryu:2006ef,Myers:2012ed}.} Roughly speaking, strong sub-additivity is holographically dual to the Null Energy Condition (NEC)~\cite{Headrick:2007km,Wall:2012uf}. All of our holographic examples will obey the NEC.

Whenever a quantity is proven to decrease monotonically along an RG flow, a number of questions naturally arise. For example, does the quantity count degrees of freedom in any precise sense? Does the monotonicity extend to other types of deformations, such as $T$, $\mu$, operators or sources that break Lorentz invariance, etc.~\cite{Swingle:2013zla}? We will answer some of these questions for $\Delta \a$, in holographic systems, using our $\s$. In particular, eq.~\eqref{eq:EElargeL} implies that when $L \to \infty$ relative to all other scales, $\s$'s leading behavior is
\beq
\label{eq:EDlargeL}
\s = s - \Delta \a \, \frac{A}{V} + \ldots,
\eeq
where the difference $-\Delta \a=\air-\auv$ appears because in eq.~\eqref{eq:eddef} we subtract the UV CFT vacuum contribution, $S-\scft$. For both the sphere and strip $A/V \propto 1/L$. In sec.~\ref{general} we borrow techniques from refs.~\cite{Liu:2012eea,Liu:2013una,Kundu:2016dyk} to show that for geometries with a horizon the leading large-$L$ correction to $\s$ is $\propto A/V$ for both the strip and sphere, as expected.

Eq.~\eqref{eq:EDlargeL} shows how we can easily extract the sign of $\Delta \a$ from $\s$'s large-$L$ behavior: as $L\to \infty$, if $\s$ approaches $s$ from below ($\s \to s^-$) then $\Delta \a>0$, while if $\s$ approaches $s$ from above ($\s\to s^+$) then $\Delta \a<0$. The sign of $\Delta \a$ will therefore be immediately obvious to the naked eye, as our examples will illustrate. Dividing by $V$ in eq.~\eqref{eq:eddef} is thus technically trivial but practically useful: otherwise, to obtain $\Delta \a$'s sign we would have to extract (typically by numerical fitting) a subtle correction in $1/L$ from the EE itself. In sec.~\ref{general}, we write the coefficient of the $1/L$ correction as an integral over bulk metric components, which typically must be performed numerically. This integral's sign gives us $\Delta \a$'s sign.

\subsection{Summary of Results}

Table~\ref{tab:summary} summarizes our main results, which we discuss in detail in this subsection.

\begin{table}[ht!]
\begin{tabular}[c]{|c|c|c|c|c|}
\hline
Section & System & Deformation(s) & FLEE? & Area Theorem Violation? \\ \hline
\ref{rg} &  $(d+1)$ AdS-to-AdS & $\mathcal{O}$ & No & No \\ \hline
\ref{adssc} & $(d+1)$ AdS-SCH & $T$ & Yes & Yes, for $d>\dcrit$ \\ \hline
\ref{adsrn} & $(d+1)$ AdS-RN & $T$, $\mu$ & Yes & Yes, for $d>\dcrit$ or low $T$ \\ \hline
\ref{hyper} & $(d+1)$ AdS-to-HV & $\mathcal{O}$, $T$, $\mu$ & No & Yes, for some $d$, $\zeta$, $\theta$ \\ \hline
\ref{translation} & AdS$_4$-Linear Axion & $\gamma \, x \, \mathcal{O}$, $T$, $\mu$ & No & Yes, for low $T$ \\ \hline
\ref{soliton} & $(d+1)$ AdS Soliton & compact $x$ & No & No \\ \hline
\end{tabular}
\caption{\label{tab:summary} Summary of our main results, discussed in detail in this subsection.}
\end{table}

In sec.~\ref{rg} we consider Lorentz-invariant RG flows, described holographically by gravity coupled to a single real scalar field with self-interaction potential designed to produce a ``domain wall'' solution interpolating between an $AdS_{d+1}$ near the boundary and another $AdS_{d+1}$ deep in the bulk~\cite{Freedman:1999gp}. Lorentz invariance implies $\langle T_{\mu\nu}\rangle = 0$ and $s=0$. We mostly focus on $d=4$, and consider flows driven either by a source for the relevant scalar operator $\mathcal{O}$ dual to the bulk scalar field, or driven by $\langle \mathcal{O} \rangle \neq 0$ with zero source. As mentioned above, the FLEE does not apply in these cases, and $\mathcal{O}$'s dimension $\Delta$ controls the leading power of $L$ in $\s$ at small $L$. We find that $\s<0$ for all $L$, and in particular $\s \to 0^-$ as $L \to \infty$, as required by the area theorem. To connect the small- and large-$L$ limits, $\s$ must have one or more minima as a function of $L$. We show how various scalar potentials, all consistent with the NEC, can produce various behaviors in $\s$ at intermediate $L$, such as multiple minima or a discontinuous first derivative. We thus learn that, although universal principles such as the area theorem may govern $\s$'s asymptotics, no universality is immediately obvious at intermediate $L$. We expect similar results for other $d$.

\begin{wrapfigure}{R}{0.5\textwidth}
{\centering
\includegraphics[width=0.5\textwidth]{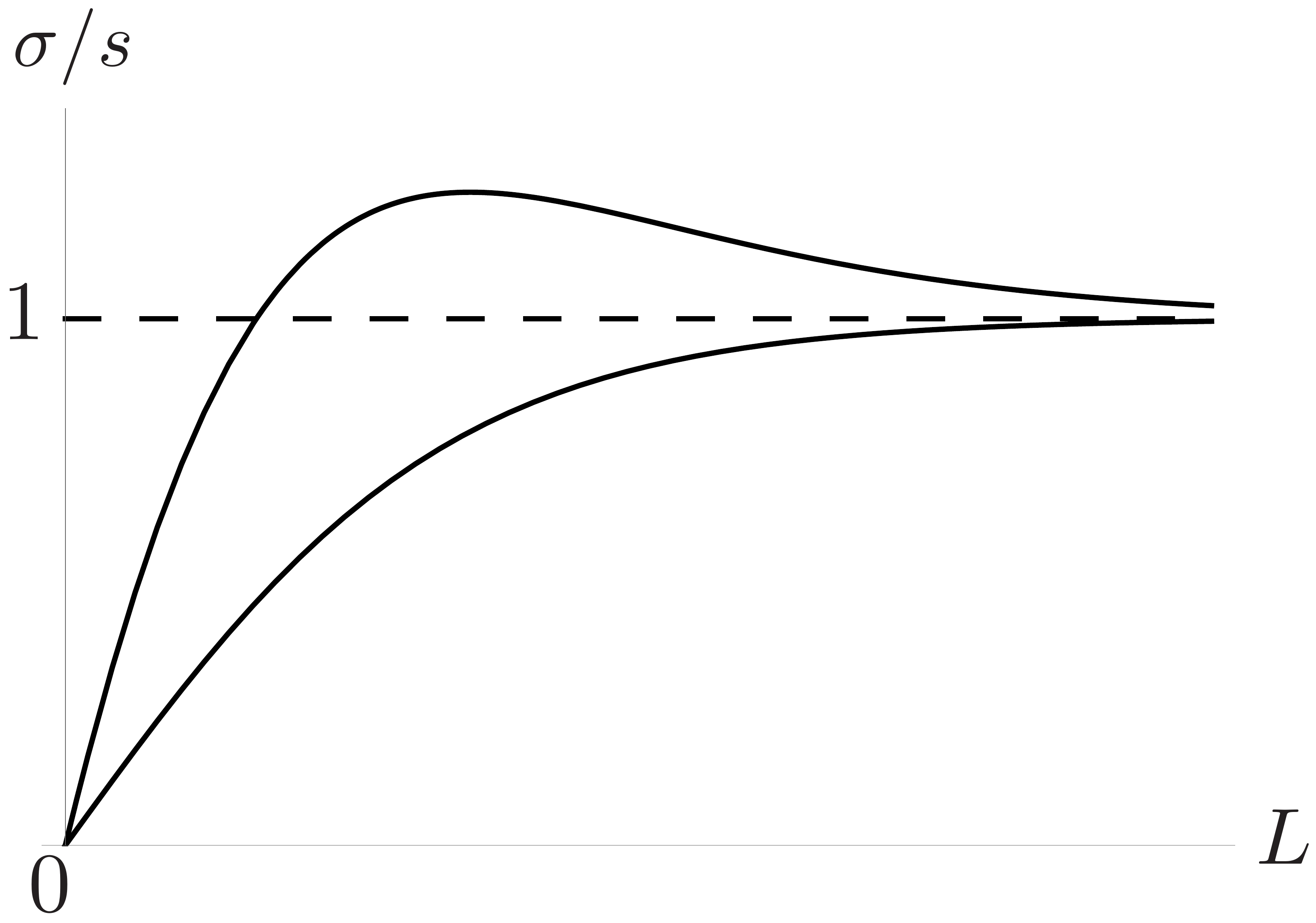}
\caption{\label{fig:schematic} For a CFT excited state in which the FLEE applies and $s \neq 0$, we schematically depict the two simplest possibile behaviors of $\sigma$, in units of $s$, versus $L$, in arbitrary units. The FLEE implies $\sigma \propto L$ at small $L$. As $L\to\infty$, either $\s \to s^-$ (lower curve), consistent with the area theorem, or $\s \to s^+$ (upper curve), violating the area theorem. The latter case necessarily has at least one maximum, as shown. \vspace{-0.4cm}}
}
\end{wrapfigure}

In sec.~\ref{adssc} we consider the AdS-Schwarzschild (AdS-SCH) black brane, dual to a translationally and rotationally invariant state of a holographic CFT deformed by $T$. In this case, the FLEE requires $\s \propto \Ttt L$ at small $L$. Sec.~\ref{adssc}'s main result is the existence of a critical dimension, $\dcrit \approx 6.7$, such that if $d<\dcrit$ then as $L$ increases $\s$ rises monotonically, and $\s \to s^-$ as $L \to \infty$, so that $\Delta \a >0$, consistent with the area theorem. However, if $d>\dcrit$ then $\s$ increases to a single global maximum, which by dimensional analysis is at an $L \propto 1/T$, and then $\s \to s^+$ as $L \to \infty$, so that $\Delta \a<0$, violating the area theorem. Figure~\ref{fig:schematic} depicts these two behaviors schematically. (These results have also been obtained using the \textit{exact} results for EE of a strip in AdS-SCH, \textit{i.e.} without numerics, in ref.~\cite{Erdmenger:2017pfh}.) More generally, for any CFT excited state in which the FLEE applies and $s \neq 0$, these are the two simplest ways to connect $\s \propto \Ttt L$ at small $L$ to $\sigma \to s^{\pm}$ at large $L$.

In sec.~\ref{adsrn} we consider an AdS-Reissner-Nordstr\"om (AdS-RN) charged black brane, dual to a translationally and rotationally invariant state of a holographic CFT deformed by $T$ and $\mu$, in which only $T_{\mu\nu}$ and the charge density have non-zero expectation values~\cite{Hartnoll:2009sz}. When $T/\mu \to \infty$, so that $\mu$ is negligible, AdS-RN approaches AdS-SCH, and we recover the results of sec.~\ref{adssc}, including in particular the existence of $\dcrit$. However, when $T/\mu \to 0$, so that $T$ is negligible, AdS-RN is dual to a ``semi-local quantum liquid'' state~\cite{Iqbal:2011in}, which at $T/\mu=0$ has a mysterious extensive ground state entropy $s \neq 0$. If $d>\dcrit$ then for all $T/\mu$, $\s$ resembles the upper curve in fig.~\ref{fig:schematic}, with a single maximum, whose position changes as $T/\mu$ decreases, and $\s \to s^+$ as $L \to \infty$. In particular, when $d>\dcrit$ the area theorem is always violated. On the other hand, if $d<\dcrit$, then at high $T/\mu$ we recover the result of sec.~\ref{adssc}, where $\s$ resembles the lower curve in fig.~\ref{fig:schematic}, with no maximum and $\s \to s^-$ as $L \to \infty$. However, as we lower $T/\mu$, a transition occurs at a critical value of $T/\mu$ from the lower curve in fig.~\ref{fig:schematic} to the upper curve, \textit{i.e.} a peak appears. In particular, at the critical $T/\mu$, $\Delta \a$ changes sign and the area theorem is violated. In short, for any $d$, at sufficiently low $T/\mu$, $\s$ resembles the upper curve in fig.~\ref{fig:schematic}, with a single maximum, $\s \to s^+$ as $L \to \infty$, and area theorem violation.

In sec.~\ref{hyper} we consider the model of ref.~\cite{Lucas:2014sba}, namely gravity in $AdS_{d+1}$ coupled to a real scalar field and two $U(1)$ gauge fields, which at $T=0$ yields domain-wall solutions from $AdS_{d+1}$ to HV geometries~\cite{Huijse:2011ef}. Such solutions are dual to CFTs in which $\mu$ and $\Op$ produce an IR fixed point with HV exponent $\theta$ and Lifshitz scaling $t \to \lambda^{\zeta} t$, $\vec{x} \to \lambda \vec{x}$, with $\lambda \in \mathbb{R}^+$, spatial coordinates $\vec{x}$, and dynamical exponent\footnote{The dynamical exponent is usually called $z$, but our $z$ is the coordinate normal to the $AdS_{d+1}$ boundary.} $\zeta$. Similarly to sec.~\ref{rg}, in general the FLEE does not apply in these cases, and $\mathcal{O}$'s dimension $\Delta$ controls the leading power of $L$ in $\s$ at small $L$. We consider only the three examples of ref.~\cite{Lucas:2014sba}, which all have $d<\dcrit$, and find several different behaviors as $T/\mu$ decreases, including both $\s \to s^{\pm}$ as $L \to \infty$, depending on the values of $\theta$ and $\zeta$, and area law violation at $T/\mu=0$, when $\theta = d-2$~\cite{Ogawa:2011bz,Huijse:2011ef}.

In sec.~\ref{translation} we consider the solution of ref.~\cite{Andrade:2013gsa}, namely gravity in $AdS_4$ coupled to a $U(1)$ gauge field and real, massless scalar ``axion'' fields scaling as $\g x$ with constant $\g$, which at $T=0$ is dual to an RG flow from a $d=3$ UV CFT driven by $\mu$ and a marginal $\Op$ with source $\g \, x$. The FLEE does not apply in this case, and at small $L$ we find $\s$ is a linear function of $L$ with slope $\propto\Ttt$ and \textit{non-zero} intercept $\propto \g^2$. When $\g=0$ the geometry reduces to AdS-RN, and we recover the results of sec.~\ref{adsrn} with $d=3<\dcrit$. When $\g/\mu \neq 0$ but $T/\mu=0$ the solutions of ref.~\cite{Andrade:2013gsa} are dual to a semi-local quantum liquid state, similar to AdS-RN with $T/\mu=0$, with $s \neq 0$. Indeed, as $T/\g$ decreases we find a transition similar to that of AdS-RN, from the lower curve in fig.~\ref{fig:schematic} to the upper curve.

Finally, in sec.~\ref{soliton} we consider the AdS soliton, namely $AdS_{d+1}$ with one direction $x$ compactified into a circle, with anti-periodic boundary conditions for fermions~\cite{Witten:1998zw,Horowitz:1998ha}. The compact direction shrinks to zero deep in the bulk, producing a ``hard wall,'' signaling mass gap generation and confinement in the dual QFT~\cite{Witten:1998zw}. The QFT also has negative Casimir energy, $\langle T_{tt} \rangle <0$~\cite{Horowitz:1998ha}. The FLEE does not apply in this case, nevertheless we find $\s \propto \langle T_{tt} \rangle L$ at small $L$. We find $\s<0$ for all $L$, and in particular as $L$ increases, $\s$ decreases to a minimum and then $\s \to 0^-$ as $L \to \infty$, similar to the relativistic RG flows of sec.~\ref{rg}.

In summary, we find area theorem violation in AdS-SCH at large $d$, AdS-RN at low $T/\mu$, some models with HV geometries, and the model of ref.~\cite{Andrade:2013gsa} at small $T/\g$. What do these all have in common? One obvious answer is: an IR fixed point that is not a $d$-dimensional CFT like the UV fixed point. In particular, the solutions of sec.~\ref{hyper} describe HV IR fixed points at $T/\mu=0$, while the other cases describe $(0+1)$-dimensional IR fixed points, meaning invariance under rescaling of $t$ but not $\vec{x}$~\cite{Iqbal:2011in,Iqbal:2011ae,Emparan:2013moa,Emparan:2013xia}, which can be interpreted as HV in the limit $\zeta \to \infty$ with $-\theta/\zeta$ fixed~\cite{Hartnoll:2012wm}. More precisely, in AdS-SCH when $d\to \infty$, in the near-horizon region $t$ and the holographic radial coordinate, $z$, form the $SL(2,\mathbb{R})/U(1)$ group manifold, while $\vec{x}$ forms $\mathbb{R}^{d-1}$~\cite{Emparan:2013xia}. In AdS-RN at $T/\mu=0$ or the model of ref.~\cite{Andrade:2013gsa} at $T/\g=0$, in the near-horizon region $t$ and $z$ form $AdS_2$ while the $\vec{x}$ form $\mathbb{R}^{d-1}$. As a result, in each near-horizon region, linearized fluctuations of fields transform covariantly under rescalings that act on $t$ but not $\vec{x}$~\cite{Emparan:2013moa,Emparan:2013xia}. Strictly speaking, such non-relativistic scale invariance occurs only for a limiting value of some parameter: $d = \infty$, $T/\mu=0$, etc. However, in our examples area theorem violation occurs at intermediate values of these parameters, as we dial them towards the limits. In other words, area theorem violation first occurs while the non-relativistic scale invariance is \textit{nascent}, \textit{i.e.} not yet exact, and hence signals the emergence of non-relativistic massless degrees of freedom.

\subsection{Outlook}
\label{sec:outlook}

Our results raise various questions for future research. For example, when does area theorem violation occur in holography? Is some version of non-relativistic scale invariance deep in the bulk necessary? If so, then for exactly what values of $d$, $\zeta$, and $\theta$? The near-horizon regions of extremal black branes generically have either $AdS_2$ or $AdS_3$~\cite{Kunduri:2007vf,Figueras:2008qh}. Do they always exhibit area theorem violation? We considered examples of $AdS_2$, but not $AdS_3$, which is dual to a CFT in $d=2$, which typically produces area law violation~\cite{Swingle:2010bt}. What about a more general holographic classification? Can the properties of the bulk metric that produce area theorem violation be fully characterized?

What about examples outside of holography? For example, what about SYK-type models~\cite{Sachdev:2015efa,kitaev}, which have $s\neq0$ at $T=0$ and $AdS_2$ IR scaling, similar to some of our examples? More generally, what about a complete classification? Can the conditions for area theorem violation be fully characterized? Is some form of non-relativistic scale invariance in the IR necessary? If so, does area theorem violation imply that degrees of freedom with non-relativistic scale invariance somehow count as ``more'' degrees of freedom than in a CFT? Even more generally, our results fit into a larger pattern, that various measures of quantum entanglement do not monotonically decrease under RG flow when Lorentz symmetry is broken~\cite{Swingle:2013zla}. Can the conditions for a measure of entanglement to be monotonic or not, in the absence Lorentz symmetry, be fully characterized?

Returning to our initial questions, our results suggests that $\s$ may indeed help characterize states of matter. For example, using $\s$'s small- and large $L$ behavior, we can classify states of matter into those in which the FLEE or area theorem applies or not, respectively. More generally, we can divide states of matter into those where $\s$ is monotonic, like the bottom curve in fig.~\ref{fig:schematic}, and those where $\s$ has one or more extrema, like the top curve in fig.~\ref{fig:schematic}. In the latter case, the location of the global extremum provides a characteristic length scale, namely the scale where the EE per unit volume is maximal or minimal. Such a characteristic length scale has various potential uses.

For example, in QFT length scales are typically defined as correlation lengths, extracted from correlators of local operators, and therefore cannot always be compared between QFTs, since the spectrum of operators is not universal. However, $\s$ can be compared between QFTs with different operator spectra. Consider for instance two holographic systems that each obey the FLEE and have a near-horizon $AdS_2 \times \mathbb{R}^{d-1}$, similar to AdS-RN at $T/\mu=0$. In each, $\s$ as a function of $L$ must have at least one maximum, one of which we assume is a global maximum, as in the top curve in fig.~\ref{fig:schematic}. Each dual field theory is in a semi-local quantum liquid state~\cite{Iqbal:2011in}, wherein space divides into ``patches'' of characteristic size $\ell$, defined from the behavior of local correlators: at separations $<\ell$, correlators exhibit the $(0+1)$-dimensional scale invariance of $AdS_2$, and at separations $>\ell$ they exhibit exponential decay~\cite{Iqbal:2011in}. (In extremal AdS-RN, $\ell \propto 1/\mu$.) If the two systems have different operators, then we cannot compare $\ell$ precisely. If we instead define $\ell$ from the maximum in $\s$, then we can.

Turning the holographic duality around, $\s$ can also help characterize geometries. For example, in a solution such as extremal AdS-RN, $\s$'s global maximum could provide a precise division between near- and far-horizon regions. We can also use $\s$ to characterize scaling geometries deep in the bulk or near a horizon, even away from the strict limit in which the geometry is scale invariant. Imagine for instance that we did not know the AdS-RN solution at $T/\mu=0$ (as often occurs when numerically solving for a metric). Area theorem violation would occur at finite $T/\mu$, not just at $T/\mu=0$, already suggesting that the extremal near-horizon geometry may have scale invariance, but cannot be $AdS_{d+1}$.

In sum, $\s$ is clearly useful for ``fingerprinting'' states of QFTs, holographic or otherwise. We therefore believe $\s$ deserves further exploration in future research.

\section{General Analysis}
\label{general} 

In most of our examples we can use the symmetries of translations in $t$ and translations and rotations in $\vec{x}$ to write the bulk metric in the form
\beq
\label{eq: 1}
ds^{2} = g_{MN} dx^M dx^N = \frac{R^{2}}{z^{2}}\left(-f(z)dt^{2}+d\vec{x}^2+\frac{dz^{2}}{g(z)}\right),
\eeq
where $M,N=0,1,\ldots d$. As $z \to 0$ the metric in eq.~\eqref{eq: 1} asymptotically approaches $AdS_{d+1}$ of radius $R$. More precisely, as $z \to 0$,
\beq
\label{eq:fgasymp}
f(z) = 1 - m z^d + \ldots, \qquad g(z) = 1 - m z^d + \ldots
\eeq
where $m$ is a constant and $\ldots$ represents powers of $z$ that go to zero as $z \to 0$ faster than those shown, and which in general are different in $f(z)$ and $g(z)$. Holographic renormalization~\cite{Balasubramanian:1999re,deHaro:2000xn} shows that $f(z)$'s asymptotic expansion determines the dual field theory's energy density:
\beq
\label{eq:energy_density}
\langle T_{tt} \rangle=\frac{(d-1)R^{d-1}}{16\pi G}\, m,
\eeq
where $G$ is the $(d+1)$-dimensional Newton's constant. The $AdS_{d+1}$ metric has $f(z)=1$ and $g(z)=1$, so in particular $m=0$ and hence $\Ttt=0$, as expected for a CFT vacuum state. As $z$ increases, \textit{i.e.}~as we move away from the boundary and into the bulk, the metric may approach that of another $AdS_{d+1}$, generically with different $R$ (sec.~\ref{rg}), or an HV geometry (sec.~\ref{hyper}), or a horizon, where $f(z_H)=0$, etc. In the case of a non-extremal horizon, the horizon's Hawking temperature and Bekenstein-Hawking entropy density determine the dual field theory's temperature and entropy density:
\beq
\label{eq:thermo_entropy}
T = \frac{\sqrt{f'(z_H)g'(z_H)}}{4\pi}, \qquad s = \frac{R^{d-1}}{4 G}\frac{1}{z_H^{d-1}},
\eeq
where $f'(z) \equiv \partial_z f(z)$, etc. Roughly speaking, $z$ corresponds to the RG scale in the dual field theory, with $z \to 0$ dual to the UV and large $z$ corresponding to the IR~\cite{Susskind:1998dq,Peet:1998wn}.

All our examples conform to the above, with the following exceptions. In the AdS-to-AdS domain walls of sec.~\ref{rg}, in $f(z)$ and $g(z)$'s expansions  the leading power of $z$ depends on $\Delta$, and in some cases is smaller than $z^d$. In secs.~\ref{adsrn} and ~\ref{translation}, the AdS-RN and AdS linear axion metrics, respectively, have \textit{extremal} horizons at $T=0$. Moreover, although the AdS linear axion metric is of the form in eq.~\eqref{eq: 1}, the linear axion itself breaks rotational and translational symmetry in $\vec{x}$. In sec.~\ref{soliton}, in the AdS soliton metric one coordinate of $\vec{x}$ is compactified, which breaks rotational symmetry, hence the metric is not of the form in eq.~\eqref{eq: 1}. We address each of these exceptions on a case-by-case basis.

As mentioned in sec.~\ref{intro}, for spacetimes with metrics of the form in eq.~\eqref{eq: 1}, we compute EE holographically via~\cite{Ryu:2006bv,Ryu:2006ef,Lewkowycz:2013nqa}
\beq
S=\frac{\amin}{4G},
\eeq
where $\amin$ is the area of the minimal surface in the spacetime at a fixed $t$ that approaches the entangling surface at the asymptotically $AdS_{d+1}$ boundary $z \to 0$.

\begin{figure}[t!]
    \centering
    \begin{subfigure}[b]{0.5\textwidth}
        \centering
        \includegraphics[width=\textwidth]{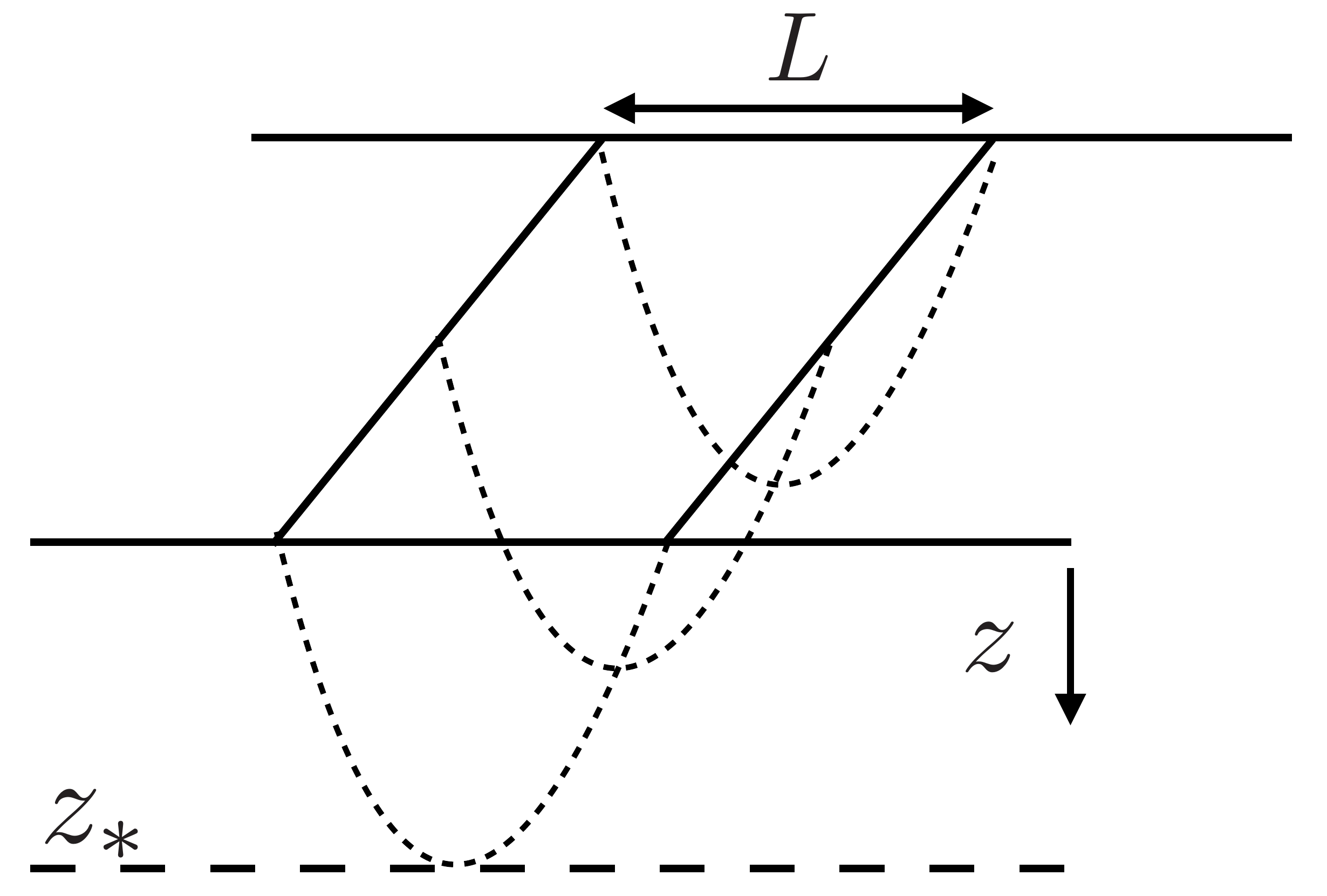}
        \caption{}
    \end{subfigure}%
    ~ 
    \begin{subfigure}[b]{0.5\textwidth}
        \centering
        \includegraphics[width=\textwidth]{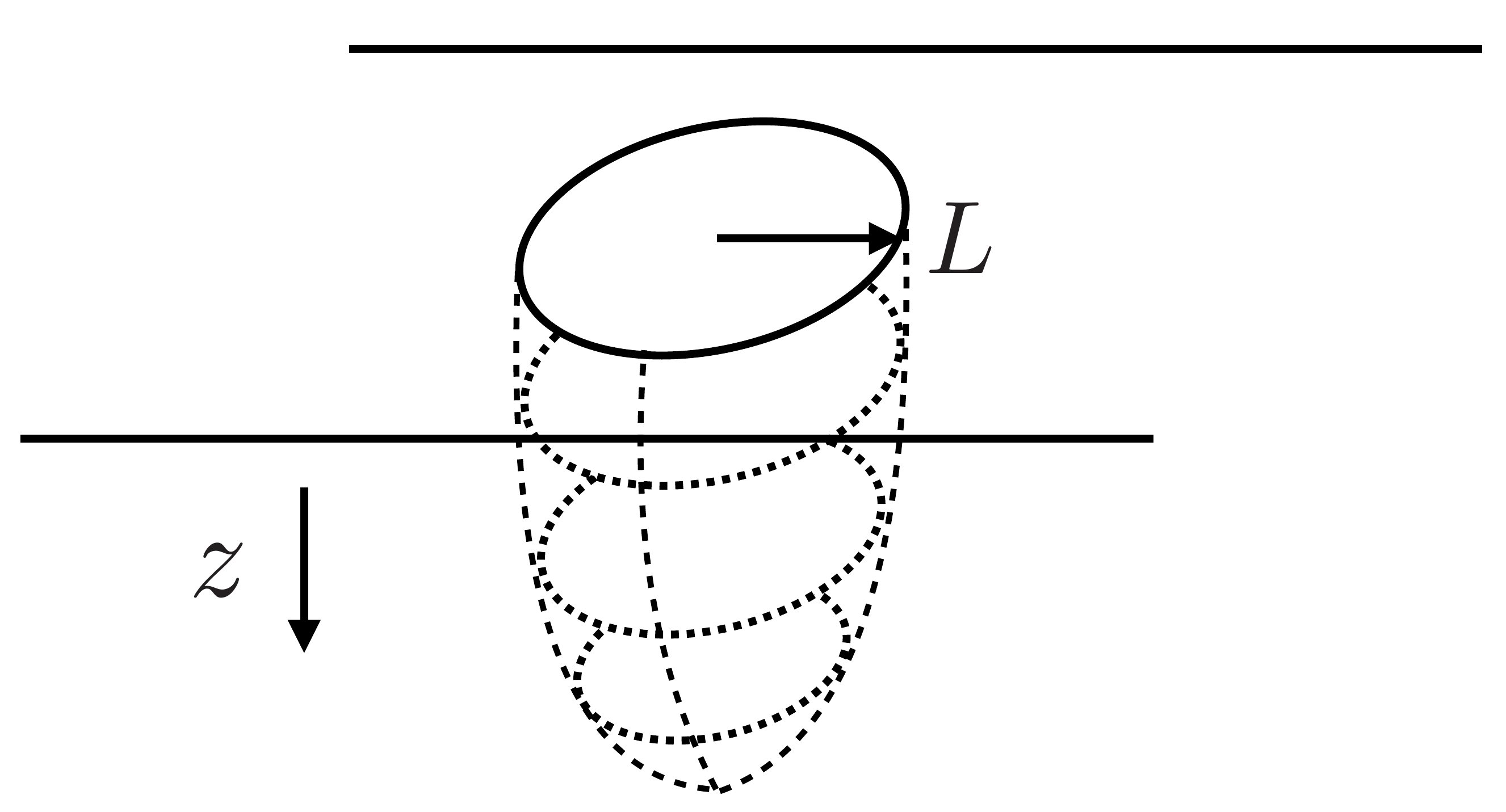}
        \caption{}
    \end{subfigure}
    \caption{\label{fig:minimalsurfaces}(a) Schematic depiction of the minimal surface for a strip sub-region of width $L$. The asymptotically $AdS_{d+1}$ boundary is at $z \to 0$. The minimal surface, depicted by the dashed lines, ``hangs down'' from the strip at the boundary to a maximal $z$ value, $z_*$. (b) Similar schematic depiction of the minimal surface for a sphere sub-region of radius $L$.}
\end{figure}

As also mentioned in sec.~\ref{intro}, we consider only strip and sphere sub-regions. The strip's entangling surface consists of two infinite parallel planes of spatial co-dimension one, \textit{i.e.} two copies of $\mathbb{R}^{d-2}$, separated by a distance $L$ in the remaining spatial direction, $x$. As is well-known~\cite{Ryu:2006bv,Ryu:2006ef}, using the translational and rotational symmetry of $\mathbb{R}^{d-2}$ we can parameterize the minimal surface as $x(z)$, and for metrics of the form in eq.~\eqref{eq: 1}, the area functional $\mathcal{A}$ depends only on $x'(z)^2$, leading to a first integral of motion. We can then solve for $x'(z)$ in terms of the first integral. The minimal surfaces ``hang down'' into the bulk to a largest $z$ value, $z_*$, the turn-around point where $x'(z)$ diverges, as depicted in fig.~\ref{fig:minimalsurfaces} (a). In short, we find a one-parameter family of solutions, where we can choose the one parameter to be either the first integral or $z_*$. We choose the latter. We then obtain $L$ by integrating $x'(z)$ from $z_*$ to the boundary,
\beq
\label{eq:ldef}
L=2\int_{0}^{z_{*}}dz\,\frac{z^{d-1}}{z_{*}^{d-1}}\frac{1}{\sqrt{1-(z/z_*)^{2(d-1)}}}\frac{1}{\sqrt{g(z)}},
\eeq
where the overall factor of $2$ appears because the solutions are invariant under the reflection $x(z) \to -x(z)$. The corresponding minimal area is
\beq
\label{eq:strip_area}
\amin^\mathrm{strip} = R^{d-1} \, 2\mathrm{Vol}(\mathbb{R}^{d-2}) \int_{\varepsilon}^{z_{*}}\frac{dz}{z^{d-1}}\frac{1}{\sqrt{1-(z/z_*)^{2(d-1)}}}\frac{1}{\sqrt{g(z)}},
\eeq
where the lower endpoint is a cutoff, $z = \varepsilon$, holographically dual to a UV cutoff. For $AdS_{d+1}$, where $g(z)=1$, we can perform the integrals in eqs.~\eqref{eq:ldef} and~\eqref{eq:strip_area} exactly, leading to
\begin{subequations}
\beq
\label{eq:strip_width}
L=\frac{\Gamma\left[\frac{d}{2(d-1)}\right]}{\Gamma\left[\frac{1}{2(d-1)}\right]}\, 2\sqrt{\pi}\,z_{*}.
\eeq
\beq
\label{eq:strip_vacuum_ee}
S^{\mathrm{strip}}_{\textrm{CFT}}=\frac{R^{d-1}}{4G}\left[\frac{1}{(d-2)}\frac{2\mathrm{Vol}(\mathbb{R}^{d-2})}{\varepsilon^{d-2}}-\frac{2^{d-2}\pi^{\frac{d-1}{2}}}{(d-2)}\left(\frac{\Gamma\left(\frac{d}{2(d-1)}\right)}{\Gamma\left(\frac{1}{2(d-1)}\right)}\right)^{d-1}\frac{2\mathrm{Vol}(\mathbb{R}^{d-2})}{L^{d-2}}\right].
\eeq
\end{subequations}
In eq.~\eqref{eq:strip_vacuum_ee} we see the form described below eq.~\eqref{eq:EElargeL}: an area law with $A = 2\textrm{Vol}\left(\mathbb{R}^{d-2}\right)$, where $\a$ is a sum of two terms, one $\propto 1/\varepsilon^{d-2}$ and the other $\propto 1/L^{d-2}$.

For the sphere sub-region we first write $d\vec{x}^2 = dr^2 + r^2 ds^2_{S^{d-2}}$, where $r$ is the radial coordinate and $ds^2_{S^{d-2}}$ is the metric of a round unit-radius $S^{d-2}$, and then parameterize the minimal surface as $r(z)$. The resulting area functional is
\beq
\label{eq:sphere_area}
\mathcal{A}^{\textrm{sphere}} = R^{d-1}\mathrm{Vol}(\mathrm S^{d-2}) \int_{\varepsilon}^{z_{*}} dz \, \frac{r(z)^{d-2}}{z^{d-1}}\sqrt{r'(z)^{2}+ \frac{1}{g(z)}},
\eeq
where \(\mathrm{Vol}(S^{d-2})\) is the volume of $S^{d-2}$. Extremizing $\mathcal{A}^\mathrm{sphere}$ leads to a non-linear second order ordinary differential equation for $r(z)$. For $AdS_{d+1}$, where $g(z)=1$, the exact solution is $r(z) = \sqrt{L^2 - z^2}$, leading to
\beq
\label{eq:sphere_vacuum_ee}
	S^\mathrm{sphere}_{\textrm{CFT}} = \begin{cases}
	\displaystyle{
		\frac{R^{d-1}\mathrm{Vol}(S^{d-2})}{4G} \left[
			\sum_{j=1}^{(d-2)/2} c_j \left(\frac{L}{\varepsilon}\right)^{d-2j}
			+ c_L \log \left(\frac{L}{\varepsilon}\right)
			+ c_0
			+ \mathcal{O}\left(\frac{\varepsilon^2}{L^2}\right)
		\right]
	}, &(d~\mathrm{even})
	\\[1.5em]
	\displaystyle{
		\frac{R^{d-1}\mathrm{Vol}(S^{d-2})}{4G}\left[
			\sum_{j=1}^{(d-1)/2} c_j \left(\frac{L}{\varepsilon}\right)^{d-2j}
			+ \tilde{c}_0
			+ \mathcal{O}\left(\frac{\varepsilon}{L}\right)
		\right]
	}, &(d~\mathrm{odd})
	\end{cases} \nn
\eeq
\beq
c_j=\frac{(-1)^{j-1}\Gamma\left[\frac{d-1}{2}\right]}{(d-2j)\Gamma\left[\frac{d-2j+1}{2}\right]\Gamma[j]},\qquad c_L = \frac{(-1)^{\frac{d-2}{2}}\Gamma\left[\frac{d-1}{2}\right]}{\sqrt{\pi}\,\Gamma\left[\frac{d}{2}\right]},
\eeq
\beq
c_0=\frac{(-1)^{\frac{d-1}{2}}\sqrt{\pi}\Gamma\left[\frac{d-1}{2}\right]}{2\,\Gamma\left[\frac{d}{2}\right]},\qquad \tilde{c}_0=\frac{(-1)^{\frac{d-2}{2}}\Gamma\left[\frac{d-1}{2}\right]}{2\sqrt{\pi}\,\Gamma\left[\frac{d}{2}\right]}\left(\psi\left[\frac{d}{2}\right]+\gamma_{E}+2\log[2]\right), \nn
\eeq
where $\psi[d/2]$ is a Digamma function and $\gamma_{E}\approx0.577$ is the Euler-Mascheroni constant. For the $g(z)$ in our examples, we have only been able to solve $r(z)$'s equation of motion numerically, using straightforward shooting algorithms. A schematic depiction of the resulting minimal surfaces appears in fig.~\ref{fig:minimalsurfaces} (b).

More generally, for a given $g(z)$ in one of our examples we compute $\s$ as follows. First, we compute $S$ numerically, meaning for the strip we choose $z_*$ and then integrate eqs.~\eqref{eq:ldef} and~\eqref{eq:strip_area} numerically, while for the sphere we solve for $r(z)$ numerically and then plug the solution into eq.~\eqref{eq:sphere_area} and integrate numerically. Next, we subtract the corresponding $S_{\textrm{CFT}}$ from eq.~\eqref{eq:strip_vacuum_ee} or~\eqref{eq:sphere_vacuum_ee}. Finally, we divide by
\beq
V = \begin{cases} \textrm{Vol}\left(\mathbb{R}^{d-2}\right) L,  & \textrm{(strip)} \\ \frac{\pi^{\frac{d-1}{2}}}{\Gamma\left(\frac{d+1}{2}\right)} L^{d-1}. & \textrm{(sphere)}\end{cases}
\eeq

We can determine $\s$'s small-$L$ behavior following ref.~\cite{Bhattacharya:2012mi}. If $L$ is small compared to all other length scales except $\varepsilon$, and in particular if $m L^d \ll 1$, then we can solve for the minimal surface order-by-order in a small-$(m L^d)$ expansion, and expand the integrands in eqs.~\eqref{eq:ldef}, \eqref{eq:strip_area}, and \eqref{eq:sphere_area} in $m L^d$ and integrate order-by-order, ultimately leading to an expansion of $S$ in powers of $mL^d$. Via eq.~\eqref{eq:eddef} we then find
\beq
\s = \Ttt \, \tent^{-1} + \mathcal{O}\left(\Ttt^2 L^{d+1}\right)
\eeq
where for the strip
\beq
\label{eq:striptent}
\tent = \frac{2 (d^2-1) \Gamma\left(\frac{d}{2(d-1)}\right)^2 \Gamma\left(\frac{d+1}{2(d-1)}\right)}{\sqrt{\pi} \, \Gamma\left(\frac{1}{2(d-1)}\right)^2 \Gamma\left(\frac{1}{d-1}\right)} \, \frac{1}{L},
\eeq
and for the sphere $\tent = \frac{d+1}{2\pi} \, \frac{1}{L}$. In short, $\s \propto \Ttt \, L$ at small $L$.

For bulk spacetimes with a horizon, we can determine $\s$'s large-$L$ behavior following refs.~\cite{Liu:2012eea,Liu:2013una,Kundu:2016dyk}. In eq.~\eqref{eq:strip_area} for $\amin^{\textrm{strip}}$, in order to extract the terms that diverge as $\varepsilon \to 0$, we add and subtract $1/z^{d-1}$ to the integrand, and integate over $1/z^{d-1}$,
\beq
\amin^{\textrm{strip}} = R^{d-1} \, 2\mathrm{Vol}(\mathbb{R}^{d-2}) \left [ \frac{1}{d-2} \left(\frac{1}{\varepsilon^{d-2}}-\frac{1}{z_*^{d-2}}\right)+ \int_0^{z_{*}}\frac{dz}{z^{d-1}}\left(\frac{1}{\sqrt{1-(z/z_*)^{2(d-1)}}}\frac{1}{\sqrt{g(z)}}-1\right)\right], \nn
\eeq
where we took $\varepsilon \to 0$ at the lower endpoint of the integral, which is now finite (because $g(z)$ obeys eq.~\eqref{eq:fgasymp}). We next change the integration variable from $z$ to $u = z/z_*$,
\beq
\label{eq:aminstripucoord}
\amin^{\textrm{strip}} = R^{d-1} \, 2\mathrm{Vol}(\mathbb{R}^{d-2}) \left [ \frac{1}{d-2} \left(\frac{1}{\varepsilon^{d-2}}-\frac{1}{z_*^{d-2}}\right)+ \frac{1}{z_*^{d-2}}\int_0^1\frac{du}{u^{d-1}}\left(\frac{1}{\sqrt{1-u^{2(d-1)}}}\frac{1}{\sqrt{g(z_*u)}}-1\right)\right].
\eeq
Our immediate goal is now to re-write the integral, as much as possible, in terms of that for $L$ from eq.~\eqref{eq:ldef}, written with the coordinate $u$,
\beq
L = 2 z_* \int_0^1 du \frac{u^{d-1}}{\sqrt{1-u^{2(d-1)}}} \frac{1}{\sqrt{g(z_*u)}}.
\eeq
To do so, in the integrand of eq.~\eqref{eq:aminstripucoord} we take
\bea
\frac{u^{-(d-1)}}{\sqrt{1-u^{2(d-1)}}} & = & \frac{u^{-(d-1)}-u^{d-1}+u^{d-1}}{\sqrt{1-u^{2(d-1)}}} = \frac{u^{-(d-1)}(1-u^{2(d-1)})+u^{d-1}}{\sqrt{1-u^{2(d-1)}}} \nn \\ & = & u^{-(d-1)}\sqrt{1-u^{2(d-1)}}+\frac{u^{d-1}}{\sqrt{1-u^{2(d-1)}}}, \label{eq:umanip}
\eea
which allows us to re-write eq.~\eqref{eq:aminstripucoord} as
\beq
\label{eq:stripamin}
\amin^{\textrm{strip}} = R^{d-1} \, 2\mathrm{Vol}(\mathbb{R}^{d-2}) \left [ \frac{1}{d-2} \left(\frac{1}{\varepsilon^{d-2}}-\frac{1}{z_*^{d-2}}\right) + \frac{1}{2}\frac{L}{z_*^{d-1}} + \frac{1}{z_*^{d-2}}\int_0^1\frac{du}{u^{d-1}}\left(\sqrt{\frac{1-u^{2(d-1)}}{g(z_* u)}}-1\right)\right].
\eeq
Collecting the $1/z_*^{d-2}$ terms, we find
\beq
\amin^{\textrm{strip}} = R^{d-1} \, 2\mathrm{Vol}(\mathbb{R}^{d-2}) \left [ \frac{1}{d-2} \frac{1}{\varepsilon^{d-2}}+  \frac{1}{2}\frac{L}{z_*^{d-1}} + \frac{C(z_*)}{z_*^{d-2}}\right], \nn
\eeq
with the dimensionless coefficient
\beq
\label{eq:cdef}
C(z_*) \equiv - \frac{1}{d-2} + \int_0^1\frac{du}{u^{d-1}} \left(\sqrt{\frac{1-u^{2(d-1)}}{g(z_* u)}} - 1 \right).
\eeq
Dividing by $4G$ to obtain $S^{\textrm{strip}}$, subtracting $S^{\textrm{strip}}_{\textrm{CFT}}$ in eq.~\eqref{eq:strip_vacuum_ee}, and dividing by $V = \textrm{Vol}\left(\mathbb{R}^{d-2}\right) L$, we obtain the ED,
\beq
\label{eq:strip_ent_density_alt}
\sigma^{\textrm{strip}} = \frac{R^{d-1}}{4 G} \left[ \frac{1}{z_*^{d-1}} + \frac{C(z_*)}{z_*^{d-2}}\frac{2}{L} + \frac{2^{d-2}\pi^{\frac{d-1}{2}}}{(d-2)}\left(\frac{\Gamma\left(\frac{d}{2(d-1)}\right)}{\Gamma\left(\frac{1}{2(d-1)}\right)}\right)^{d-1}\frac{2}{L^{d-1}} \right].
\eeq
So far we took no limits of $L$, \textit{i.e.} eq.~\eqref{eq:strip_ent_density_alt} is valid for any $L$. As $L \to \infty$, we expect the minimal surface to probe deep into the bulk, and eventually to lie flat along the horizon,\footnote{In fact, for sufficiently large $L$ two solutions for $x(z)$ may exist. The first is our solution, described above. The second consists of two segments with constant $x(z)$, stretching from the boundary to the horizon, which must be connected by a third segment along the horizon, since minimal surfaces cannot cross a horizon~\cite{Hubeny:2012ry}. The third segment contributes zero to the area. However, in all our examples with horizons we have checked explicitly that the latter solution always has larger area than our solution, \textit{i.e.} is not the global minimum of the area functional, and hence may be safely ignored.} so that in particular $\lim_{L \to \infty} z_* = z_H$. In that case eq.~\eqref{eq:strip_ent_density_alt} gives, using eq.~\eqref{eq:thermo_entropy},
\beq
\lim_{L \to \infty} \s^{\textrm{strip}} = \frac{R^{d-1}}{4 G} \frac{1}{z_H^{d-1}} = s.
\eeq
We thus find that the leading term in $\s$'s large-$L$ expansion is the entropy density $s$, as expected. The leading $1/L$ correction is also straightforward to obtain: our examples have $d \geq 3$, so the final term in eq.~\eqref{eq:strip_ent_density_alt} is sub-leading, and thus
\beq
\label{eq:striplargeL}
\s^{\textrm{strip}} = s + s\,z_H \, C(z_H)\, \frac{2}{L} + \mathcal{O}\left(\frac{1}{L^2}\right).
\eeq
For the strip, $A/V = 2/L$, hence eq.~\eqref{eq:striplargeL} is of the form in eq.~\eqref{eq:EDlargeL},
\beq
\label{eq:EDlargeL2}
\s = s - \Delta \a \, \frac{A}{V} + \ldots,
\eeq
where we identify
\beq
\label{eq:alphaholo}
\Delta \a = -s\,z_H \, C(z_H).
\eeq

For the sphere, following ref.~\cite{Liu:2013una}, we solve for the minimal surface in two regimes, $r(z) \approx L$ and then, switching parameterization to $z(r)$, also $z(r) \gtrsim z_H$, and match the solutions at large $L$, where the two regimes overlap. The details are practically identical to those in ref.~\cite{Liu:2013una}, so for brevity we omit them. Ultimately, we again find the form of $\s$ in eq.~\eqref{eq:EDlargeL2}, with $\Delta \a$ again given by eq.~\eqref{eq:alphaholo}. To be clear, $z_*$ is not defined for the sphere, hence $C(z_*)$ in eq.~\eqref{eq:cdef}  is not defined. However, in $\s$'s leading large-$L$ correction, for the sphere we find exactly the same integral as $C(z_H)$, and hence exactly eq.~\eqref{eq:alphaholo}. Such agreement between the strip and sphere at $L\to\infty$ is intuitive, since we expect the $L \to \infty$ limit to suppress any effects from the entangling surface's curvature. In short, $C(z_H)$ determines whether $\s \to s^{\pm}$ as $L\to \infty$, for both the strip and sphere. 

The area theorem of refs.~\cite{Casini:2012ei,Casini:2016udt} requires $\Delta \a >0$ and hence $C(z_H) \propto - \Delta \a <0$. Strictly speaking, the proofs of the area theorem in refs.~\cite{Casini:2012ei,Casini:2016udt} were only for Lorentz-invariant RG flows, and only for the sphere. However, $C(z_H)$ is identical for the sphere and the strip, so the proofs in refs.~\cite{Casini:2012ei,Casini:2016udt} imply an area theorem for the strip as well, in holographic systems describing Lorentz-invariant RG flows.

Eq.~\eqref{eq:alphaholo} is the main novel result of this section, and allows us to test for area theorem violation simply by computing $C(z_H)$'s sign: if $C(z_H)<0$ then $\Delta \a >0$ and the area theorem is obeyed, while if $C(z_H)>0$ then $\Delta \a <0$ and the area theorem is violated.

\section{AdS-to-AdS Domain Walls}
\label{rg}

In this section we consider a bulk action
\beq
\label{eq:holo_rg_action}
\sbulk = \frac{1}{16 \pi G} \int d^{d+1}x \sqrt{-\detg} \left[\mathcal{R} -  \frac{1}{2} \partial_M \phi \partial^M \phi - V(\phi) \right].
\eeq
where $\mathcal{R}$ is the Ricci scalar and $\phi$ is a real scalar field with potential $V(\phi)$. We want solutions to the equations of motion derived from $\sbulk$ that describe Lorentz-invariant RG flows between CFTs, driven by the scalar operator $\mathcal{O}$ holographically dual to $\phi$. We thus assume $V(\phi)$ has (at least) two stationary points, at which the equations of motion reduce to those of pure $AdS_{d+1}$ with radius of curvature $R$ given by
\beq
8 \pi G \left . V(\f) \right |_{\textrm{stationary}}= - \frac{d(d-1)}{2 R^2}.
\eeq
Domain-wall solutions that interpolate between an asymptotic $AdS_{d+1}$, dual to the UV CFT, and another $AdS_{d+1}$ deep in the bulk, dual to the IR CFT, have the form
\beq
ds^2 = g_{MN}dx^Mdx^N=\frac{R^2}{z^2} \left( -dt^2 + d\vec{x}^2 + \frac{dz^2}{g(z)} \right), \qquad \phi=\phi(z),
\eeq
with $0 \leq z < \infty$. Following refs.~\cite{Freedman:1999gp,Ammon:2015wua}, if we introduce a ``superpotential'' $W$ via
\beq
\label{eq:VW}
V(\phi) = \frac{1}{16 \pi G} \left(\partial_{\phi}W\right)^2 - \frac{1}{2} \frac{d}{d-1} W^2,
\eeq
then any solution to the equations of motion derived from $\sbulk$ is also a solution to~\cite{Freedman:1999gp}
\beq
\label{eq:superpotential}
\phi' = \frac{d-1}{8 \pi G} \frac{1}{z W} \, \partial_{\phi}W, \qquad g(z) = \frac{8 \pi G}{(d-1)^2} \, R^2 \, W^2.
\eeq
We therefore only need to solve the first-order eq.~\eqref{eq:superpotential}. In fact, for our purposes, we can \textit{choose} $g(z)$, which then determines $W$ and hence $\phi(z)$ via eq.~\eqref{eq:superpotential}, which in turn is guaranteed to solve the equations of motion for the corresponding potential $V(\phi)$ in eq.~\eqref{eq:VW}.

Crucially, $g(z)$ obeys several constraints. For instance, eq.~\eqref{eq:superpotential} implies
\beq
\label{eq:rg_flow_g_condition_1}
\phi'(z)^2 = \frac{d-1}{16 \pi G} \frac{g'(z)}{z g(z)},
\eeq
so that $g'(z)\geq 0$, since by assumption $g(z)>0$. The NEC also requires $g'(z)\geq 0$, so any solution of eq.~\eqref{eq:superpotential} is guaranteed to obey the NEC. We also want $\mathcal{O}$ to be relevant, $\D<d$, and unitary, $\Delta \geq \frac{d-2}{2}$, and moreover we want to avoid poorly-understood UV divergences in the EE that the subtraction $S - \scft$ does not cancel, hence we restrict to $\Delta < (d+2)/2$~\cite{Liu:2013una,Casini:2016udt}. We demand that asymptotically $\phi(z) = \phi_0 z^{\Delta_-}+\ldots$, where $\Delta_-=\mathrm{Min}(d-\D,\Delta)$, $\phi_0$ is proportional either to $\Op$'s source ($\Delta_-=d-\Delta$) or to $\Opv$ ($\Delta_-=\Delta$), and $\ldots$ represents terms with higher powers of $z$. Via eq.~\eqref{eq:rg_flow_g_condition_1}, $g(z)$'s asymptotic expansion is then
\beq
\label{eq:rg_flow_g_condition_2}
g(z) = 1 + \frac{8 \pi G \D_-}{d-1} \, \phi_0^2 \, z^{2\D_{-}} + \ldots,
\eeq
where again the $\ldots$ represents terms with higher powers of $z$.

The FLEE does not apply to these solutions because on the gravity side $g(z)$ does not have the asymptotics in eq.~\eqref{eq:fgasymp}, and on the field theory side we introduce a source for $\mathcal{O}$. However, with the assumptions above, for the strip we can determine $\s$'s small-$L$ behavior by expanding eq.~\eqref{eq:strip_area} for $\mathcal{A}_{\textrm{min}}^{\textrm{strip}}$ in small $z_*$, that is, for a minimal surface close to the asymptotic $AdS_{d+1}$ boundary. Expanding also eq.~\eqref{eq:strip_width} for $L$ in small $z_*$, inverting order-by-order, and plugging the result into the expansion for $\mathcal{A}_{\textrm{min}}^{\textrm{strip}}$ gives the leading small-$L$ behavior
\beq
\label{eq:rgsmallL}
\s = \frac{2 \pi^{3/2}\D_- \G \left(\frac{2\D_- + d}{2d-2} \right)}{(2 \D_- + 1)(2 \D_-  + 2 - d) \G \left(\frac{2\D_- + 1}{2d-2}\right)} \left(\frac{ \G \left(\frac{1}{2d-2} \right) }{2 \sqrt{\pi} \G \left( \frac{d}{2d-2} \right) }\right)^{2\D_- + 2 - d} \f_0^2 \, R^{d-1} \, L^{2\D_- + 1 - d}+\ldots,
\eeq
where $\ldots$ represents terms with higher powers of $L$. When $\phi_0$ is proportional to $\Op$'s source, the area theorem requires $\s \to 0^-$ as $L \to \infty$, for both the strip and sphere. Our examples will conform to these limits.

EE in holographic RG flows has been studied in detail before, for example in refs.~\cite{Albash:2011nq,Myers:2012ed,Liu:2012eea,Liu:2013una}, so we focus only on a few cases that illustrate some of $\s$'s possible behaviors in $L$. In particular, we restrict to $d=4$ and choose
\begin{subnumcases}{\label{eq:rgchoices} g(z)=} 1 + \tanh^4(\beta z), \label{eq:rg_flow_single_peak} \\  1 + \tanh^4(\beta z)  + \frac{3}{2}\tanh(\beta z - 2) \tanh^5(\beta z),\label{eq:holo_rg_double_trough} \\ 1 + \tanh^4(\beta z) + \frac{20(\beta z - 1)^2 + 1}{(\beta z - 1)^2 + 1} \left[1 + \tanh(\beta z) \right]\tanh^4(\beta z),\label{eq:holo_rg_transition} \\ 1 + \tanh^3(\beta z), \label{eq:rg_flow_constant} \\ 1 + \tanh^{7/2}(\beta z), \label{eq:rg_flow_fractional_power}
\end{subnumcases}
where in each case $\beta$ is a constant of mass dimension one, which may be related to $\phi_0$ via eq.~\eqref{eq:rg_flow_g_condition_2}. Table~\ref{tab:rgchoices} summarizes some properties of our choices of $g(z)$. In table~\ref{tab:rgchoices}, the second column is $R_{\textrm{IR}}$, the value of the $AdS_5$ radius at $z \to \infty$, determined by the value of $\lim_{z \to \infty} g(z)$. The holographic $c$-theorem~\cite{Freedman:1999gp} requires $R_{\textrm{IR}}\leq R$. The third column shows $g(z)$'s leading asymptotic powers of $z$, which via eq.~\eqref{eq:rg_flow_g_condition_2} determines $\Delta_-$, listed in the fourth column, with the corresponding $\Delta$ in the fifth column. The sixth column indicates whether $\phi_0$ is proportional to $\Op$'s source or to $\Opv$. For $g(z)$ in eqs.~\eqref{eq:rg_flow_single_peak} to~\eqref{eq:holo_rg_transition}, $\phi(z)$ saturates the Breitenlohner-Freedman bound, hence $\phi(z)$'s leading asymptotic terms are $z^{\Delta_-}$ and $z^{\Delta_-}\log(z)$, however, we demand that the coefficient of the $\log(z)$ term vanish, so that in standard quantization $\phi_0 \propto \Opv$. In these cases, the RG flow is driven by $\Opv\neq 0$ alone, with zero source, similar to the RG flow on the moduli space of a supersymmetric theory.

\begin{table}[ht!]
\centering
\begin{tabular}[c]{|c|c|c|c|c|c|}
\hline
$g(z)$ & $R_{\textrm{IR}}$ & Asymptotics & $\Delta_-$ & $\Delta$ & $\phi_0$ \\ \hline
\eqref{eq:rg_flow_single_peak} & $R/\sqrt{2}$ & $1+\left(\beta z\right)^4+\ldots$ & $2$ & $2$ & $\propto \Opv$ \\ \hline
\eqref{eq:holo_rg_double_trough} & $R/\sqrt{7/2}$ & $1+\left(\beta z\right)^4+\ldots$ & $2$ & $2$ & $\propto \Opv$ \\ \hline
\eqref{eq:holo_rg_transition} & $R/\sqrt{42}$ & $1+\frac{23}{2}\left(\beta z\right)^4+\ldots$ & $2$ & $2$ & $\propto \Opv$ \\ \hline
\eqref{eq:rg_flow_constant} & $R/\sqrt{2}$ & $1+\left(\beta z\right)^3+\ldots$ & $3/2$ & $5/2$ & source \\ \hline
\eqref{eq:rg_flow_fractional_power} & $R/\sqrt{2}$ & $1+\left(\beta z\right)^{7/2}+\ldots$ & $7/4$ & $9/4$ & source \\ \hline
\end{tabular}
\caption{\label{tab:rgchoices} Summary of properties of our choices of $g(z)$ in eq.~\eqref{eq:rgchoices}.}
\end{table}

\FIGURE{
\begin{tabular}{c c}
\includegraphics[width=0.4\textwidth]{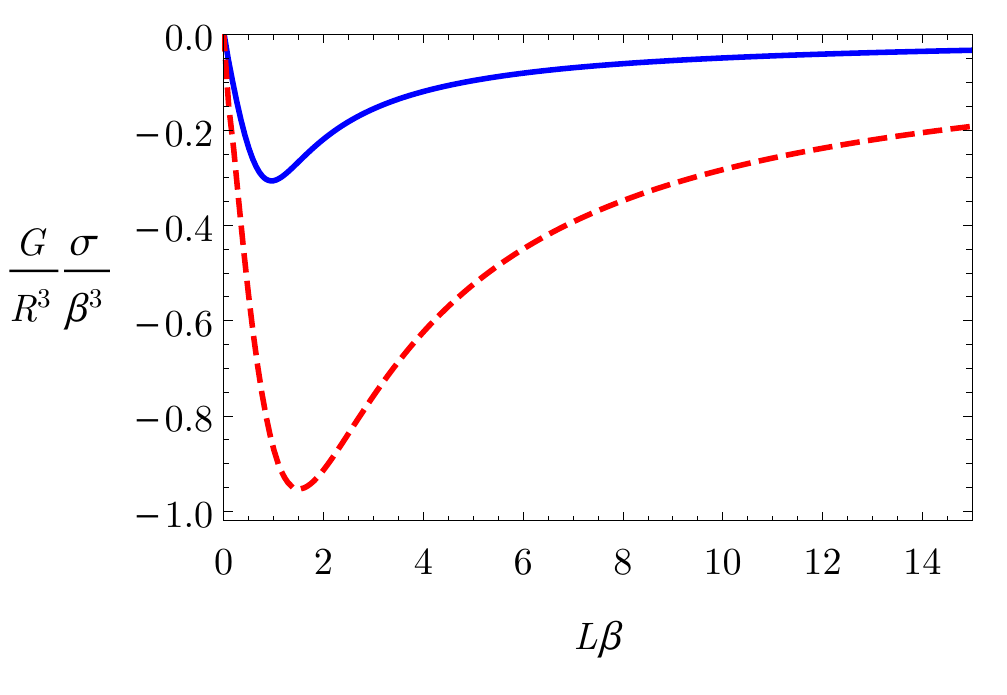}
&
\includegraphics[width=0.4\textwidth]{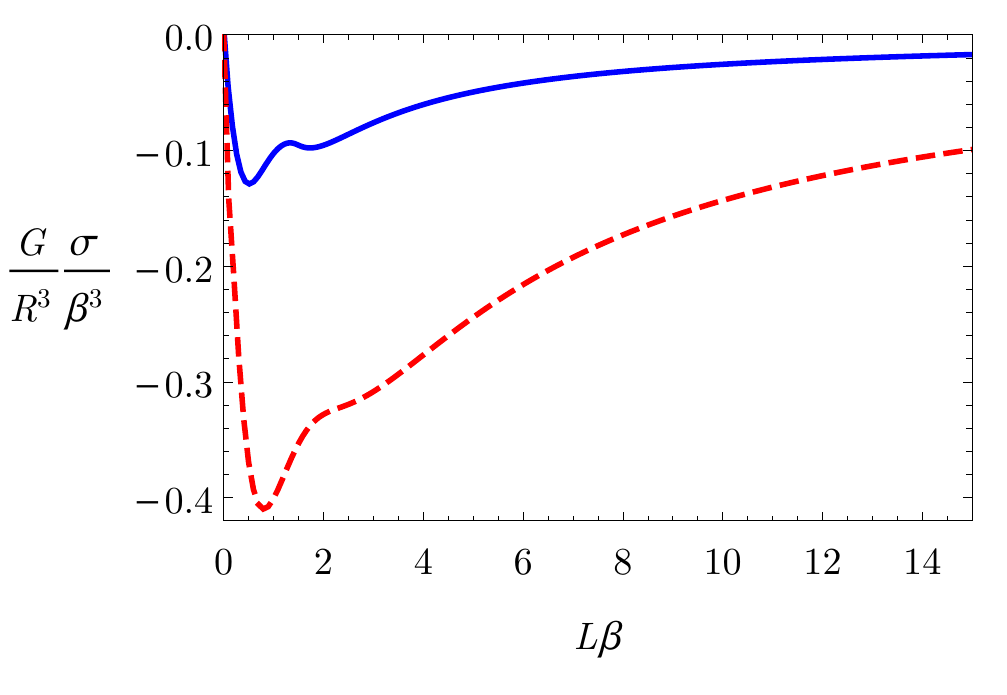}
\\
(a) eq.~\eqref{eq:rg_flow_single_peak}
&
(b) eq.~\eqref{eq:holo_rg_double_trough}
\\
\includegraphics[width=0.4\textwidth]{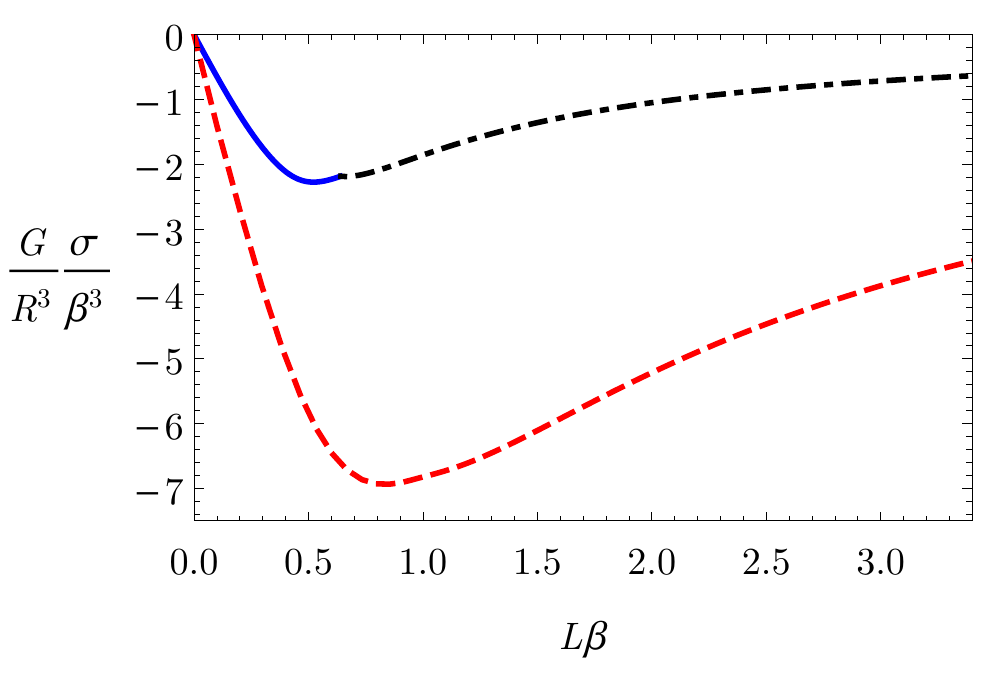}
&
\includegraphics[width=0.4\textwidth]{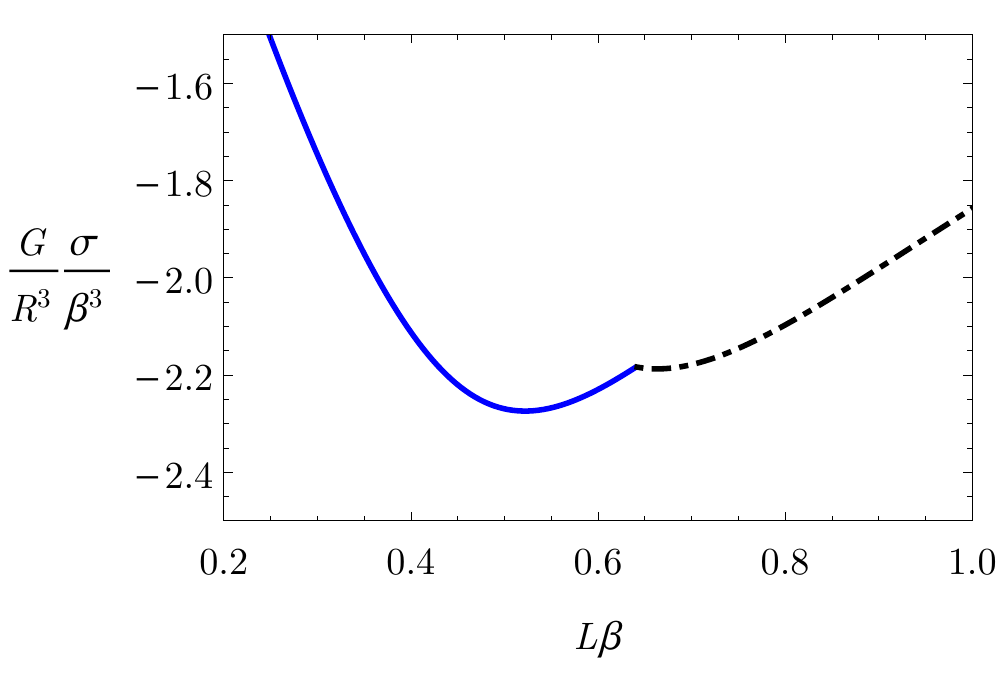}
\\
(c) eq.~\eqref{eq:holo_rg_transition}
&
(d) eq.~\eqref{eq:holo_rg_transition}
\\
\includegraphics[width=0.4\textwidth]{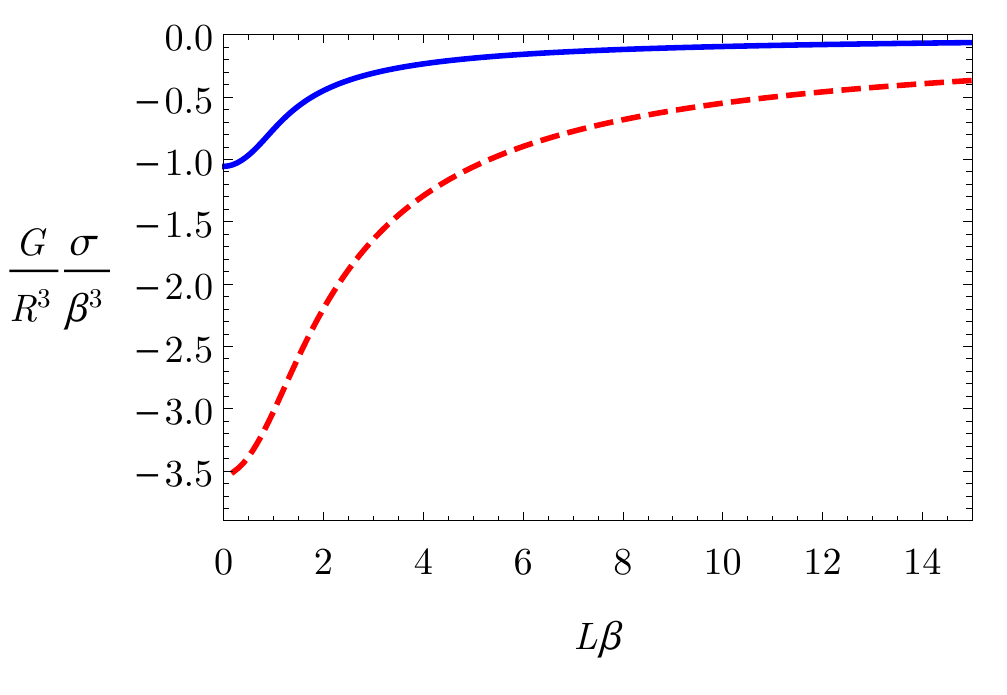}
&
\includegraphics[width=0.4\textwidth]{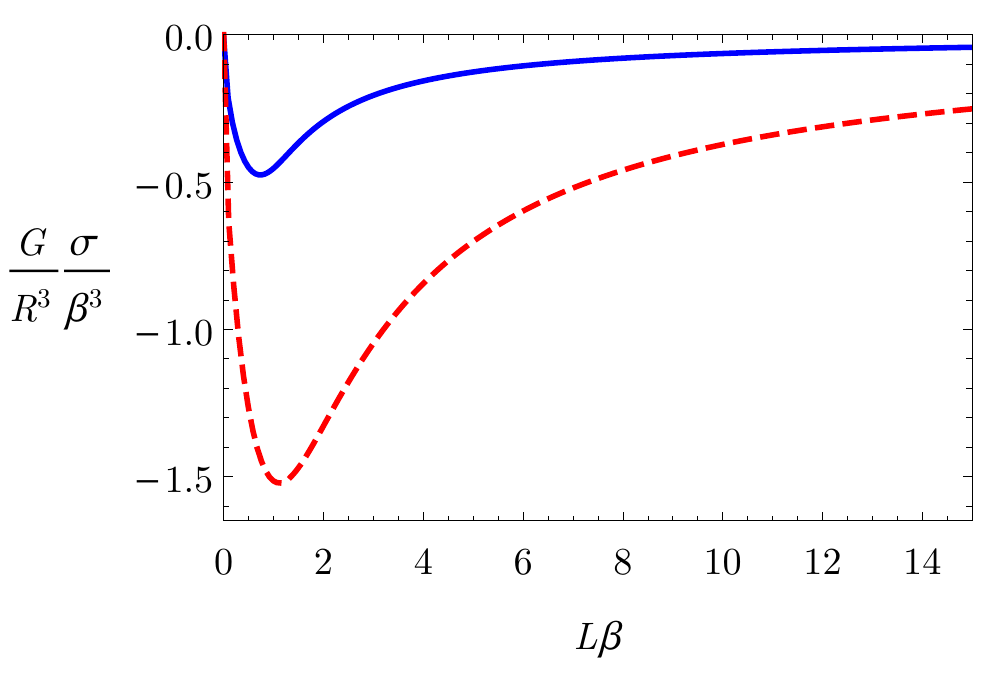}
\\
(e) eq.~\eqref{eq:rg_flow_constant}
&
(f) eq.~\eqref{eq:rg_flow_fractional_power}
\end{tabular}
\caption{\label{fig:rged} The ED, $\s$, in units of $\beta^3 R^3/G$, versus $L\beta$ for RG flows between holographic CFTs in $d=4$. In each plot, the blue solid line is for the strip and the red dashed line is for the sphere. The label below each plot indicates the $g(z)$ we chose from eq.~\eqref{eq:rgchoices}. For the $g(z)$ in eq.~\eqref{eq:holo_rg_transition}, and for the strip only, a ``first-order phase transition'' occurs between different extremal surfaces in the bulk when $L\beta\approx0.65$, leading to the kink in $\s$ shown in (c), and in close-up in (d), where the blue solid and black dot-dashed curves meet.}
}

Fig.~\ref{fig:rged} shows our numerical results for $\s$ as a function of $L$. More specifically, we plot $\s$ in units of $\beta^3 R^3/G$, where $R^3/G$ is the UV CFT's central charge~\cite{Ammon:2015wua}, versus $L$ in units of $\beta$. In all cases, $\s<0$ for all $L\beta$, with $\s \to 0^-$ as $L\beta \to \infty$, as required by the area theorem.

Fig.~\ref{fig:rged} (a) shows the simplest behavior, for the $g(z)$ in eq.~\eqref{eq:rg_flow_single_peak}, in which $\sigma \propto - L$ at small $L$, and then a single minimum appears before $\s \to 0^-$ as $L\beta \to \infty$, for both the strip and sphere. Fig.~\ref{fig:rged} (b), for the $g(z)$ in eq.~\eqref{eq:holo_rg_double_trough}, is similar, but with a second, local minimum, and corresponding local maximum, at intermediate $L$, for both the strip and sphere.

\begin{figure}[t]
  \centering
  $\begin{array}{cc}
  \includegraphics[width=0.5\textwidth]{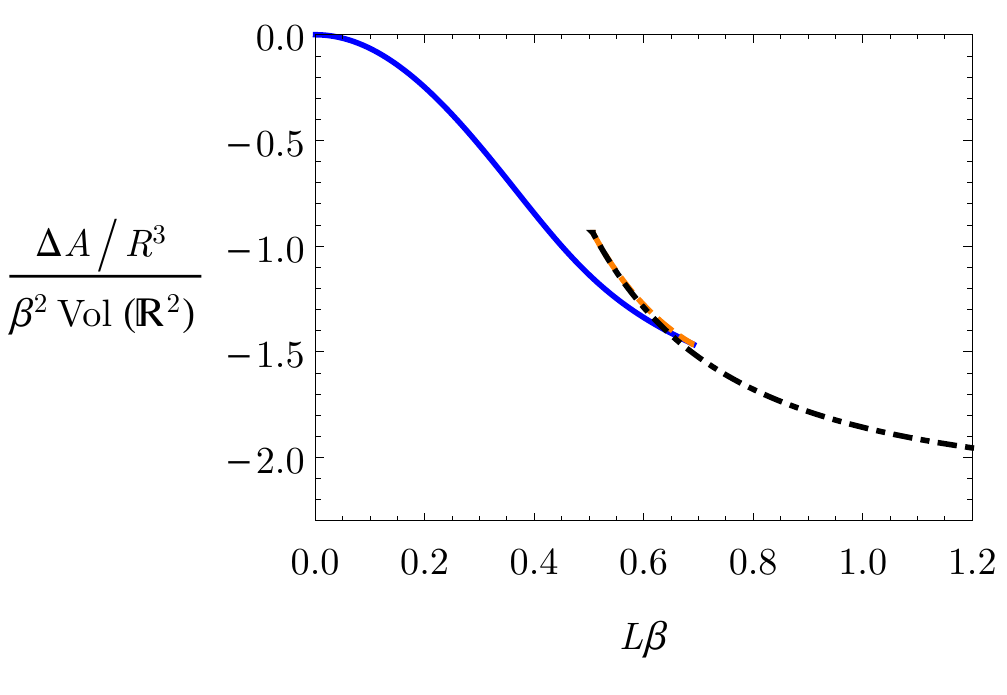} & \includegraphics[width=0.5\textwidth]{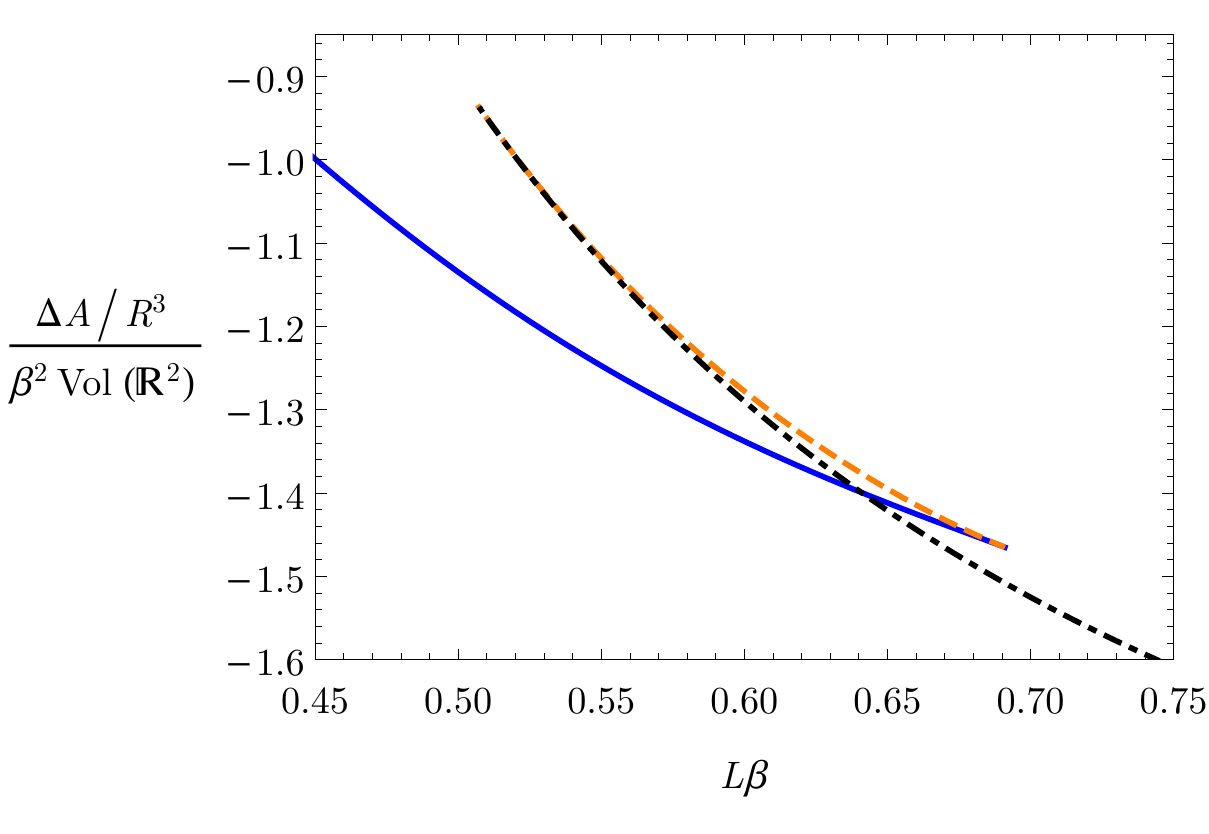} \\
  (a) & (b)
  \end{array}$
    \caption{For the $g(z)$ in eq.~\eqref{eq:holo_rg_transition}, and for the strip, (a) shows the differences in area, $\Delta A$, normalized by $\beta^2 \textrm{Vol}\left(\mathbb{R}^2\right)/R^3$, between three extremal surfaces in the bulk (blue solid, orange dashed, black dot-dashed) and the minimal surface in $AdS_5$ with the same $L$, indicating a ``first-order phase transition'' at $L\beta\approx 0.65$ from one (blue solid) to the other (black dot-dashed) as the area functional's global minimum. (b) Close-up of (a).}\label{fig:rgphasetrans}
\end{figure}

For the $g(z)$ in eq.~\eqref{eq:holo_rg_transition}, for the strip three extremal surfaces exist over a range of $L$. Fig.~\ref{fig:rgphasetrans} shows the difference in area, $\Delta A$, between each of these three surfaces and the minimal surface in $AdS_5$ with the same $L$, indicating a ``first-order phase transition'' from one to the other as the global minimum of the area functional, at the critical value $L\beta\approx 0.65$. Correspondingly, $\s$ for the strip exhibits a kink (discontinuous first derivative) at the critical $L$, shown in figs.~\ref{fig:rged} (c) and (d). In contrast, for the sphere, no transition occurs, and therefore $\s$ exhibits no kink, as shown in fig.~\ref{fig:rged} (c).

The $g(z)$ in eq.~\eqref{eq:rg_flow_constant} yields $\Delta_-=3/2$, hence eq.~\eqref{eq:rgsmallL} implies $\s \propto - L^0$ at small $L$, that is, $\s$ starts at a negative constant value at $L=0$, before monotonically rising as $L$ increases, and then $\s \to 0^-$ as $L\beta \to \infty$, as shown in fig.~\ref{fig:rged} (e). The $g(z)$ in eq.~\eqref{eq:rg_flow_fractional_power} yields $\Delta_-=7/4$, hence eq.~\eqref{eq:rgsmallL} implies $\s \propto - L^{1/2}$ at small $L$. However, aside from the fractional power of $L$ at small $L$, fig.~\ref{fig:rged} (f) shows that $\s$ behaves similarly to that in fig.~\ref{fig:rged} (a), with a single global minimum before $\s \to 0^-$ as $L\beta \to \infty$.

In summary, $\s$ can clearly exhibit a variety of behaviors as a function of $L$, depending on details of the RG flow. However, $\s$ often exhibits a unique \textit{global} minimum, which by dimensional analysis must be at an $L \propto 1/\beta$. As discussed in section~\ref{intro}, that $L$ can be used to characterize and compare RG flows. For example, the $L$ of $\s$'s global minimum could provide a precise definition of the crossover scale from the UV to IR.

\section{AdS-Schwarzschild}
\label{adssc}

In this section we consider a bulk action
\beq
\label{eq:adsscaction}
S_{\textrm{bulk}} = \frac{1}{16 \pi G} \int d^{d+1} x \, \sqrt{-\det g_{MN}} \left( \mathcal{R} + \frac{d(d-1)}{R^2} \right).
\eeq
The corresponding Einstein equation admits the $(d+1)$-dimensional AdS-SCH black brane solution, of the form in eq.~\eqref{eq: 1} with
\beq
\label{eq:schwarzschild_metric}
f(z) = g(z) = 1 - m \, z^d,
\eeq
and hence a horizon at $z_H = m^{-1/d}$, with $\Ttt$, $T$, and $s$ given by eqs.~\eqref{eq:energy_density} and~\eqref{eq:thermo_entropy}.

As mentioned in sec.~\ref{intro}, for AdS-SCH the FLEE requires $\s \propto \Ttt L$ at small $L$. We also expect $\lim_{L\to\infty}\s = s$. Our main result for AdS-SCH is the existence of a critical dimension, $\dcrit$, such that $\s \to s^-$ as $L \to \infty$ when $d<\dcrit$, while $\s \to s^+$ as $L \to \infty$ when $d>\dcrit$.

\begin{figure}[t]
  \centering
  $\begin{array}{cc}
  \includegraphics[width=0.5\textwidth]{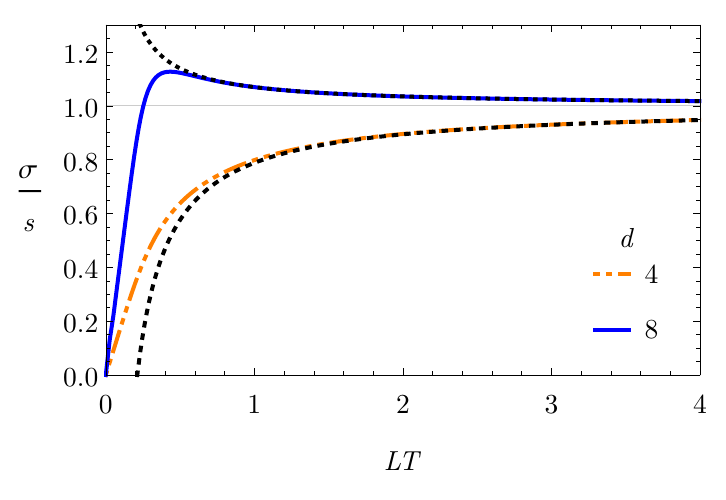} & \includegraphics[width=0.5\textwidth]{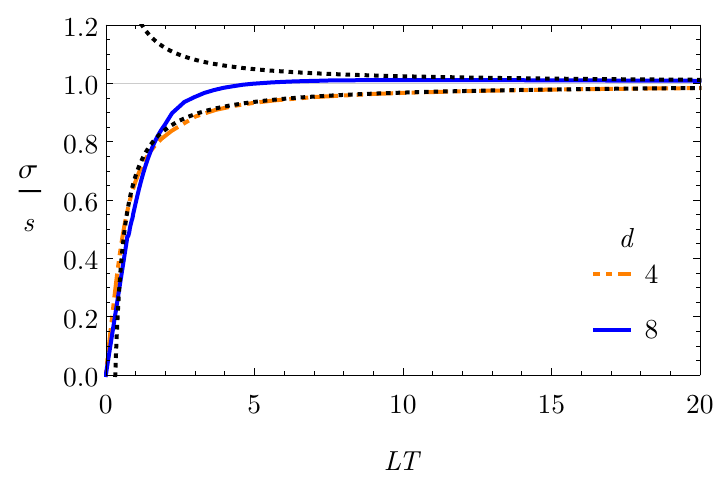} \\
  (a) & (b)
  \end{array}$
    \caption{The ED, $\s$, in units of entropy density, $s$, versus $LT$ for (a) the strip and (b) the sphere, for AdS-SCH in $d=4$ (orange dashed) and $d=8$ (blue solid). The dotted lines are $1 - \frac{\Delta \a}{s} \frac{A}{V}$, with $\Delta \a$ in eq.~\eqref{eq:alphaholo}, representing the $LT\to\infty$ limit and first $1/L$ correction.}
    \label{fig:adsscd48}
\end{figure}

For example, fig.~\ref{fig:adsscd48} shows our numerical results for $\s/s$ as a function of $LT$ for (a) the strip and (b) the sphere in AdS-SCH with $d=4$ and $8$. In all cases we find $\s/s \propto (\Ttt/s) L$ at small $LT$, as expected. For $d=4$ and for both the strip and sphere, we find $\s/s$ increases monotonically and $\s/s \to 1^-$ as $LT \to \infty$, whereas for $d=8$, $\s/s$ rises to a global maximum at an $L$ that by dimensional analysis must be $\propto 1/T$, and then $\s/s \to 1^+$ as $LT \to \infty$.

The dotted lines in fig.~\ref{fig:adsscd48} show $s - \Delta \a \frac{A}{V}$ divided by $s$, with $\Delta \a$ from eq.~\eqref{eq:alphaholo}. In other words, the dotted curves show the leading large-$L$ behavior, $s$, plus the first correction, which scales as $A/V\propto 1/L$ . The dotted curves agree with $\s/s$ not only at large $LT$, as expected, but over a surprisingly large range of $LT$, down to $LT\approx 1$. Crucially, the dotted lines reveal that the transition between $\s \to s^{\pm}$ as $L \to \infty$ occurs when the coefficient $-\Delta \a$ of the $1/L$ correction changes sign, from $-\Delta \a <0$ for $d=4$ to $-\Delta \a>0$ for $d=8$.

\begin{figure}[t]
\centering
\includegraphics[width=0.5\textwidth]{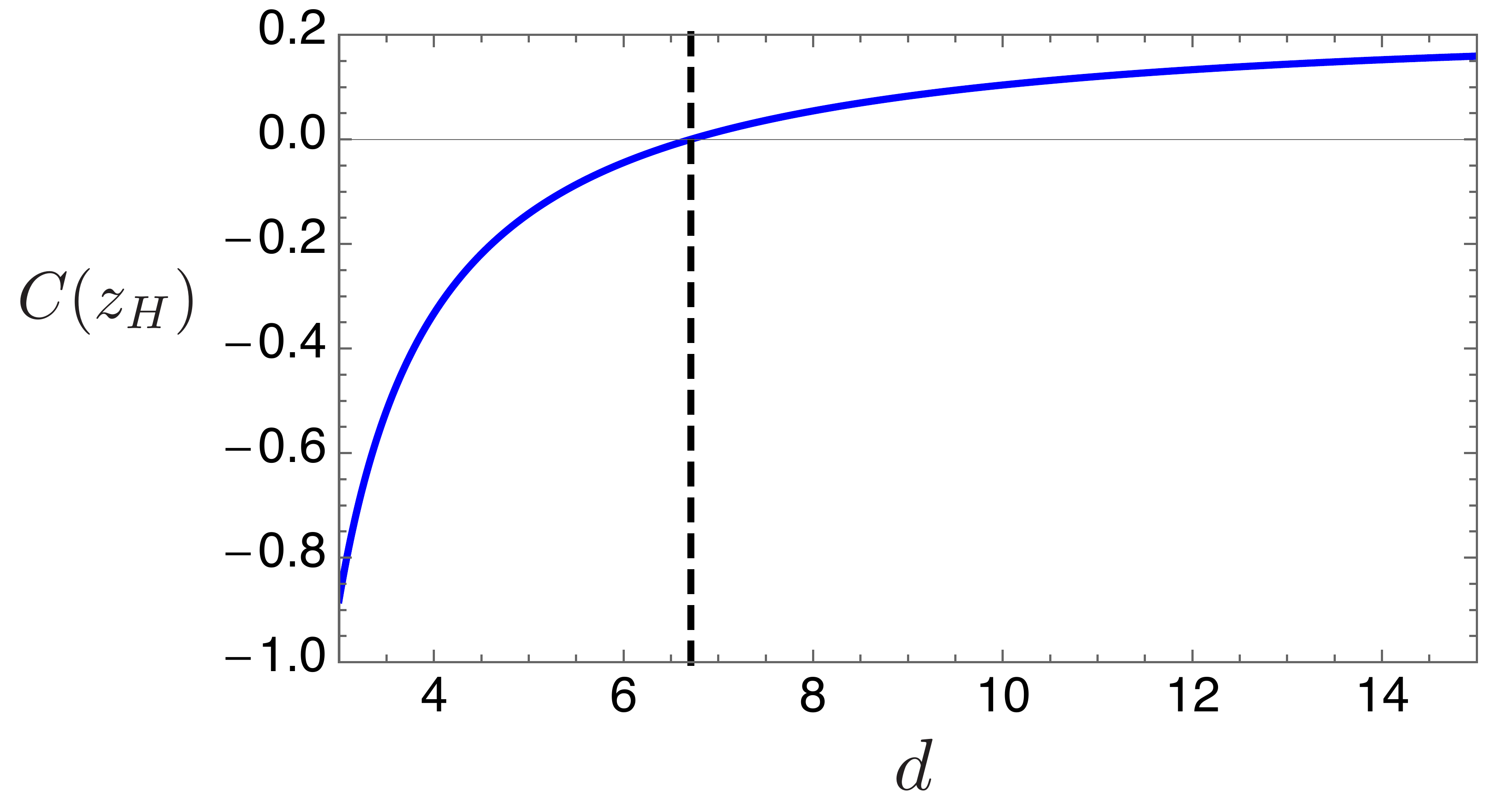}
\caption{The dimensionless coefficient $C(z_H)$ from eq.~\eqref{eq:alphaholo} for AdS-SCH, versus dimension $d$. At $d=3$, $C(z_H)\approx -0.88$, and $C(z_H)$ then increases monotonically with $d$, reaching zero at $\dcrit \approx 6.7$, indicated by the dashed black vertical line.}\label{fig:adssc_cplot}
\end{figure}

Indeed, fig.~\ref{fig:adssc_cplot} shows the dimensionless coefficient $C(z_H) \propto - \Delta \a$ from eq.~\eqref{eq:alphaholo} as a function of $d$, which begins at $C(z_H)\approx -0.88$ when $d=3$ and then monotonically increases as $d$ increases, eventually crossing through zero, which defines the critical dimension, $\dcrit \approx 6.7$. We can easily show that $C(z_H)$ is monotonically increasing for all $d$, and hence has only the single zero at $\dcrit$, by showing $\partial C(z_H)/\partial d \geq 0$, as follows. The $\frac{\partial}{\partial d}$ of eq.~\eqref{eq:cdef} gives
\beq
\label{eq:dnschwarz_dd}
\frac{\partial C(z_H)}{\partial d} = \frac{1}{(d-2)^2} + \int_0^1 du \, \frac{\log(u)}{u^{d-1}} \left(1 - \frac{1}{2} \frac{2 + u^{3d-2} -  3u^{d}}{(1-u^d)^{3/2}(1-u^{2d-2})^{1/2}}\right).
\eeq
Since $\frac{\log(u)}{u^{d-1}}\leq 0$ for $u \in [0,1]$, we need to show that
\beq
\label{eq:condition}
\frac{1}{2} \frac{2 + u^{3d-2} -  3u^{d}}{(1-u^d)^{3/2}(1-u^{2d-2})^{1/2}} \geq 1,
\eeq
for $u \in [0,1]$. The denominator in eq.~\eqref{eq:condition} is positive, so multiplying both sides of eq.~\eqref{eq:condition} by $(1-u^d)^{3/2}(1-u^{2d-2})^{1/2}$, squaring, and re-arranging, we find
\beq
\label{eq:monotonicity_inequality}
\left(1 + \frac{1}{2} u^{3d-2} - \frac{3}{2} u^d\right)^2 - (1-u^d)^3 (1-u^{2d-2}) \geq 0.
\eeq
Since $u^{2d-2} \geq u^{2d}$ for $u \in [0,1]$, eq.~\eqref{eq:monotonicity_inequality} implies
\begin{align}
\left(1 + \frac{1}{2} u^{3d-2} - \frac{3}{2} u^d\right)^2 -	(1-u^d)^3 (1-u^{2d-2}) &\geq \left(1 + \frac{1}{2} u^{3d} - \frac{3}{2} u^d\right)^2 - (1-u^d)^3 (1-u^{2d}) \nonumber \\ &= \frac{1}{4} u^{2d} (1 - u^d)^4 \geq 0,
\end{align}
and thus $\partial C(z_H)/\partial d \geq 0$, as advertised.

\begin{figure}[t!]
    \centering
    \begin{subfigure}[b]{0.48\textwidth}
        \centering
        \includegraphics[width=\textwidth]{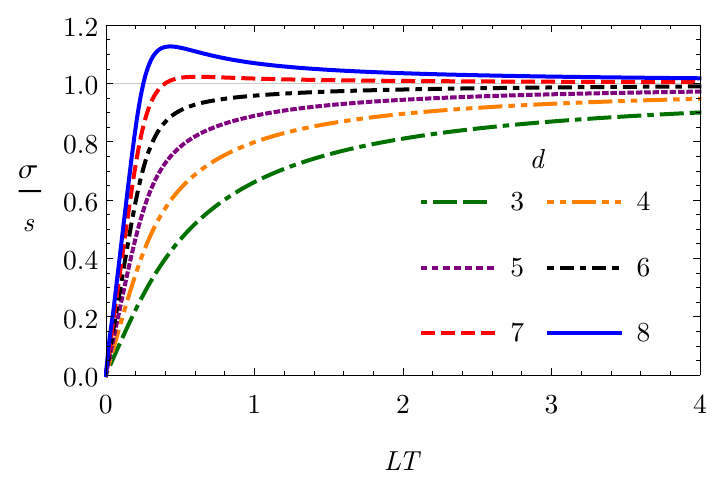}
        \caption{}
    \end{subfigure}
    ~ 
    \begin{subfigure}[b]{0.48\textwidth}
        \centering
        \includegraphics[width=\textwidth]{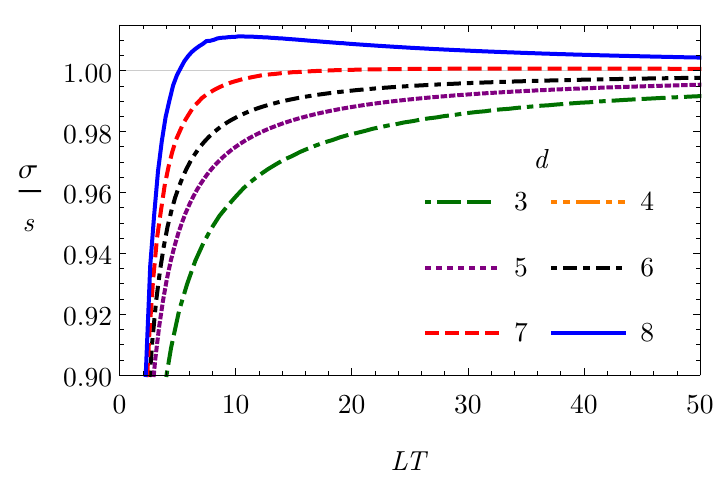}
        \caption{}
    \end{subfigure}
    \caption{\label{fig:schwarzschild_dimensions} The ED, $\s$, in units of entropy density $s$, versus $LT$ for (a) the strip, and (b) the sphere, for AdS-SCH in $d=3,4,\ldots,8$. For each entangling surface, when $d=3,4,5,6<\dcrit \approx 6.7$, $\s/s$ monotonically increases with $LT$ and $\s/s \to 1^-$ as $L\to\infty$, whereas when $d=7,8>\dcrit$, $\s/s$ rises to a global maximum before $\s/s \to 1^+$ as $LT \to \infty$.}
\end{figure}

Fig.~\ref{fig:schwarzschild_dimensions} shows $\s/s$ versus $LT$ for (a) the strip and (b) the sphere for $d=3,4,\ldots, 8$, illustrating the change of behavior at $\dcrit$. For both entangling surfaces, when $d=3,4,5,6<\dcrit$, we find $\s/s$ increases monotonically and $\s/s \to 1^-$ as $LT \to \infty$. When $d=7,8>\dcrit$, we find $\s/s$ rises to a global maximum before $\s/s \to 1^+$ as $LT \to \infty$. The maximum occurs at an $LT$ on the order of $10^0$ to $10^2$.

The above pattern extends also to CFTs at non-zero $T$ in $d=2$, where $S$ for an interval of length $L$ is known exactly~\cite{Calabrese:2004eu}. Given $d=2<\dcrit$, we expect $\s \to s^-$ as $LT \to \infty$. Indeed, the result of ref.~\cite{Calabrese:2004eu} leads to
\beq
\s = \frac{c}{3} \frac{1}{L}\ln\left[\frac{\sinh\left(\pi LT\right)}{\pi LT}\right]\nn = \frac{c}{3} \pi T - \frac{c}{3} \frac{\ln\left(2 \pi LT\right)}{L} + \mathcal{O}\left(e^{-2\pi TL}/L\right), \nn
\eeq
where $c$ is the CFT's central charge, and in the second equality we performed the $1/L$ expansion. In that expansion, the first term is Cardy's result for $s$~\cite{Cardy:1986ie}, while the second term exhibits the area law violating factor $\ln\left(2 \pi LT\right)$. Our key observation is: the leading correction has \textit{negative} coefficient, so that indeed $\s \to s^-$ as $LT \to \infty$.

The results above have also been obtained using the \textit{exact} form for EE of a strip in AdS-SCH derived in ref.~\cite{Erdmenger:2017pfh}.

As mentioned in sec.~\ref{intro}, the change from $C(z_H)<0$ and $\Delta \a >0$ when $d<\dcrit$ to $C(z_H)>0$ and $\Delta \a <0$ when $d>\dcrit$ represents area theorem violation~\cite{Casini:2012ei,Casini:2016udt}. Why does AdS-SCH violate the area theorem while relativistic RG flows do not? On the gravity side of the correspondence, the key difference is the behavior of $g(z)$. As mentioned below eq.~\eqref{eq:rg_flow_g_condition_1}, for relativistic RG flows the NEC implies $g'(z)\geq 0$, that is, $g(z)$ is strictly non-decreasing as $z$ increases. However, for AdS-SCH the NEC imposes no such constraint, and indeed $g(z) = 1-mz^d$ \textit{decreases} monotonically as $z$ increases, from $g(z=0)=1$ to $g(z=z_H)=0$. Apparently, as $d$ increases, eventually $g(z)$ decreases quickly enough to render $C(z_H)>0$.

How does AdS-SCH evades the field theory proofs in refs.~\cite{Casini:2012ei,Casini:2016udt} of the area theorem for the sphere in relativistic RG flows? The proofs of refs.~\cite{Casini:2012ei,Casini:2016udt} relied crucially on Lorentz invariance, which non-zero $T$ clearly breaks. In fact, in the $d\to\infty$ limit AdS-SCH is dual to an RG flow from a $(d+1)$-dimensional UV CFT to a $(0+1)$-dimensional IR CFT, which is clearly only possible when Lorentz symmetry is broken. More specifically, when $d \to \infty$ the AdS-SCH near-horizon geometry becomes $SL(2,\mathbb{R})/U(1) \times \mathbb{R}^{d-1}$, where the latter factor represents the spatial directions $\vec{x}$~\cite{Emparan:2013moa,Emparan:2013xia}. After a mode decomposition on $\mathbb{R}^{d-1}$, the action in eq.~\eqref{eq:adsscaction} gives rise to a string theory with target space $SL(2,\mathbb{R})/U(1)$~\cite{Emparan:2013xia}. Linearized fluctuations in the near-horizon region then exhibit scale invariance in $t$ and $z$ but not $\vec{x}$~\cite{Emparan:2013xia,Castro:2010fd}. AdS-SCH thus provides our our first hint that area theorem violation can occur as we dial a parameter towards a limiting value in which an IR fixed point emerges with scaling different from the UV fixed point. We will find further examples of such behavior in the following.

\section{AdS-Reissner-Nordstr\"om}
\label{adsrn} 

In this section we consider the bulk action
\beq
S_\mathrm{bulk} = \frac{1}{16 \pi G} \int d^{d+1} x \, \sqrt{-\det g_{MN}} \left( \mathcal{R} + \frac{d(d-1)}{R^2} - R^2 F_{MN}F^{MN}\right),
\eeq
where $F_{MN}=\partial_M A_N - \partial_N A_M$ is the field strength for a U(1) gauge field, $A_M$, dual to a conserved $U(1)$ current. The corresponding equations of motion admit the $(d+1)$-dimensional AdS-RN charged black brane solution~\cite{Ammon:2015wua}, with metric of the form in eq.~\eqref{eq: 1}, with 
\beq
f(z) = g(z)=1-mz^{d}+q^{2}z^{2(d-1)},
\eeq
where $q$ is proportional to the black brane's charge density. The solution has a horizon at the smallest positive root of $f(z_H)=0$. The gauge field solution's only non-zero component is
\beq
A_t = \mu \left(1 - \frac{z^{d-2}}{z_H^{d-2}}\right), \qquad \mu = \sqrt{\frac{d-1}{2(d-2)}} \, z_H^{d-2} \, q.
\eeq
AdS-RN is dual to a CFT with non-zero $\Ttt$, $T$, and $s$, given by eqs.~\eqref{eq:energy_density} and~\eqref{eq:thermo_entropy}, and non-zero chemical potential $\mu$ and charge density, proportional to $q$. In particular,
\beq
T = \frac{d}{4\pi R^2} \frac{1}{z_H} \left(1-\frac{d-2}{d} \, z_H^{2(d-2)}\,q^2\right),
\eeq
so that $T\geq 0$ implies $q^2 \leq \frac{d}{d-2} z_H^{-2(d-1)}$. In the extremal limit, where $q$ saturates the upper bound and $T=0$, an extremal horizon is present, so that $s \neq 0$. Moreover, when $T=0$ the near-horizon geometry becomes $AdS_2 \times \mathbb{R}^{d-2}$, with $AdS_2$ of radius $R/\sqrt{d(d-1)}$ in the $t$ and $z$ directions. When $T=0$, the dual is in a semi-local quantum liquid state~\cite{Iqbal:2011in}, describing an RG flow from a $(d+1)$-dimensional UV CFT to a $(0+1)$-dimensional IR CFT.

\begin{figure}[t!]
\centering
 \includegraphics[width=0.6\textwidth]{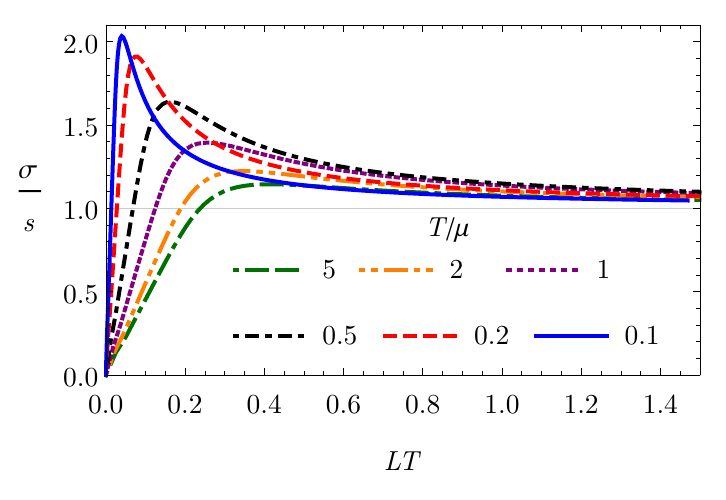}
 \caption{The ED, $\s$, in units of entropy density, $s$, versus $LT$ for the strip in AdS-RN with $d=8$, for $T/\mu=5$ down to $0.1$.}\label{fig:adsrnd8}
\end{figure}

The CFT states are parameterized by $T/\mu$, which determines $\Ttt$ and $q$. When $T/\mu \gg 1$, AdS-RN approaches AdS-SCH, and we recover the results of sec.~\ref{adssc}, including the existence of the critical dimension $\dcrit\approx 6.7$. For example, fig.~\ref{fig:adsrnd8} shows $\s/s$ versus $LT$ for the strip in AdS-RN with $d=8>\dcrit$ for various $T/\mu$. We find $\s/s \propto LT$ at small $L$ for all $T/\mu$, as required by the FLEE. For $T/\mu \gg 1$ we find $\s/s$ rises monotonically to a global maximum, and then $\s/s\to 1^+$ as $LT \to \infty$, consistent with our results from sec.~\ref{adssc}. As $T/\mu$ decreases and AdS-RN increasingly deviates from AdS-SCH, the global maximum persists, moving to smaller $LT$ while growing taller and narrower, such that $\s/s\to1^+$ for all $T/\mu$.

\FIGURE{
\begin{tabular}{c c}
\includegraphics[width=0.5\textwidth]{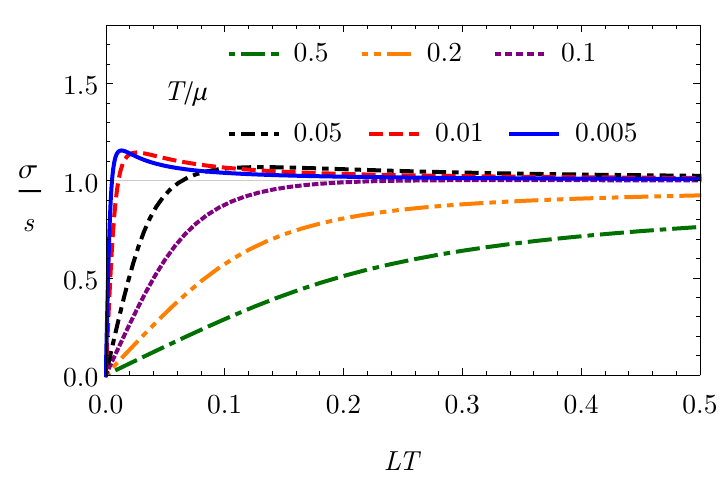}
&
\includegraphics[width=0.5\textwidth]{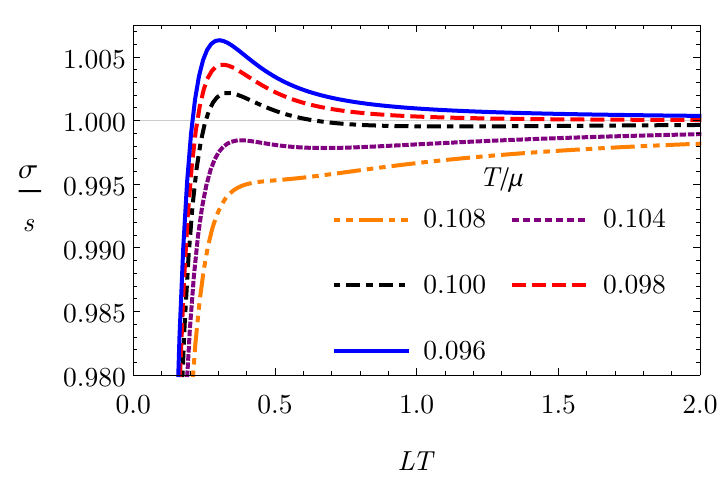}
\\
(a) & (b)
\\
\includegraphics[width=0.5\textwidth]{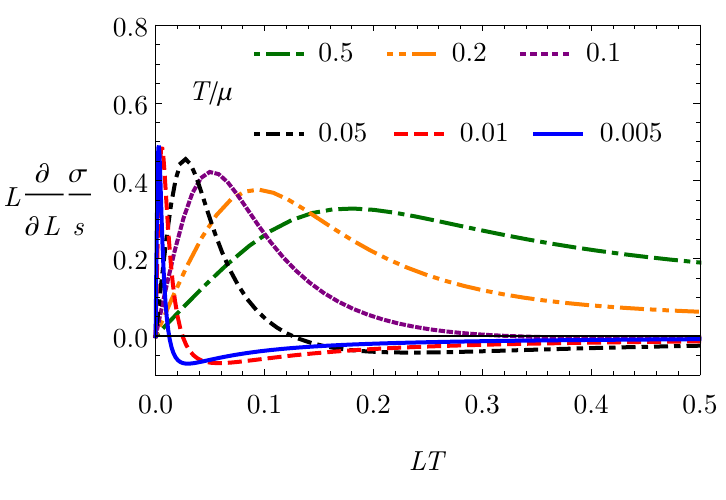}
&
\includegraphics[width=0.5\textwidth]{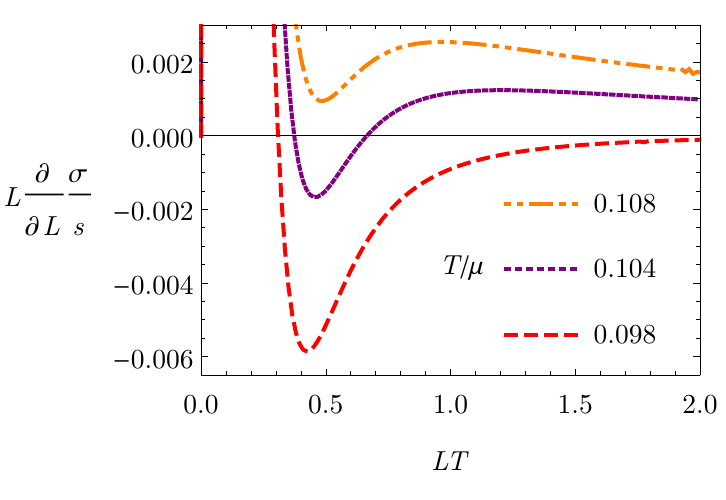}
\\
(c) & (d)
\end{tabular}
\caption{\label{fig:adsrnstripd4} (a) The ED, $\s$, in units of entropy density $s$, versus $LT$ for the strip in AdS-RN with $d=4$, showing the formation of a global maximum as $T/\mu$ decreases from $T/\mu=0.5$ down to $0.005$. (b) Close-up of (a), showing the formation of a local minimum and maximum before the formation of the global maximum, as $T/\mu$ decreases from $T/\mu=0.108$  down to $0.096$. (c) Logarithmic derivative $L\frac{\partial}{\partial L}$ of (a). (d) Close-up of (c) showing the logarithmic derivative for $T/\mu=0.108$, which has no zero, indicating $\s/s$ is monotonic in $LT$, then $T/\mu=0.104$, which has two zeroes, indicating a local minimum and maximum in $\s/s$, and finally $T/\mu=0.098$, which has a single zero, indicating a global maximum in $\s/s$.}
}

Fig.~\ref{fig:adsrnstripd4} shows $\s/s$ versus $LT$ for the strip in AdS-RN with $d=4<\dcrit$ for various $T/\mu$. When $T/\mu \gg 1$ we find $\s/s$ rises monotonically as $LT$ increases, and eventually $\s/s \to 1^-$ as $LT \to \infty$, consistent with our results from sec.~\ref{adssc}. However, as $T/\mu$ decreases we find a transition in which a global maximum appears and $\s/s\to1^+$ as $LT \to \infty$, shown in fig.~\ref{fig:adsrnstripd4} (a). The transition actually occurs in stages, as shown in fig.~\ref{fig:adsrnstripd4} (b). First, at $T/\mu\approx 0.107$, a local minimum and maximum appear, with $\s/s<1$ for all $LT$. Second, at $T/\mu \approx 0.102$, the maximum rises above $\s/s=1$, becoming a global maximum, but a local minimum persists at $\s/s<1$, and then $\s/s\to 1^-$ as $LT \to \infty$. Third and finally, at $T/\mu \approx 0.098$, a transition occurs from $\s/s\to1^-$ to $\s/s\to1^+$ as $LT \to \infty$, and the local minimum disappears. Figs.~\ref{fig:adsrnstripd4} (c) and (d) show the logarithmic derivative $L\frac{\partial}{\partial L} \frac{\s}{s}$, which clearly has no zero for $T/\mu> 0.107$, indicating $\s/s$ is monotonic in $LT$, then develops two zeroes for $0.107 > T/\mu > 0.102$, indicating a local minimum and maximum in $\s/s$, and then develops a single zero for $T/\mu < 0.098$, indicating a global maximum in $\s/s$.

We find qualitatively similar behavior for the strip in all $d<\dcrit$: at some $(T/\mu)_1$ a local minimum and maximum appear, but $\s/s$ remains below one for all $LT$, at some $(T/\mu)_2<(T/\mu)_1$ a global maximum emerges, but still $\s/s\to1^-$ for $LT\to\infty$, and finally at some $(T/\mu)_3<(T/\mu)_2$ the transition occurs to $\s/s\to1^+$ as $LT \to \infty$. Our numerical estimates for $(T/\mu)_1$, $(T/\mu)_2$, and $(T/\mu)_3$ for $d=3,4,5,6<\dcrit$ appear in table~\ref{tab:adsrn}.

\begin{table}[ht!]
\centering
\begin{tabular}[c]{|c|c|c|c|}
\hline
$d$ & $(T/\mu)_1$ & $(T/\mu)_2$ & $(T/\mu)_3$ \\ \hline
$3$ & $6.343 \times 10^{-4}$ & $4.858 \times 10^{-4}$ & $2.967 \times 10^{-4}$ \\ \hline
$4$ & $0.107$ & $0.102$ & $0.098$  \\ \hline
$5$ & $0.407$ & $0.403$ & $0.399$ \\ \hline
$6$ & $1.219$ & $1.215$ & $1.213$ \\ \hline
\end{tabular}
\caption{\label{tab:adsrn} For the strip in AdS-RN with $d<\dcrit\approx 6.7$, as $(T/\mu)$ decreases, at $(T/\mu)_1$ a local minimum and maximum appear in $\s/s$ as a function of $LT$, at $(T/\mu)_2<(T/\mu)_1$ the local maximum becomes a global maximum, but a local minimum remains, and $\s/s<1$ for all $LT$, and then at $(T/\mu)_3<(T/\mu)_2$ the global maximum rises above one, and the transition occurs to $\s/s\to1^+$ as $LT \to \infty$. }
\end{table}

\FIGURE{
\begin{tabular}{c c}
\includegraphics[width=0.5\textwidth]{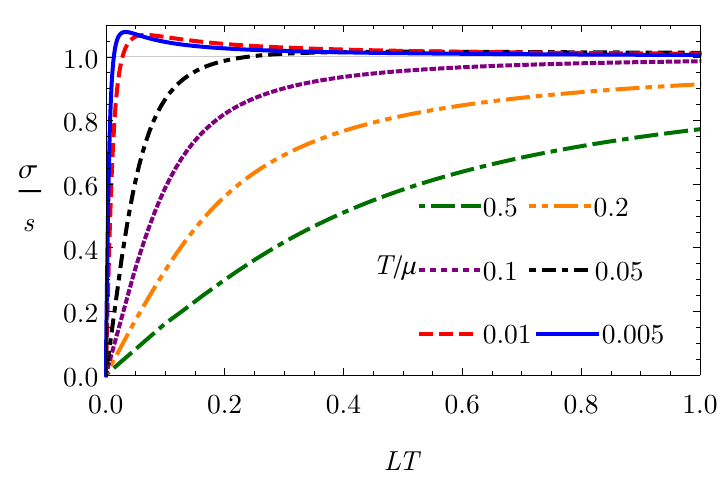}
&
\includegraphics[width=0.5\textwidth]{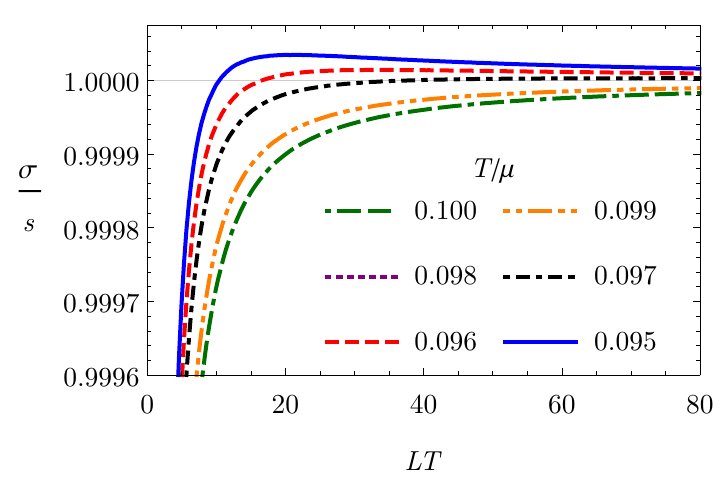}
\\
(a) & (b)
\end{tabular}
\caption{\label{fig:adsrnsphered4} (a) The ED, $\s$, in units of entropy density $s$, versus $LT$ for the sphere in AdS-RN with $d=4$, showing the formation of a global maximum as $T/\mu$ decreases from $T/\mu=0.5$ down to $0.005$. (b) Close-up of (a) for $T/\mu=0.1$ down to $0.095$, including the critical value $T/\mu\approx0.097$ where the maximum forms.}
}

In contrast,we find no evidence of such a multi-stage transition for the sphere in AdS-RN with $d< \dcrit$. For example, fig.~\ref{fig:adsrnsphered4} shows $\s/s$ versus $LT$ for the sphere in AdS-RN with $d=4<\dcrit$. When $T/\mu \gg 1$ we find $\s/s$ rises monotonically as $LT$ increases, and eventually $\s/s \to 1^-$ as $LT \to \infty$, consistent with our results from sec.~\ref{adssc}. As $T/\mu$ decreases we find a transition in which a global maximum appears and $\s/s\to1^+$ as $LT \to \infty$, shown in fig.~\ref{fig:adsrnsphered4} (a).  Fig.~\ref{fig:adsrnsphered4} (b) shows a close-up for $T/\mu$ near the transition, which shows no sign of a local minimum and maximum forming before the global minimum forms. Crucially, however, we cannot rule out a multi-stage transition like the strip's, but on scales of $LT$ and $T/\mu$ smaller than our numerical precision, \textit{i.e.} between $T/\mu$ steps smaller than those in fig.~\ref{fig:adsrnsphered4} (b).

\begin{figure}[t!]
\centering
 \includegraphics[width=0.6\textwidth]{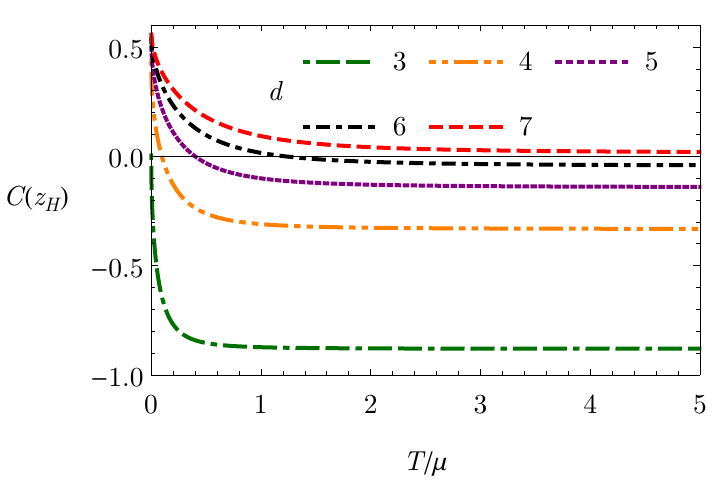}
 \caption{The dimensionless coefficient $C(z_H)$ from eq.~\eqref{eq:alphaholo} versus $T/\mu$ for AdS-RN, for $d=3,4,5,6,7,8$. For $d>\dcrit \approx 6.7$, $C(z_H)>0$ for all $T/\mu$, indicating area theorem violation. For $d<\dcrit$, as $T/\mu$ decreases $C(z_H)$ changes sign from negative to positive at the $(T/\mu)_3$ in table~\ref{tab:adsrn}, indicating area theorem violation for $T/\mu<(T/\mu)_3$.}\label{fig:rnc}
\end{figure}

In all cases above, the transition between $\s/s\to 1^{\pm}$ as $LT \to \infty$ indicates area theorem violation. Indeed, fig.~\ref{fig:rnc} shows the dimensionless coefficient $C(z_H)$ as a function of $T/\mu$ for $d=3,4,5,6,7$. For all $d>\dcrit \approx 6.7$, at all $T/\mu$ we find $C(z_H)>0$, indicating $\Delta \a <0$ and hence the area theorem is violated. For all $d<\dcrit$, at high $T/\mu$ we find $C(z_H) < 0$, indicating $\Delta \a > 0$ and the area theorem is obeyed, but as $T/\mu$ decreases $C(z_H)$ eventually passes through zero, so that at low $T/\mu$ we find $C(z_H)>0$, indicating $\Delta \a >0$ and the area theorem is violated. In each case, the critical $T/\mu$ where $C(z_H)=0$ is precisely the $(T/\mu)_3$ for the strip in table~\ref{tab:adsrn}, as expected.

\FIGURE{
\begin{tabular}{c c}
\includegraphics[width=0.5\textwidth]{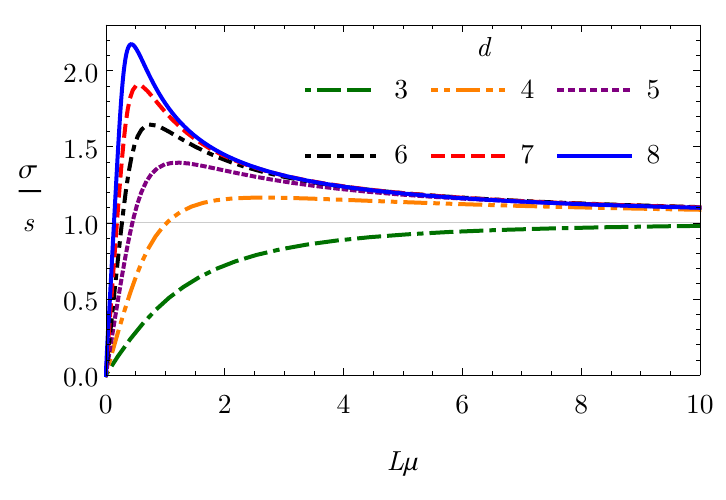}
&
\includegraphics[width=0.5\textwidth]{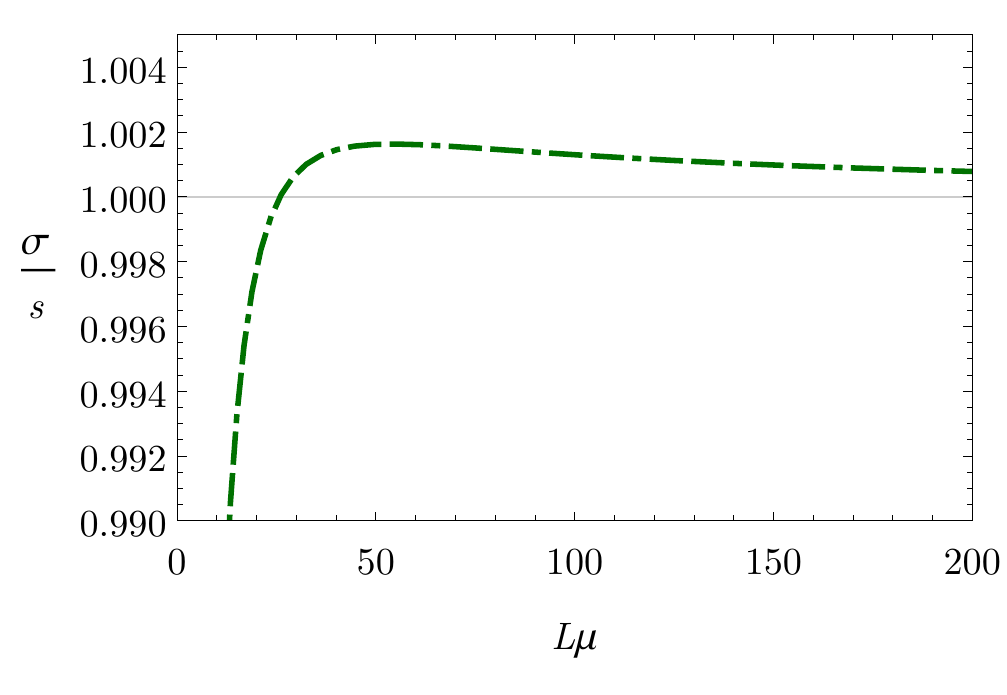}
\\
(a) & (b)
\\
\includegraphics[width=0.5\textwidth]{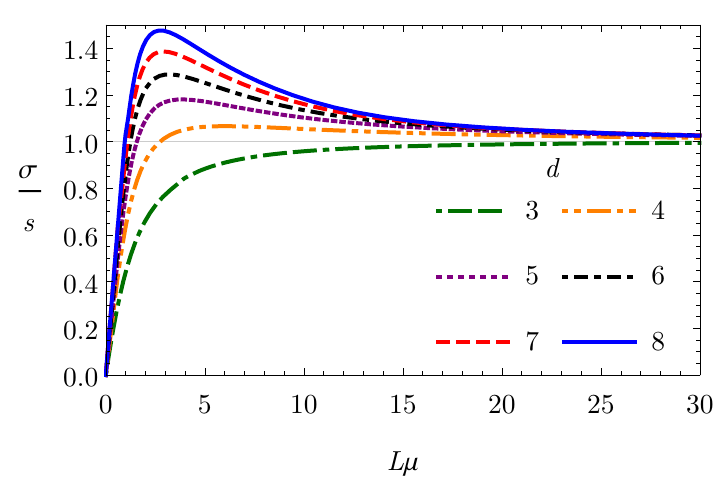}
&
\includegraphics[width=0.5\textwidth]{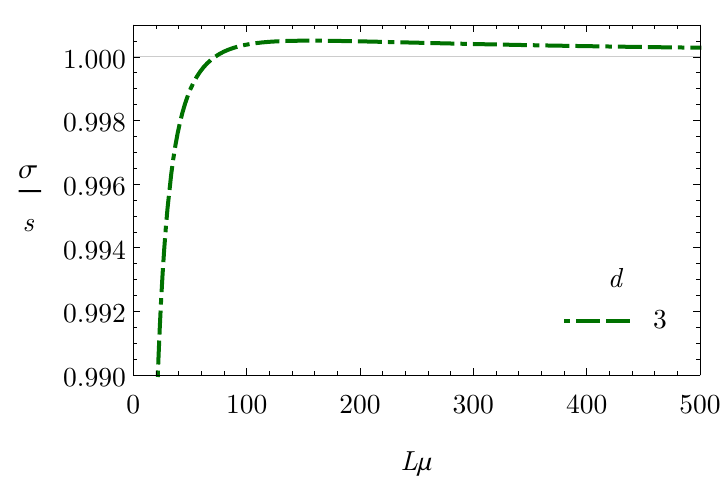}
\\
(c) & (d)
\end{tabular}
\caption{\label{fig:adsrnextremal} (a) The ED, $\s$, in units of entropy density $s$, versus $L\mu$ for the strip in AdS-RN with $d=3,4,5,6,7,8$ and $T/\mu=0$. In all cases $\s/s$ has a single global maximum and $\s/s\to1^+$ as $L\mu \to \infty$. (b) Close-up of (a) for $d=3$, showing the maximum. (c) The same as (a), but for the sphere. (d) Close-up of (c) for $d=3$, showing the maximum.}
}

Ultimately, when $T/\mu=0$, where the bulk metric is extremal AdS-RN and the near-horizon geometry is $AdS_2 \times \mathbb{R}^{d-1}$, for all $d$ we find $C(z_H)>0$, so that for both the strip and sphere $\s/s \to 1^+$ as $L\mu \to \infty$ and the area theorem is violated. Fig.~\ref{fig:adsrnextremal} shows $\s/s$ versus $L\mu$ in AdS-RN with $d=3,4,5,6,7,8$ and $T/\mu=0$, for the strip ((a) and (b)) and the sphere ((c) and (d)). In all cases $\s/s$ indeed has a global maximum and $\s/s\to1^+$ as $L\mu\to\infty$.

In summary, in AdS-RN for either $d>\dcrit$ at any $T/\mu$, or for any $d$ and sufficiently small $T/\mu$, we find a global maximum in $\s/s$, and in particular $\s/s \to 1^+$ as $LT \to \infty$, indicating area theorem violation. In other words, as we dial a parameter towards a limiting value in which an IR fixed point appears with different scaling from the UV CFT ($d \to \infty$ or $T/\mu \to 0$), we find area theorem violation, as we saw in AdS-SCH and as we will see in some, but not all, of the following examples.

\section{AdS-to-Hyperscaling-Violating Domain Walls}
\label{hyper}

In this section we consider the bulk action
\beq
S_{\textrm{bulk}} = \frac{1}{16 \pi G}\int d^{d+1} x \sqrt{-\det g_{MN}} \left(\mathcal{R} - 2 \left(\partial \F\right)^2 - V(\Phi)- \frac{Z(\F)}{4} F_{PQ} F^{PQ} - \frac{\tilde{Z}(\F)}{4} \tilde{F}_{RS} \tilde{F}^{RS}\right). \nn
\eeq
where $\Phi$ is a real scalar field with potential $V(\Phi)$, $F_{MN}$ and $\tilde{F}_{MN}$ are two $U(1)$ field strengths for two $U(1)$ gauge fields $A_M$ and $\tilde{A}_N$, respectively, and $Z(\Phi)$ and $\tilde{Z}(\Phi)$ are two real functions of $\Phi$. The scalar field $\Phi$ is dual to a scalar operator $\Op$ while $A_M$ and $\tilde{A}_M$ are dual to two conserved $U(1)$ currents. We will consider the solutions of ref.~\cite{Lucas:2014sba}, with metric of the form
\beq
\label{eq:sachdev_metric}
ds^2 = \frac{R^2}{z^2} \left(- a(z)b(z) dt^2 + d \vec{x}^2+\frac{a(z)}{b(z)} dz^2 \right),
\eeq
with real functions $a(z)$ and $b(z)$, which is of the form in eq.~\eqref{eq: 1} with $f(z)=a(z) b(z)$ and $g(z) = b(z)/a(z)$. If $b(z_H)=0$ then a horizon exists at $z=z_H$, with $\Ttt$, $T$, $s$ given by eqs.~\eqref{eq:energy_density} and~\eqref{eq:thermo_entropy}. The solutions of ref.~\cite{Lucas:2014sba} also include non-zero $\Phi(z)$, $F_{zt}(z)$, and $\tilde{F}_{zt}(z)$, with all other components of $F_{MN}$ and $\tilde{F}_{MN}$ vanishing.

A central result of ref.~\cite{Lucas:2014sba} is that if we split $b(z)$ as
\beq
\label{eq:bsplit}
b(z) = b_0(z) + \eta^2 \, b_2(z),
\eeq
where $b_0(z)$ and $b_2(z)$ are $T$-independent but the real parameter $\eta$ may depend on $T$, and furthermore we extract a factor of $\eta$ from one of the $U(1)$ gauge fields, say $\tilde{A}_M$, then we can simplify the equations of motion by separating terms by powers of $\eta$. Ultimately, we can obtain an entire family of solutions completely specified by a single parameter, $T/\mu$, with $\mu$ the chemical potential for the $U(1)$ current dual to $A_M$, with corresponding charge density $Q \equiv - \delta S_{\textrm{bulk}}/\delta F_{zt}$. In fact, as shown in ref.~\cite{Lucas:2014sba}, for $b(z)$ of the form in eq.~\eqref{eq:bsplit} we can solve all the equations of motion by freely choosing two functions in the solution which then determine all other functions \textit{and} the corresponding $V(\Phi)$, $Z(\Phi)$, and $\tilde{Z}(\Phi)$, leaving only a choice of boundary conditions. Following ref.~\cite{Lucas:2014sba}, we choose $b_2(z)$ and $F_{zt}(z)$, and obtain $a(z)$ by solving, from the equations of motion,
\beq
\label{eq:aeq}
\frac{\partial}{\partial\hat{z}}\left(\frac{a\,b_2}{\hat{z}^{2(d-1)}}\right) = \hat{c} \, \frac{a}{\hat{z}^{d-1}},
\eeq
with constant $\hat{c}$, and then obtain $b_0(z)$ by solving, from the equations of motion,
\beq
\label{eq:b0eq}
\frac{\partial}{\partial\hat{z}}\left(\frac{\frac{\partial}{\partial\hat{z}}(a\,b_0)}{a \hat{z}^{d-1}}\right) = -2 \hat{F}_{zt},
\eeq
where $\hat{z}$ and $\hat{F}_{zt}$ are defined by the re-scalings
\beq
z = Q^{-\frac{1}{d-1}} R^{\frac{d-2}{d-1}}\left(8\pi G\right)^{\frac{1}{1-d}} \, \hat{z}, \qquad \qquad F_{zt} = Q^{\frac{1}{d-1}} R^{\frac{1}{d-1}}\left(8\pi G\right)^{\frac{1}{d-1}}\, \hat{F}_{zt}.
\eeq

In what follows, we solve eqs.~\eqref{eq:aeq} and~\eqref{eq:b0eq} numerically. We focus on the three solutions of ref.~\cite{Lucas:2014sba} that at $T=0$ have no horizon, and describe domain walls from an asymptotic $AdS_{d+1}$ as $z \to 0$ to an HV geometry as $z \to \infty$. Specifically, as $z \to 0$ we require
\beq
\label{eq:sachdev_small_z}
a(z) = 1 + \mathcal{O}(z^{d-1}), \qquad b_0(z) = 1 + \mathcal{O}(z^{d-1}),
\eeq
and at leading order $b_2(z) \propto - z^d$. If we choose
\beq
\hat{c} = d-2 + \frac{(d-1)(\zeta-1)-\theta}{d-1-\theta},
\eeq
then when $z \to \infty$ we find the following scalings
\beq
\label{eq:hvscalings}
a(z) \sim z^{-\left((d-1)(\z-1) - \th\right)/(d-\th-1)},\qquad b_0(z) \sim z^{-\left((d-1)(\z-1) + \th\right)/(d-\th-1)},\qquad b_2(z) \sim - z^d.
\eeq
As a result, under a Lifshitz re-scaling, $t \to \lambda^{\zeta} t$, $\vec{x} \to \lambda \vec{x}$, $z \to \lambda z$, the $z \to \infty$ metric re-scales as $ds \rightarrow \lambda^{\th/(d-1)} ds$, indicating HV~\cite{Huijse:2011ef,Dong:2012se}. Roughly speaking, with HV the thermodynamics is that of a theory with dynamical exponent $\zeta$ in $d-\theta$ dimensions. When $\zeta \to \infty$ with $-\theta/\zeta$ fixed, the $z \to \infty$ metric becomes conformal to $AdS_2 \times \mathbb{R}^{d-1}$, with no horizon~\cite{Hartnoll:2012wm}. 

Specifying $b_2(z)$ and $F_{zt}(z)$ and then solving eqs.~\eqref{eq:aeq} and~\eqref{eq:b0eq} with the boundary conditions described above determines the metric completely, which is sufficient to compute $\s$. However, to interpret the results in the dual field theory we should also solve for $\tilde{F}_{zt}(z)$ and $\Phi(z)$. For a detailed discussion of their equations of motion and boundary conditions, see ref.~\cite{Lucas:2014sba}. In what follows we only need two facts about their solutions. First, as $z \to 0$ we require $\Phi(z) \propto z^{d-2}$ at leading order. As a result, $\Delta_-=d-2$, hence via eq.~\eqref{eq:rgsmallL}, $\s \propto L^{2\Delta_-+1-d} = L^{d-3}$ at small $L$, indicating FLEE violation when $d\neq4$. Second, $\tilde{F}_{tz}(z)$ and $\Phi(z)$'s solutions are generically non-trivial, indicating that the dual theory has non-zero chemical potential and charge density for the second $U(1)$, and also $\Opv \neq 0$ and possibly a non-zero source for $\Op$. However, in the approach of ref.~\cite{Lucas:2014sba} described above, all of these quantities are outputs determined by the single input, $\eta$, or equivalently $T/\mu$.

We first consider the solution of ref.~\cite{Lucas:2014sba} with $d=3$ and
\beq
\label{eq:sachdev1}
b_2 = - \hat{z}^3 \, \frac{9\hat{z}^2+20\hat{z}+80}{9\hat{z}^2+20\hat{z}+40}, \qquad \hat{F}_{zt} = - \left(1+0.891\,\hat{z}\right)^{-4},
\eeq
which at $T=0$ describes a domain wall from $AdS_4$ to a HV geometry with $\zeta = 2$ and $\theta=-2$. Asymptotically $\Phi(z) \propto z$ at leading order as $z \to 0$, so $\Op$ has a non-zero source and the FLEE may be violated. Indeed, as argued above using eq.~\eqref{eq:rgsmallL}, $\s \propto L^0$ at small $L$.

\FIGURE{
	\begin{tabular}{c c}
		\includegraphics[width=0.48\textwidth]{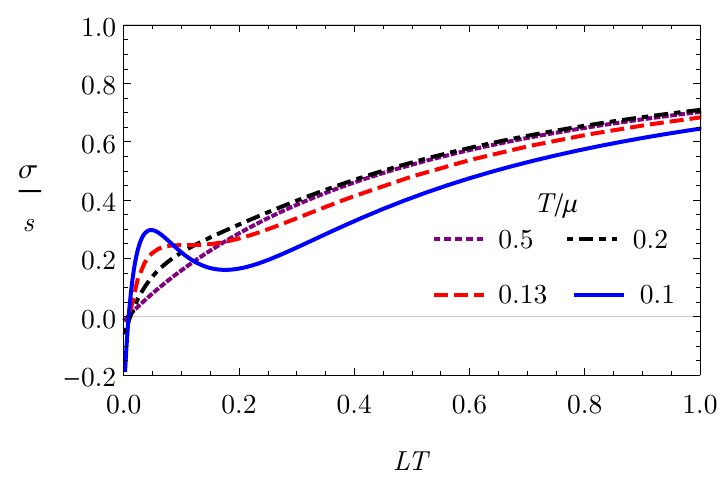}
		&
		\includegraphics[width=0.48\textwidth]{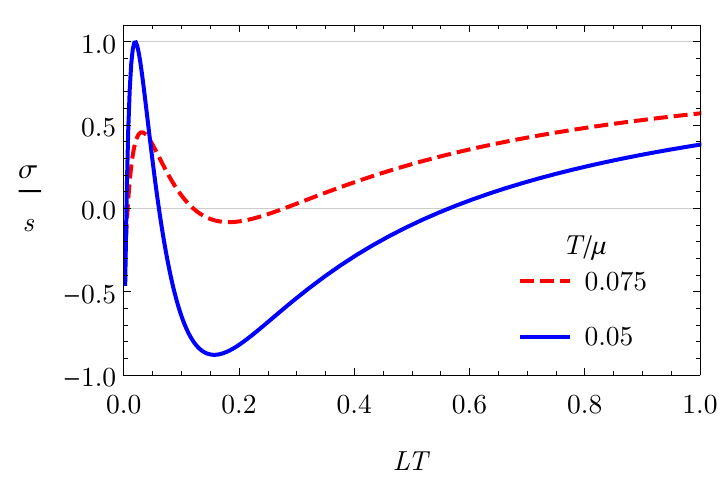}
		\\
		(a) & (b) \\
		\includegraphics[width=0.48\textwidth]{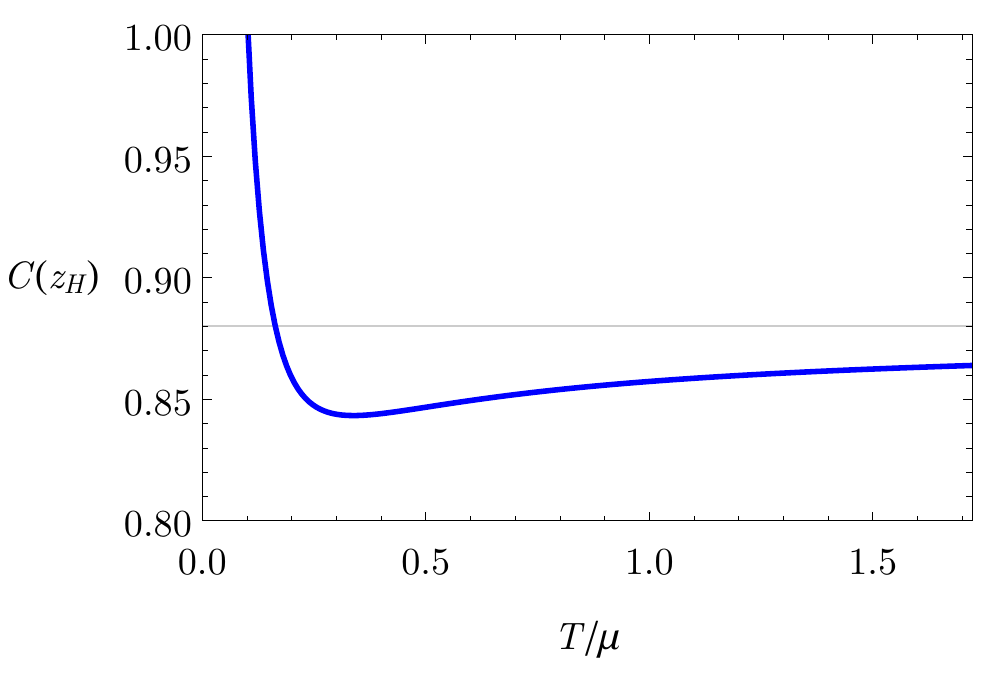}
		&
		\includegraphics[width=0.48\textwidth]{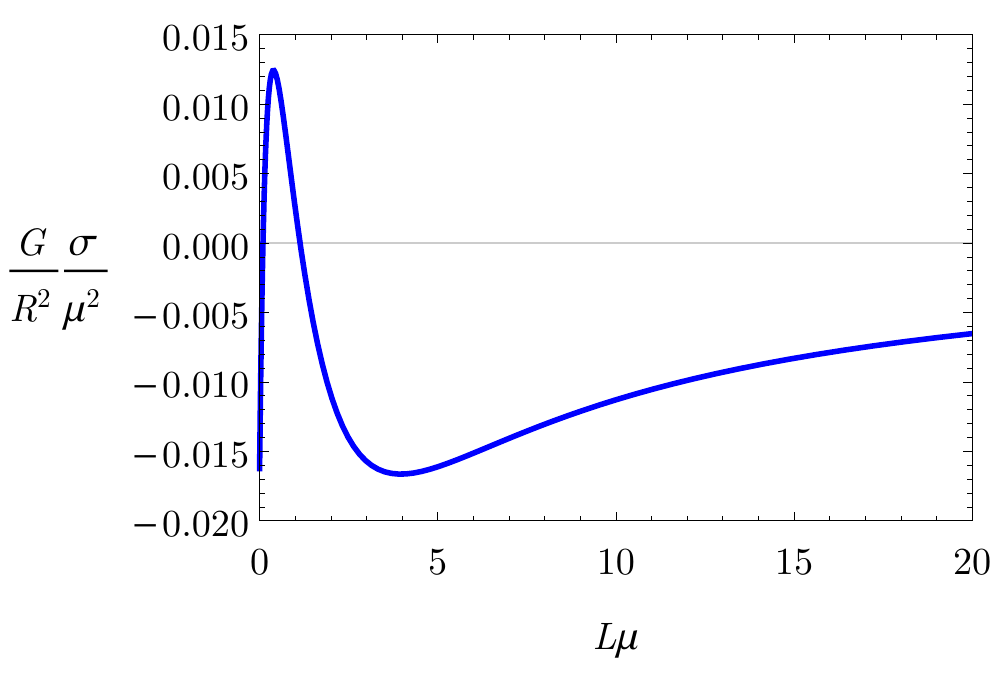}\\
		(c) & (d)
	\end{tabular}
	\caption{Plots of our results using the solution given by eq.~\eqref{eq:sachdev1} ($d=3$, $\zeta=2$, $\theta=-2$). (a) The ED, $\s$, in units of entropy density $s$, for the strip, for $T/\mu=0.5$ down to $0.1$. A local maximum and minimum appear when $T/\mu \lesssim 0.130$. (b) Close-up of (a), but showing the local maximum and minimum's growth from $T/\mu=0.075$ down to $0.05$. (c) The dimensionless coefficient $C(z_H)$ from eq.~\eqref{eq:alphaholo} versus $T/\mu$. As $T/\mu\to\infty$, $C(z_H)$ approaches the value for AdS-SCH with $d=3$ (the horizontal line). Clearly $C(z_H)<0$ for all $T/\mu$, indicating the area theorem is obeyed. (d) The ED, $\s$, in units of $\mu^2 R^2/G$, versus $L\mu$, for the strip at $T/\mu=0$. The local maximum and minimum remain, and the area theorem is still obeyed.}\label{fig:sachdev1}
}

Figs.~\ref{fig:sachdev1} (a) and (b) show $\s/s$ versus $LT$ for the strip in this solution with various $T/\mu$. For all $T/\mu$ we find $\s/s \propto - L^0$ at $LT=0$, as expected. Surprisingly, figs.~\ref{fig:sachdev1} (a) and (b) also reveal that as we lower $T/\mu$, when $T/\mu\approx0.130$ a local maximum and minimum appear at intermediate $LT$, and grow in height as $T/\mu$ continues decreasing. Also surprisingly, figs.~\ref{fig:sachdev1} (a) and (b) show that $\s/s \to 1^-$ as $LT \to \infty$ for \textit{all} $T/\mu$, indicating the area theorem is obeyed. Indeed, fig.~\ref{fig:sachdev1} (c) shows $C(z_H)<0$ for all $T/\mu$. These features persist to $T/\mu=0$. In this solution, $s=0$ when $T/\mu=0$, so fig.~\ref{fig:sachdev1} (d) shows $\s$ in units of $\mu^2 R^2 /G$ versus $L\mu$ for the strip at $T/\mu=0$. When $L\mu=0$ we find $\s/(\mu^2 R^2 /G)=-0.016$, and then as $L\mu$ increases the local maximum and minimum still appear, and finally $\s/(\mu^2 R^2 /G)\to 0^-$ as $L\mu\to\infty$, indicating the area theorem is obeyed. In fact, this is our only example of an IR fixed-point with non-relativistic scaling where the area theorem is obeyed, which provides an important lesson: non-relativistic scaling allows, but does not require, area theorem violation.

We next consider the solution of ref.~\cite{Lucas:2014sba} with $d=4$ and
\beq
\label{eq:sachdev2}
b_2 = - \hat{z}^4 \, \frac{\hat{z}^2+12}{\hat{z}^2+6}, \qquad \hat{F}_{zt} = - \hat{z}\left(1+0.852\,\hat{z}^2\right)^{-3},
\eeq
which at $T=0$ describes a domain wall from $AdS_5$ to a HV geometry with $\zeta \to \infty$ and $-\theta/\zeta=3$. Asymptotically $\Phi(z) \propto z^2$ at leading order as $z \to 0$, saturating the Breitenlohner-Freedman bound, but the absence of a $z^2 \log z$ term indicates that $\Op$'s source vanishes, and hence the FLEE is obeyed. Indeed, as argued above using eq.~\eqref{eq:rgsmallL}, $\s \propto L$ at small $L$.

\FIGURE{
	\begin{tabular}{c c}
		\includegraphics[width=0.48\textwidth]{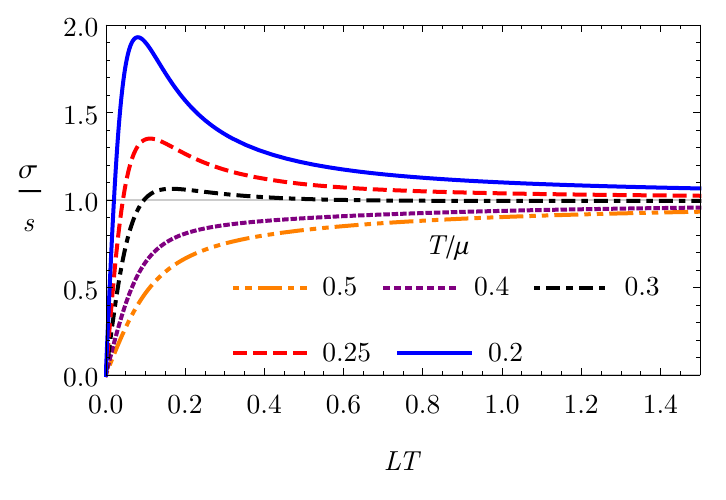}
		&
		\includegraphics[width=0.48\textwidth]{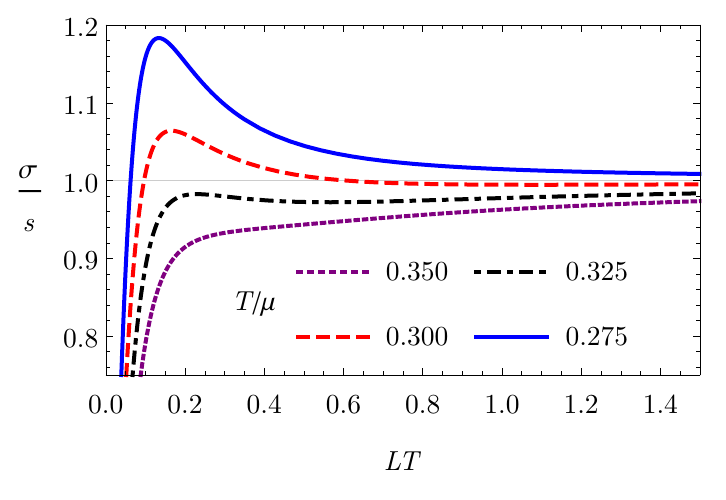}
		\\
		(a) & (b)\\
		\includegraphics[width=0.48\textwidth]{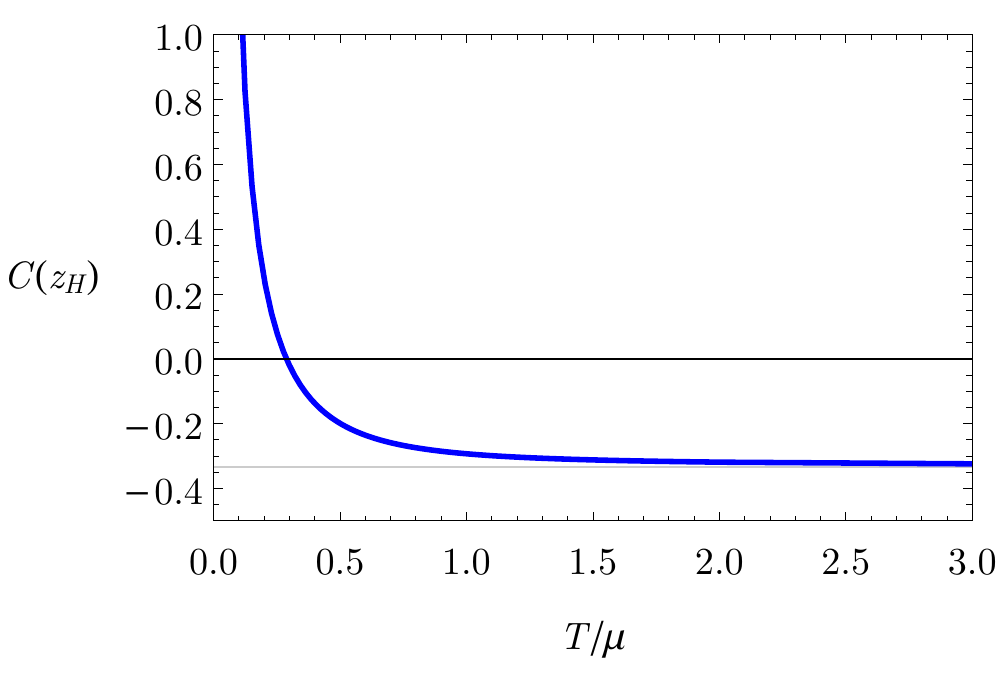}
		&
		\includegraphics[width=0.48\textwidth]{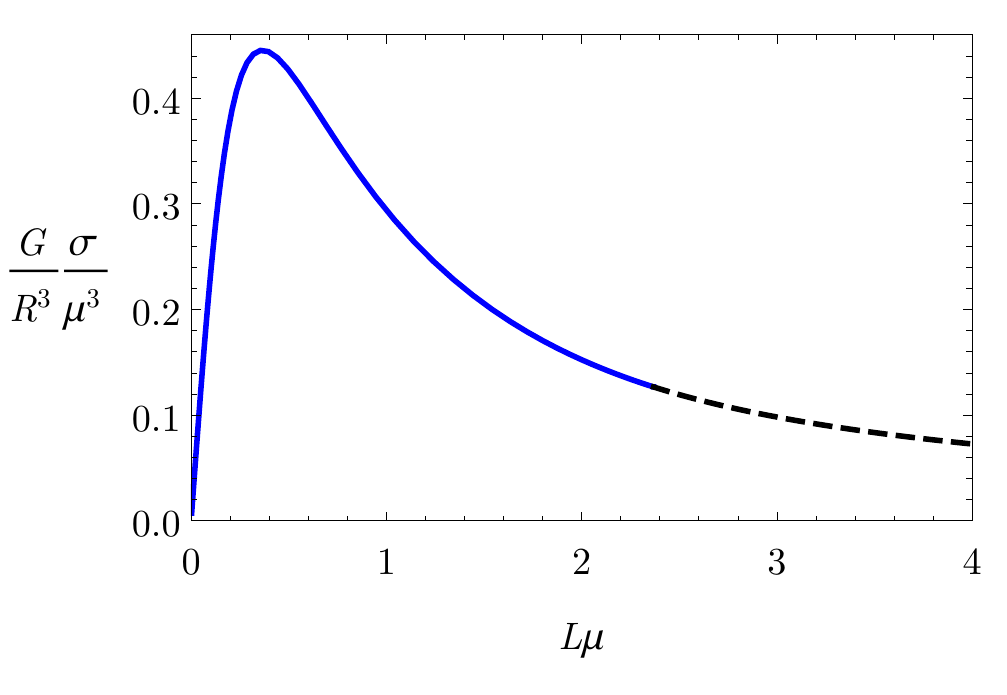}\\
		(c) & (d)
	\end{tabular}
	\caption{Plots of our results using the solution given by eq.~\eqref{eq:sachdev2} ($d=4$, $\zeta=\infty$, $\theta=-3\zeta$). (a) The ED, $\s$, in units of entropy density $s$, for the strip, for $T/\mu=0.5$ down to $0.2$. (b) Close-up of (a), showing the formation of a local maximum and minimum before the formation of the global maximum as $T/\mu$ decreases from $0.350$ down to $0.275$. (c) The dimensionless coefficient $C(z_H)$ from eq.~\eqref{eq:alphaholo} versus $T/\mu$. As $T/\mu \to \infty$, $C(z_H)$ approaches the value for AdS-SCH with $d=4$ (the horizontal line). As $T/\mu$ decreases, $C(z_H)$ changes sign from negative to positive at the $(T/\mu)_3$ from tab.~\ref{tab:hyper}, indicating area theorem violation when $T/\mu<(T/\mu)_3$. (d) The ED, $\s$, in units of $\mu^2 R^2/G$, versus $L\mu$, at $T/\mu=0$. The global maximum and area theorem violation remain, although a ``phase transition'' occurs at $L\mu\approx2.37$ between extremal surfaces in the bulk (from the blue solid to the black dashed).}\label{fig:sachdev2}
}

Figs.~\ref{fig:sachdev2} (a) and (b) show $\s/s$ versus $LT$ for the strip in this solution with various $T/\mu$. For all $T/\mu$ we find $\s/s \propto L$ at small $LT$, as expected, and in particular $\lim_{LT\to0}\s/s=0$. At sufficiently high $T/\mu$, as $LT$ increases $\s/s$ increases monotonically, and $\s/s \to 1^-$ as $LT \to \infty$, indicating the area theorem is obeyed. However, as we decrease $T/\mu$ we find a transition very similar to that of AdS-RN with $d<\dcrit$, discussed in sec.~\ref{adsrn}. Specifically, at some $(T/\mu)_1$ a local minimum and maximum appear, but $\s/s$ remains below one for all $LT$, then at some $(T/\mu)_2<(T/\mu)_1$ the local maximum rises above one to become a global maximum, while the local minimum remains and $\s/s\to1^-$ for $LT\to\infty$, and finally at some $(T/\mu)_3<(T/\mu)_2$ the local minimum disappears and the transition occurs to $\s/s\to1^+$ as $LT \to \infty$. Our numerical estimates for $(T/\mu)_1$, $(T/\mu)_2$, and $(T/\mu)_3$ appear in tab.~\ref{tab:hyper}. Correspondingly, fig.~\ref{fig:sachdev2} (c) shows $C(z_H)$ versus $T/\mu$, where $C(z_H)<0$ for $T/\mu>(T/\mu)_3$, indicating the area theorem is obeyed, while $C(z_H)>0$ for $T/\mu<(T/\mu)_3$, indicating area theorem violation.

As mentioned above, for a solution such as this, with $\zeta \to \infty$, when $T/\mu=0$ the $z \to \infty$ geometry is conformal to $AdS_2 \times \mathbb{R}^{d-2}$, with no horizon. In particular, $s=0$ when $T/\mu=0$, so fig.~\ref{fig:sachdev2} (d) shows $\s$ in units of $\mu^2 R^2 /G$ versus $L\mu$ at $T/\mu=0$. As $L\mu$ increases, we find a transition at $L\mu\approx 2.37$, from a connected to disconnected minimal surface, similar to the transition in fig.~\ref{fig:rgphasetrans}, and the transitions in various geometries conformal to $AdS_2 \times \mathbb{R}^{d-2}$ in refs.~\cite{Kulaxizi:2012gy,Erdmenger:2013rca}. Otherwise, however, the overall behavior is the natural extrapolation from $T/\mu>0$, with $\s \propto L$ at small $L\mu$, then as $L\mu$ increases a global maximum appears, and finally $\s \to 0^+$ as $L\mu \to \infty$, indicating area theorem violation.

\begin{table}[ht!]
\centering
\begin{tabular}[c]{|c|c|c|c|c|c|}
\hline
$d$ & $\zeta$ & $\theta$ & $(T/\mu)_1$ & $(T/\mu)_2$ & $(T/\mu)_3$ \\ \hline
$4$ & $\infty$ & $-3\zeta$ & $0.336$ & $0.319$ & $0.289$ \\ \hline
$3$ & $3$ & $1$ & $0.0629$ & $0.0516$ & $0.0334$  \\ \hline
\end{tabular}
\caption{\label{tab:hyper} For the strip in the solution of eq.~\eqref{eq:sachdev2} ($d=4$, $\zeta=\infty$, $\theta=-3\zeta$) or eq.~\eqref{eq:sachdev3} ($d=3$, $\zeta=3$, $\theta=1$), as $(T/\mu)$ decreases, at $(T/\mu)_1$ a local minimum and maximum appear in $\s/s$ as a function of $LT$, at $(T/\mu)_2<(T/\mu)_1$ the local maximum becomes a global maximum, but the local minimum remains and still $\s/s\to1^-$ as $LT\to\infty$, and then at $(T/\mu)_3<(T/\mu)_2$ the local minimum disappears, and the transition occurs to $\s/s\to1^+$ as $LT \to \infty$. }
\end{table}

Finally we consider the solution of ref.~\cite{Lucas:2014sba} with $d=3$ and
\beq
\label{eq:sachdev3}
b_2 = - \hat{z}^3 \, \frac{\hat{z}^2+12\hat{z}+288}{\hat{z}^2+12\hat{z}+72}, \qquad \hat{F}_{zt} = - \left(1+0.891\,\hat{z}\right)^{-4},
\eeq
which at $T=0$ describes a domain wall from $AdS_4$ to a HV geometry with $\zeta =3$ and $\theta=1$. Asymptotically $\Phi(z) \propto z$ at leading order as $z \to 0$, so $\Op$ has a non-zero source and the FLEE may be violated. Indeed, as argued above using eq.~\eqref{eq:rgsmallL}, $\s \propto L^0$ at small $L$.

Aside from the small-$L$ behavior, our results for this solution are very similar to those of the previous solution, and those of AdS-RN with $d<\dcrit$: as we lower $T/\mu$, we find a multi-stage transition to area theorem violation. Fig.~\ref{fig:sachdev3} (a) and (b) show $\s/s$ versus $LT$ for the strip in this solution with various $T/\mu$. For all $T/\mu$ we find $\s/s \propto -L^0$ at small $LT$, as expected. For sufficiently high $T/\mu$, as $LT$ increases $\s/s$ increases monotonically and eventually $\s/s \to 1^-$ as $LT \to \infty$, indicating the area theorem is obeyed. However, as in the solution of eq.~\eqref{eq:sachdev3} and AdS-RN with $d<\dcrit$, as we lower $T/\mu$ a local minimum and maximum appear at some $(T/\mu)_1$, then the maximum rises above $\s/s=1$ but still $\s/s\to1^-$ as $LT \to \infty$ at some $(T/\mu)_2$, and ultimately the local minimum disappears and $\s/s \to 1^+$ as $LT \to \infty$ at some $(T/\mu)_3$, indicating area theorem violation. Our numerical estimates for $(T/\mu)_1$, $(T/\mu)_2$, and $(T/\mu)_3$ appear in tab.~\ref{tab:hyper}. Correspondingly, fig.~\ref{fig:sachdev3} (c) shows $C(z_H)$ versus $T/\mu$, where $C(z_H)<0$ for $T/\mu>(T/\mu)_3$, indicating the area theorem is obeyed, while $C(z_H)>0$ for $T/\mu<(T/\mu)_3$, indicating area theorem violation.

\FIGURE{
	\begin{tabular}{c c}
		\includegraphics[width=0.48\textwidth]{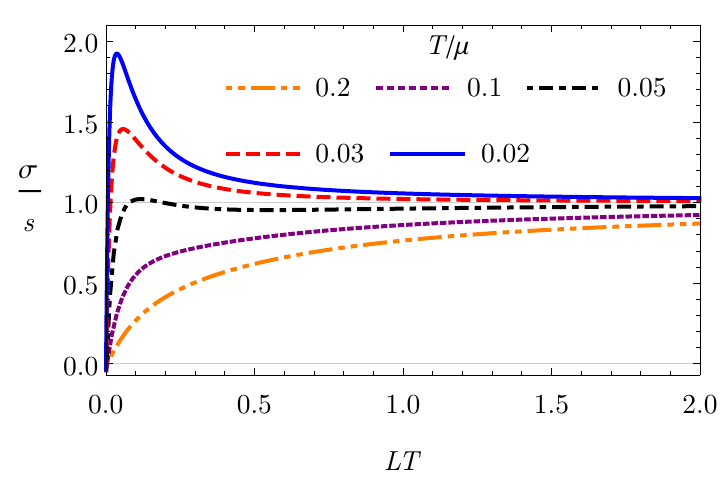}
		&
		\includegraphics[width=0.48\textwidth]{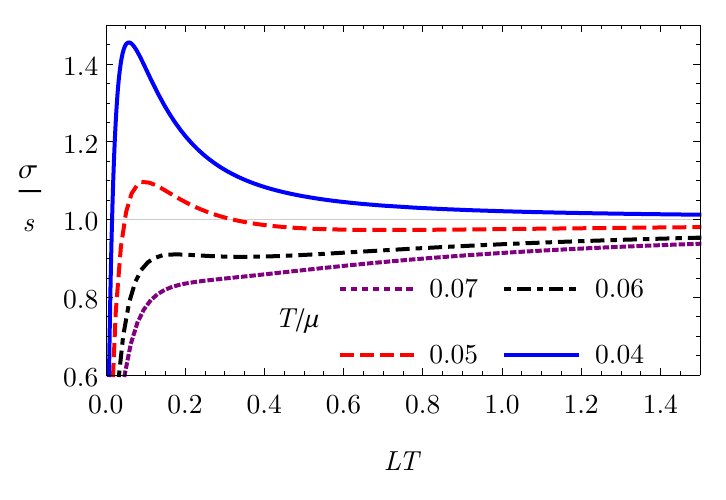}
		\\ (a) & (b) \\
		\includegraphics[width=0.48\textwidth]{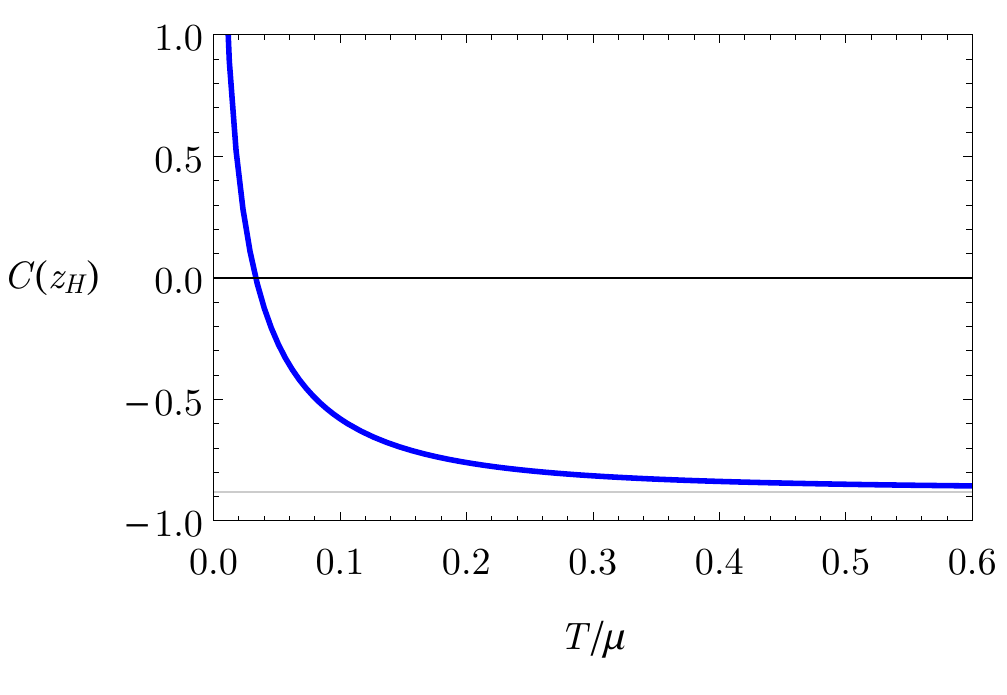}
		&
		\includegraphics[width=0.48\textwidth]{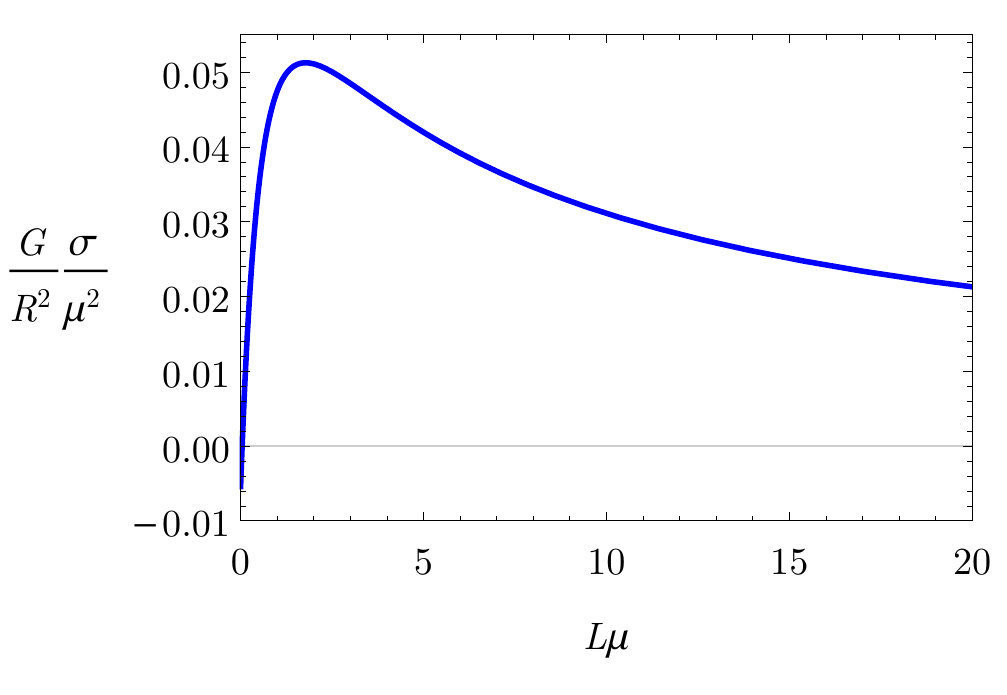}
		\\ (c) & (d)
	\end{tabular}
	\caption{Plots of our results using the solution given by eq.~\eqref{eq:sachdev3} ($d=3$, $\zeta=3$, $\theta=1$). (a) The ED, $\s$, in units of entropy density $s$, for the strip, for $T/\mu=0.2$ down to $0.02$. (b) Close-up of (a), showing the formation of a local maximum and minimum before the formation of the global maximum as $T/\mu$ decreases from $0.07$ down to $0.02$. (c) The dimensionless coefficient $C(z_H)$ from eq.~\eqref{eq:alphaholo} versus $T/\mu$. As $T/\mu \to \infty$, $C(z_H)$ approaches the value for AdS-SCH with $d=3$ (the horizontal line). As $T/\mu$ decreases, $C(z_H)$ changes sign from negative to positive at the $(T/\mu)_3$ from tab.~\ref{tab:hyper}, indicating area theorem violation when $T/\mu<(T/\mu)_3$. (d) The ED, $\s$, in units of $\mu^2 R^2/G$, versus $L\mu$, at $T/\mu=0$. The global maximum and area theorem violation remain, albeit with the logarithmic area law violation in eq.~\eqref{eq:arealawviolation}.}\label{fig:sachdev3}
}

As in the solutions above, when $T/\mu=0$ this solution has $s=0$, so fig.~\ref{fig:sachdev3} (d) shows $\s$ in units of $\mu^2 R^2/G$ versus $L\mu$ for the strip at $T/\mu=0$. When $L\mu=0$ we find $\s/(\mu^2 R^2 /G)=-0.006$, and then as $L\mu$ increases the global maximum remains, and $\s \to 0^+$ as $L\mu \to \infty$, indicating that the area theorem violation remains.

However, this solution has a key difference from the others at $T/\mu=0$. The solution in eq.~\eqref{eq:sachdev3} produces a HV geometry with $d=3$ and $\theta=1$, hence $\theta=d-2$, which produces logarithmic area law violation, possibly signaling a ``hidden'' (\textit{i.e.} not gauge-invariant) Fermi surface~\cite{Ogawa:2011bz,Huijse:2011ef}. To see the origin of the logarithm, we use $g(z)=b(z)/a(z)$ and the scalings of $a(z)$ and $b(z)$ at $T/\mu=0$ and $z \to \infty$ in eq.~\eqref{eq:hvscalings}, leading to $g(z)\sim z^{2(2-d)}$. Plugging that into the definition of $C(z_*)$ in eq.~\eqref{eq:cdef}, we find at large $u=z/z_*$ an integral $\int_0^1 du/u$, producing a logarithm. More precisely, at large $L$ we find
\beq
\label{eq:arealawviolation}
\sigma \approx 0.17 \, \frac{\mu^4 R^2}{G} \frac{\log \left(\mu L\right)}{L} + \mathcal{O}(1/L),
\eeq
so the leading term is $\frac{\log \left(\mu L\right)}{L}$ not $\frac{1}{L}$, which is the definition of logarithmic area law violation.

In summary, AdS-to-HV domain walls exhibit various behaviors, depending on the values of $d$, $\zeta$, and $\theta$. In particular, in our first example, with $d=3$, $\zeta =2$, and $\theta=-2$, the area theorem was obeyed for all $T/\mu$. The lesson: non-relativistic scaling in the IR allows, but does not require, area theorem violation. Moreover, our second example, with $d=4$, $\zeta \to \infty$, and $\theta=-3\zeta$, has a $z \to \infty$ metric conformal to $AdS_2 \times \mathbb{R}^{d-1}$, hence the dual field theory describes a semi-local quantum liquid, but with $s=0$ at $T/\mu=0$, unlike extremal AdS-RN~\cite{Iqbal:2011in,Iqbal:2011ae,Hartnoll:2012wm}. However, like extremal AdS-RN, the area theorem is violated, raising the question of whether the same is true for all semi-local quantum liquids. More generally, as mentioned in sec.~\ref{intro}, a natural question is for what values of $d$, $\zeta$, and $\theta$ area theorem violation occurs. We leave a completely general (holographic) analysis to future research.

\section{AdS with Broken Translational Invariance}
\label{translation} 

In this section we consider the bulk action of ref.~\cite{Andrade:2013gsa},
\beq
	S_{\textrm{bulk}} = \frac{1}{16\pi G}\int d^{d+1} x \sqrt{-\det g_{MN}} \left(
		\mathcal{R} + \frac{d(d-1)}{R^2} - \frac{1}{4} F_{MN} F^{MN} - 8\pi G \sum_{I=1}^{d-1} (\p \psi^I)^2
	\right),
\eeq
where $F_{MN}=\partial_M A_N - \partial_N A_M$ is the field strength of a $U(1)$ gauge field $A_M$, dual to a conserved $U(1)$ current, and the $\psi^I$ are a set of massless scalar ``axion'' fields, dual to exactly marginal scalar operators $\mathcal{O}^I$. We focus on the solutions of ref.~\cite{Andrade:2013gsa} with
\beq
	A_t = \mu\left(1 - \frac{z^{d-2}}{z_H^{d-2}}\right),
	\quad
	\psi^I = \frac{1}{\sqrt{16\pi G}}\,\sum_{k=1}^{d-1}\gamma^I_k \, x_k,
\eeq
with all other components of $A_M$ vanishing, and where $x_k$ are the $d-1$ components of the spatial vector $\vec{x}$, while the $\gamma^I_k$ are dimensionful constants obeying
\beq
\sum_{I=1}^{d-1}\gamma^I_j \gamma^I_k \equiv \gamma^2 \, \delta_{jk},
\eeq
where $\gamma$ is a constant. In the solutions of ref.~\cite{Andrade:2013gsa}, the metric takes the form in eq.~\eqref{eq: 1} with
\beq
\label{eq:andrade_metric_functions}
f(z) = g(z) = 1 -\frac{1}{2(d-2)}\gamma^{2}z^{2} -m z^d +\frac{(d-2)}{2(d-1)}\mu^{2}z_{H}^{2} \left(\frac{z}{z_H}\right)^{2(d-1)},
\eeq
with a horizon at $z = z_H$. These solutions are dual to CFTs with non-zero $\Ttt$, $T$, and $s$, given by eqs.~\eqref{eq:energy_density} and~\eqref{eq:thermo_entropy}, and also non-zero chemical potential $\mu$. In particular,
\beq
\label{eq:transT}
T = \frac{d}{4\pi R^2} \frac{1}{z_H} \left(1-\left[\frac{\gamma^2}{2d}-\frac{(d-2)^2\mu^2}{2d(d-1)}\right]z_H^2\right).
\eeq
The solution also includes non-zero sources for the $\mathcal{O}^I$, but with $\langle \mathcal{O}^I\rangle=0$. The sources are $\propto x_k$, thus breaking the CFT's translational symmetry. Momentum can therefore dissipate, so the DC conductivity is finite even with non-zero $U(1)$ charge density~\cite{Andrade:2013gsa}. As in AdS-RN, when $T=0$ the near-horizon geometry is $AdS_2 \times \mathbb{R}^{d-1}$, with $AdS_2$ radius $R_{AdS_2}$ given by
\beq
\label{eq:transads2}
R^2_{AdS_2} = \frac{1}{d(d-1)} \frac{(d-1) \gamma^2 + (d-2)^2 \mu^2}{\gamma^2 + (d-2)^2\mu^2} R^2.
\eeq
The $AdS_2$ appears even when $\mu=0$, in which case $R_{AdS_2}=R/(\gamma\sqrt{d})$.

The two dimensionless ratios $\gamma/T$ and $\mu/T$ determine the solution completely. When $\gamma/T=0$ with $\mu/T$ fixed, the solution reduces to AdS-RN, and we recover the results of sec.~\ref{adsrn}. When $\gamma/T\neq 0$, because we introduce sources for the $\mathcal{O}^I$ we find some features similar to the RG flows of sec.~\ref{rg}. In particular, the term $\propto \gamma^2 z^2$ in $g(z)$ in eq.~\eqref{eq:andrade_metric_functions} produces UV divergences that are cancelled by the subtraction $S - \scft$~\cite{Liu:2013una,Casini:2016udt} only when $d=3$, to which we restrict in the rest of this section. Moreover, because $g(z)$ does not have the asymptotics in eq.~\eqref{eq:fgasymp}, the FLEE does not apply. Nevertheless, by straightforwardly modifying the methods of sec.~\ref{general}, for strip of width $L$ small compared to all other length scales we find
\beq
\label{eq:andrade_small_L}
\sigma = \frac{R^2}{4 G} \left(\frac{\Gamma\left(\frac{1}{4}\right)^2}{24 \, \Gamma\left(\frac{3}{4}\right)^2} \, \gamma^2 + \frac{\Gamma\left(\frac{1}{4}\right)^2}{32 \, \Gamma\left(\frac{3}{4}\right)^2} \, m L + \ldots\right),
\eeq
where $\ldots$ are terms with higher powers of $L$ than those shown. In other words, for the strip in these solutions, $\s$ at small $L$ is linear in $L$, but with non-zero, positive intercept $\propto \gamma^2$. An intercept $\propto \gamma^2$ also appears in $d>3$~\cite{Mozaffara:2016iwm}.

In the two-parameter solution space, we focus on two one-parameter subspaces: extremal solutions, $T/\mu=0$ with $\mu/\gamma$ fixed, and uncharged solutions, $\mu/T=0$ with $T/\gamma$ fixed.

Fig.~\ref{fig:transt0} (a) shows $\s/s$ versus $L\g$ for the strip in the extremal case for various $\mu/\g$. For sufficiently large $\mu/\g$, the effects of $\g$ are small, so $\s/s$ resembles that of extremal AdS-RN with $d=3$: as $L\g$ increases, $\s/s$ rises linearly from zero, reaches a global maximum, and then $\s/s\to1^+$ as $L\g \to \infty$, indicating area theorem violation. However, as $\mu/\gamma$ decreases, the effects of $\g$ grow prominent, especially at small $L\g$. Specifically, as $\mu/\g$ decreases the intercept, $\lim_{L\g\to0} \s/s$, increases, and moreover the slope at small $L\g$ changes sign from positive to negative. To see why, we use $\partial \s/\partial L \propto m$ from eq.~\eqref{eq:andrade_small_L}. Solving eq.~\eqref{eq:transT} (with $T=0$) for $z_H$, plugging the result into $g(z_H)=0$ in eq.~\eqref{eq:andrade_metric_functions}, and solving for $m$ gives
\beq
\frac{\partial\sigma}{\partial L} \propto m = \frac{(\mu^2 - \gamma^2)\sqrt{\mu^2 + 2 \gamma^2}}{6\sqrt{3}},
\eeq
which clearly changes from positive to negative as $\mu/\g$ decreases. Meanwhile, for all $\mu/\g$ area theorem violation occurs: $\s/s\to1^+$ as $L\g\to\infty$. That is unsurprising since the near-horizon geometry is $AdS_2 \times \mathbb{R}^{d-1}$, which we know from extremal AdS-RN exhibits area theorem violation, and changing $\mu/\g$ just changes $R_{AdS_2}$ in eq.~\eqref{eq:transads2}. As confirmation, fig.~\ref{fig:transt0} (b) shows $C(z_H)$ versus $\mu/\g$ for extremal solutions, where indeed $C(z_H)>0$ for all $\mu/\g$.

\FIGURE{
\begin{tabular}{c c}
\includegraphics[width=0.5\textwidth]{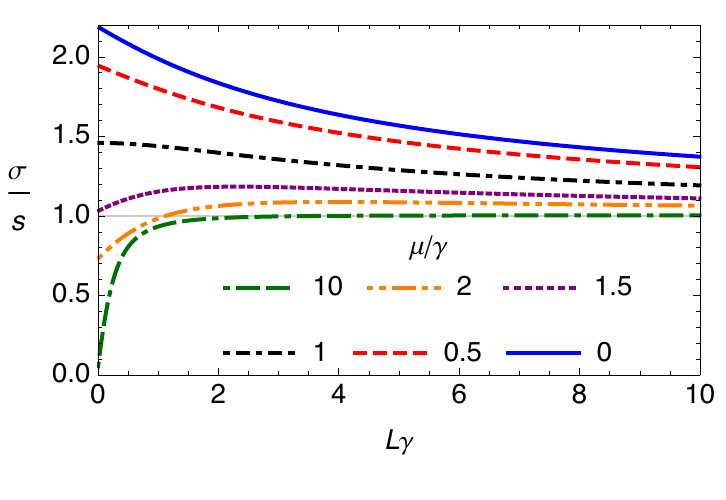}
&
\includegraphics[width=0.5\textwidth]{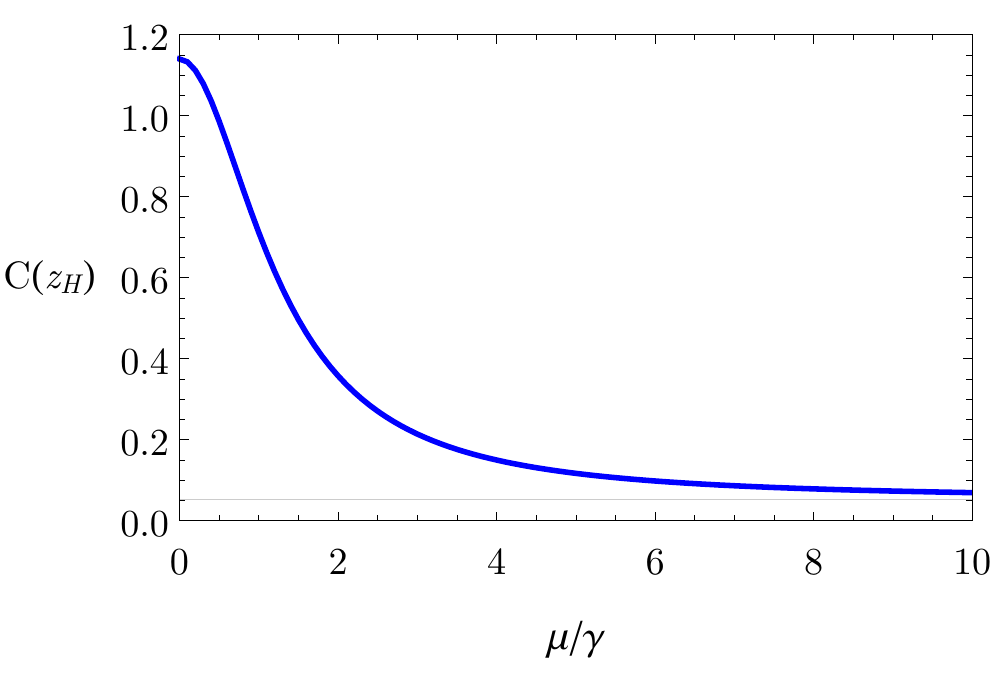}
\\
(a) & (b)
\end{tabular}
\caption{\label{fig:transt0} (a) The ED, $\s$, in units of entropy density $s$, versus $L\g$ for the strip in the solution of eq.~\eqref{eq:andrade_metric_functions} with $T/\mu=0$, showing the intercept increase and the slope change sign at small $L\g$ as $\mu/\g$ decreases from $10$ to $0$. (b) The dimensionless coefficient $C(z_H)$ from eq.~\eqref{eq:alphaholo} versus $\mu/\gamma$. As $\mu/\g \to \infty$, $C(z_H)$ approaches the value for extremal AdS-RN with $d=3$ (the horizontal line). Clearly $C(z_H)>0$ for all $\mu/\g$, indicating area theorem violation.}
}

Fig.~\ref{fig:transmu0} (a) shows $\s/s$ versus $LT$ for the strip in the uncharged case for various $T/\g$. For sufficiently large $T/\g$, the effects of $\gamma$ are small, so $\s/s$ resembles that of AdS-SCH with $d=3$: as $LT$ increases, $\s/s$ increases monotonically until $\s/s\to1^-$ as $LT \to \infty$, and the area theorem is obeyed. However, as $T/\g$ decreases, the effects of $\g$ grow prominent. In particular, as $T/\g$ decreases the intercept, $\lim_{LT \to 0} \s/s$, increases, and the slope at small $LT$ changes sign from positive to negative. To see why, we again use $\partial \s/\partial L \propto m$ from eq.~\eqref{eq:andrade_small_L}, and solve eqs.~\eqref{eq:transT} and~\eqref{eq:andrade_metric_functions} for $m$, obtaining
\beq
\label{eq:transslope}
\frac{\partial\sigma}{\partial L} \propto m = \frac{2 \gamma^4 \left(8 \pi^2 T^2 + \gamma^2 - 2 \pi T \sqrt{16 \pi^2 T^2  + 6 \gamma^2}\right)}{\left(4 \pi T - \sqrt{16 \pi^2 T^2 + 6 \gamma^2}\right)^3}.
\eeq
As $T/\g$ decreases, the $\partial \s/\partial L$ in eq.~\eqref{eq:transslope} changes sign from positive to negative at $T/\g=1/(2 \pi \sqrt{2})$. Meanwhile at large $LT$, as $T/\g$ decreases a transition to area theorem violation occurs at $T/\g \approx 0.034$. Again, that is unsurprising, since at $T/\g=0$ the near-horizon geometry is $AdS_2 \times \mathbb{R}^{d-1}$. As confirmation, fig.~\ref{fig:transmu0} (b) shows $C(z_H)$ versus $T/\g$ for uncharged solutions, where indeed $C(z_H)<0$ for $T/\g>0.034$ and $C(z_H)>0$ for $T/\g<0.034$, indicating area theorem violation.

\FIGURE{
\begin{tabular}{c c}
\includegraphics[width=0.5\textwidth]{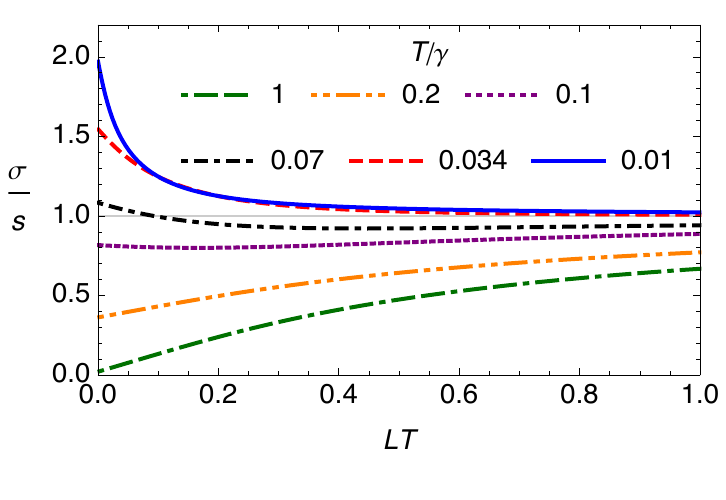}
&
\includegraphics[width=0.5\textwidth]{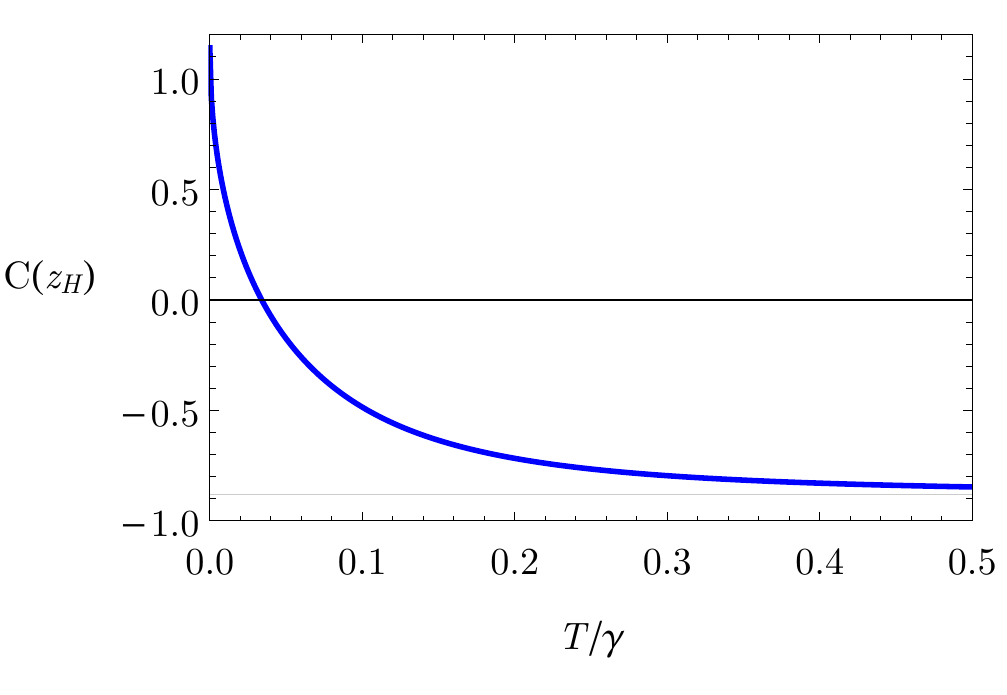}
\\
(a) & (b)
\end{tabular}
\caption{\label{fig:transmu0} (a) The ED, $\s$, in units of entropy density $s$, versus $LT$ for the strip in the solution of eq.~\eqref{eq:andrade_metric_functions} with $\mu/T=0$, showing the intercept increase and the slope change sign at small $LT$ as $T/\g$ decreases from $1$ to $0.01$. A transition to area theorem violation also occurs at $T/\g\approx0.034$ (b) The dimensionless coefficient $C(z_H)$ from eq.~\eqref{eq:alphaholo} versus $T/\gamma$. As $T/\g \to \infty$, $C(z_H)$ approaches the value for AdS-SCH with $d=3$ (the horizontal line). As $T/\g$ decreases, $C(z_H)$ becomes positive at $T/\g\approx 0.034$, indicating area theorem violation.}
}

In summary, in the parameter space we explored for the solutions of ref.~\cite{Andrade:2013gsa}, $\s/s$'s large-$L$ behavior is similar to that of AdS-RN, namely, as we approach the extremal limit, area theorem violation occurs. Crucially, the extremal solutions of ref.~\cite{Andrade:2013gsa} have a near-horizon $AdS_2 \times \mathbb{R}^{d-1}$, and hence are dual to semi-local quantum liquid states~\cite{Iqbal:2011in,Iqbal:2011ae}, similar to extremal AdS-RN and the $\zeta \to \infty$ solution of sec.~\ref{hyper}, again suggesting that semi-local quantum liquids always violate the area theorem. However, the solutions of ref.~\cite{Andrade:2013gsa} describe non-zero sources for the $\mathcal{O}^I$, which violate the FLEE, so $\s$'s small-$L$ behavior is radically different from that of AdS-RN. Specifically, as $T/\g$ or $\mu/\g$ decrease, \textit{i.e.} as $\g$ increases, the value of $\s$ at $L=0$ increases, and $\partial \sigma/\partial L$ changes sign from positive to negative. In these cases, $\s$ as a function of $L$ does not have a maximum, in stark contrast to extremal AdS-RN. As a result, although in these semi-local quantum liquids space should still divide into patches of size $\ell$, as described in sec.~\ref{sec:outlook}, we cannot define $\ell$ from a maximum in $\s$. We leave an exploration of the full parameter space of the solutions of ref.~\cite{Andrade:2013gsa} to future research.

\section{AdS Soliton}
\label{soliton}

In this section we consider the same bulk action as in sec.~\ref{adssc}, namely a $(d+1)$-dimensional Einstein-Hilbert action with negative cosmological constant, and study the AdS soliton solution~\cite{Witten:1998zw,Horowitz:1998ha,Klebanov:2007ws,Nishioka:2009zj}, obtained from AdS-SCH by double Wick-rotation, with metric
\beq
\label{eq:ads_soliton_metric}
ds^2 = \frac{R^2}{z^2} \left( - dt^2 + d\vec{x}^2 + g(z) d\chi^2 + \frac{dz^2}{g(z)}\right),
\eeq
where $g(z)=1 - z^d/z_0^d$, the coordinate $\chi$ is compact, $\chi \sim \chi + 4 \pi z_0 /d$, and $\vec{x}$ represents $(d-2)$ non-compact spatial directions. The AdS soliton has a ``hard wall'' at $z=z_0$, where $g(z_0)=0$, indicating that compactifying a spatial direction in the dual CFT, with anti-periodic boundary conditions for fermions, produces a mass gap and confinement~\cite{Witten:1998zw,Klebanov:2007ws}. The AdS soliton has $T=0$, $s=0$, and
\beq
\label{eq:casimir}
\Ttt=-\frac{1}{16\pi G} \, \frac{R^{d-1}}{z_0^d},
\eeq
that is, the CFT has a negative Casimir energy.

The metric in eq.~\eqref{eq:ads_soliton_metric} is not of the form in eq.~\eqref{eq: 1}, so the results of sec.~\ref{general} do not apply, however, the minimal area calculations generalize straightforwardly~\cite{Klebanov:2007ws,Bueno:2016rma}. Our entangling region is a strip of width $L$ with planar boundaries along a non-compact direction, so that in particular the entangling surface wraps around $\chi$. As shown in ref.~\cite{Klebanov:2007ws}, for any $L$, multiple extremal surfaces exist. In particular, for any $L$, ``disconnected'' extremal surfaces exist that drop straight from the asymptotic $AdS_{d+1}$ boundary to the hard wall, with area
\beq
\label{eq:ads_soliton_disconnected_area}
\mathcal{A}^\mathrm{strip}_\mathrm{discon.} = R^{d-1}\mathrm{Vol}(\mathbb{R}^{d-3})\frac{8\pi z_0}{d(d-2)} \left(\frac{1}{\varepsilon^{d-2}} - \frac{1}{z_0^{d-2}}
\right).
\eeq
For sufficiently small $L$, connected extremal surfaces also exist. These ``hang down'' into the bulk to a turn-around point $z_*$, as in fig.~\ref{fig:minimalsurfaces} (a), where the analogue of eq.~\eqref{eq:ldef} is
\beq
\label{eq:ads_soliton_strip_width}
L = 2 z_* \int_0^1 du \sqrt{\frac{g(z_*)}{g(z_* u)}} \frac{u^{d-1}}{\sqrt{g(z_* u) - g(z_*) u^{2(d-1)}}},
\eeq
where $u=z/z_*$ as in sec.~\ref{general}, and the analogue of eq.~\eqref{eq:stripamin} is
\begin{align}
\label{eq:ads_soliton_connected_area}
\mathcal{A}^\mathrm{strip}_{\textrm{con.}} = R^{d-1} \mathrm{Vol}(\mathbb{R}^{d-3}) \frac{8\pi z_0}{d}\Biggl[&
\frac{1}{d-2} \left(\frac{1}{\varepsilon^{d-2}} - \frac{1}{z_*^{d-2}} \right)
\nonumber \\ & + \frac{1}{z_*^{d-2}}\int_{0}^1 du \frac{1}{u^{d-1}} \left( \sqrt{\frac{g(z_* u)}{g(z_* u) - g(z_*) u^{2(d-1)}}} - 1\right) \Biggr].
\end{align}
In fact, two connected extremal surfaces exist. Fig.~\ref{fig:solitonarea} (a) shows that $L$ in eq.~\eqref{eq:ads_soliton_strip_width} is multi-valued in $z_*$, such that two different connected surfaces with different $z_*$ can have the same $L$. The maximal $L$ for which two connected solutions exists depends on $d$. For example, for $d=4$ two solutions exist when $L \lesssim 0.7 \, z_0$, as shown in fig.~\ref{fig:solitonarea} (a).

\FIGURE{
\begin{tabular}{c c}
\includegraphics[width=0.5\textwidth]{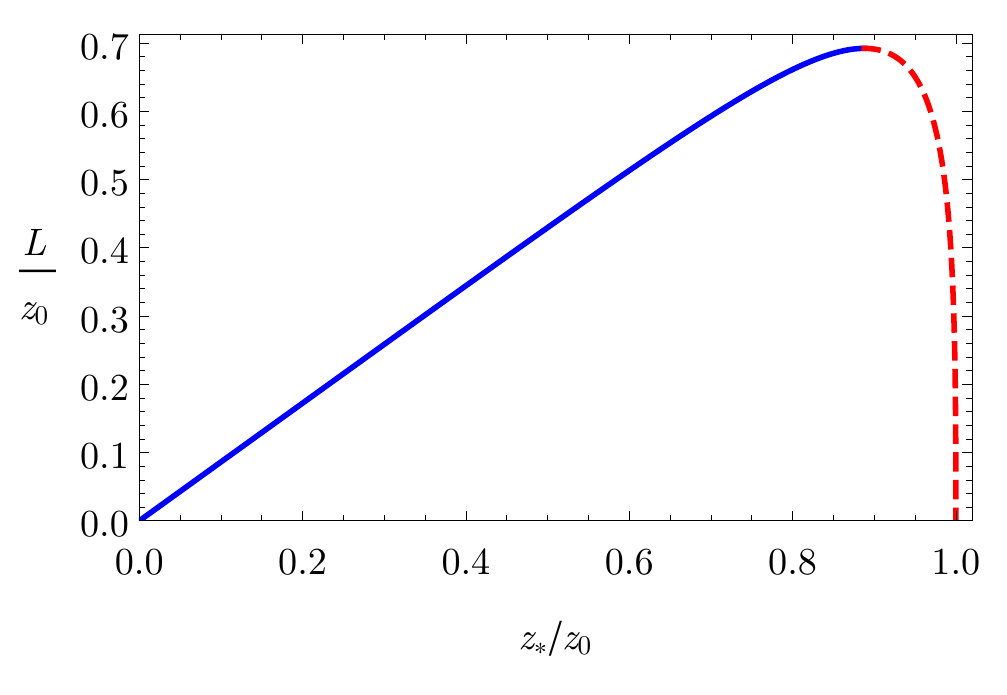}
&
\includegraphics[width=0.5\textwidth]{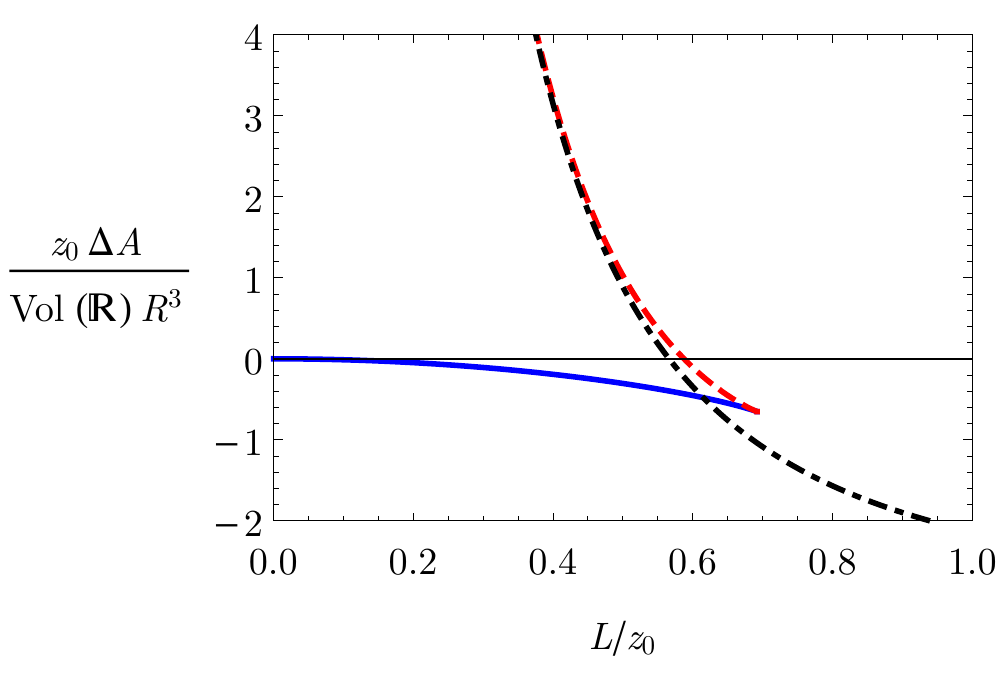}
\\
(a) & (b)
\end{tabular}
\caption{\label{fig:solitonarea} (a) Strip width $L$ versus turn-around point $z_*$, both in units of $z_0$, for connected extremal surfaces in the AdS soliton with $d=4$. Two branches of connected extremal surfaces exist, with $z_*/z_0<0.89$ (blue solid) and $z_*/z_0>0.89$ (red dashed). (b) The differences in area, $\Delta A$, normalized by $\textrm{Vol}\left(\mathbb{R}\right)R^3/z_0$, between the two connected extremal surfaces (blue solid and red dashed) and the disconnected extremal surfaces (black dot-dashed), and the minimal surface with the same $L$ in compactified $AdS_5$, versus $L/z_0$. A ``first-order phase transition'' occurs at $L/z_0\approx 0.615$~\cite{Bueno:2016rma} from a connected (blue solid) to a disconnected (black dot-dashed) extremal surface.}
}

The EE is given by the extremal surface with minimal area~\cite{Ryu:2006bv,Ryu:2006ef}. As $L$ increases a ``first order phase transition'' occurs between extremal surfaces as the area functional's global minimum, from the connected surface with smaller $z_*$ to the disconnected surface. For example, fig.~\ref{fig:solitonarea} (b) shows the transition for $d=4$, which occurs at $L\approx0.615\,z_0$~\cite{Bueno:2016rma}.

\begin{figure}[t!]
\centering
 \includegraphics[width=0.6\textwidth]{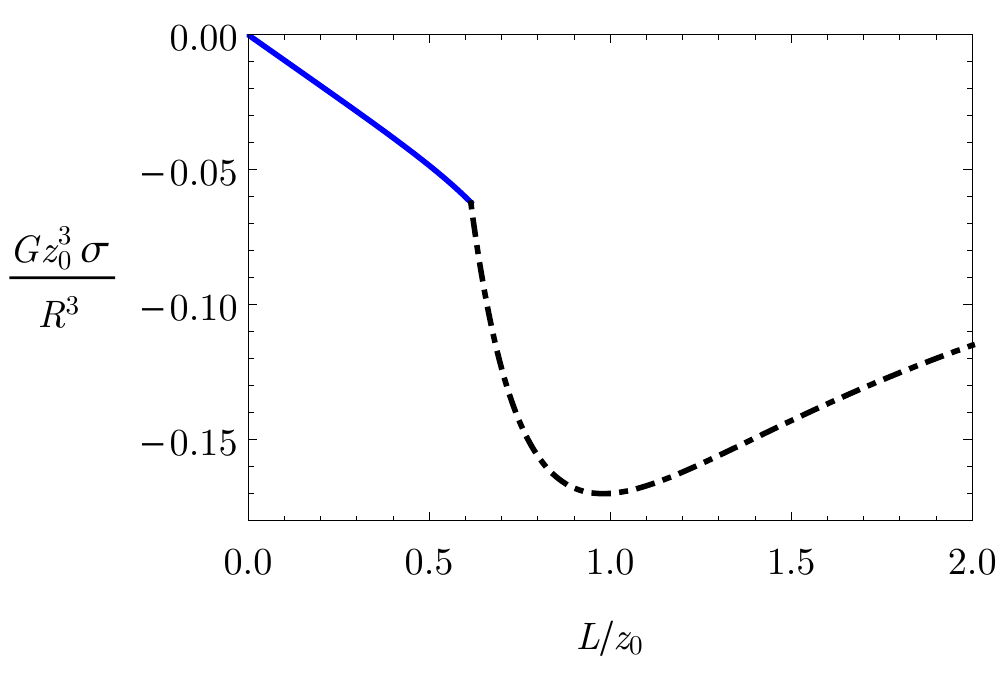}
 \caption{The ED, $\s$, normalized by $R^3/(G z_0^3)$, versus $L/z_0$ for the AdS soliton with $d=4$.}\label{fig:solitonED}
\end{figure}

The AdS soliton is asymptotically locally $AdS_{d+1}$, but has a compact spatial direction, which changes the EE's UV divergences compared to $AdS_{d+1}$. Indeed, our entangling surfaces wrap the compact direction $\chi$, so that the divergent area law term, $\propto 1/\varepsilon^{d-2}$, will include a factor of $\chi$'s length, $4\pi z_0/d$. The $AdS_{d+1}$ result in eq.~\eqref{eq:strip_vacuum_ee} has no $z_0$ dependence, hence the subtraction $S-\scft$ will not cancel the UV divergence. For a detailed discussion of $S$'s divergences in the AdS soliton, and regularization schemes, see ref.~\cite{Bueno:2016rma}. For simplicity, we will just compare the AdS soliton to $AdS_{d+1}$ with a compact direction of length $4 \pi z_0/d$, and \textit{periodic} boundary conditions for fermions, which we call ``compactified $AdS_{d+1}$.'' The compactified $AdS_{d+1}$ metric is locally identical to $AdS_{d+1}$, but produces divergences in extremal surfaces identical to those in the AdS soliton. For the AdS soliton, we thus define the area differences in fig.~\ref{fig:solitonarea} (b) and the ED, $\s$, by subtracting the result for compactified $AdS_{d+1}$.

A key caveat, however, is that compactified $AdS_{d+1}$ has a conical singularity at the Poincar\'e horizon~\cite{Gibbons:1998th}. The singularity could affect $\s$'s behavior as $L\to\infty$, the regime where the corresponding extremal surface hangs deeper and deeper into the bulk, approaching the Poincar\'e horizon. However, we have compared our subtraction to renormalization via covariant counterterms~\cite{Taylor:2016aoi,Taylor:2016kic,Taylor:2017zzo}, and found no difference at large $L$. Indeed, the counterterms ultimately subtract only the area term $\propto 1/\varepsilon^{d-2}$, and so differ from the compactified $AdS_{d+1}$ subtraction only by the area law term $\propto 1/L^{d-2}$, which primarily affects the small-$L$ behavior. Our subtraction is therefore sufficient to obtain $\s$'s large-$L$ behavior, and in particular to determine whether the area theorem is violated.

Applying our subtraction to eq.~\eqref{eq:ads_soliton_connected_area} for $\mathcal{A}^{\textrm{strip}}_{\textrm{con.}}$ thus gives $\s$ for $L$ below the transition,
\beq
\label{eq:solitonsmallL}
\s=\frac{R^{d-1}}{4 G}\left[ \frac{\sqrt{g(z_*)}}{z_*^{d-1}} + \frac{\hat{C}(z_*)}{z_*^{d-2}} \frac{2}{L} + \frac{2^{d-2} \pi^{\frac{d-1}{2}}}{d-2} \left( \frac{\G \left( \frac{d}{2(d-1)} \right)}{\G \left( \frac{1}{2(d-1)} \right)} \right)^{d-1} \frac{2}{L^{d-1}}\right],
\eeq
where $\hat{C}(z_*)$ is defined in analogy with $C(z_*)$ in eq.~\eqref{eq:cdef},
\beq
\hat{C}(z_*) = - \frac{1}{d-2} + \int_0^1 \frac{du}{u^{d-1}} \left( \sqrt{1 - \frac{g(z_*)}{g(z_*u)} \, u^{2(d-1)}} - 1\right).
\eeq
Applying our subtraction to eq.~\eqref{eq:ads_soliton_disconnected_area} for $\mathcal{A}^{\textrm{strip}}_{\textrm{discon.}}$ gives $\s$ for $L$ above the transition,
\beq
\label{eq:solitonlargeL}
\s=\frac{R^{d-1}}{4 G} \left[\frac{2^{d-1} \pi^{\frac{d-1}{2}}}{d-2} \left( \frac{\G \left( \frac{d}{2(d-1)} \right)}{\G \left( \frac{1}{2(d-1)} \right)} \right)^{d-1} \frac{1}{L^{d-1}}- \frac{2}{d-2} \frac{1}{z_0^{d-2}}\frac{1}{L}\right].
\eeq
Fig.~\ref{fig:solitonED} shows $\s$, normalized by $R^3/(G z_0^3)$, versus $L/z_0$ for the AdS soliton with $d=4$. We checked explicitly that $\s$'s qualitative behavior is the same as that in fig.~\ref{fig:solitonED} up to $d=40$.

Compactifying a spatial direction is not merely a change of state, so we do not expect the FLEE to apply. Nevertheless, a small-$L/z_0$ expansion of eq.~\eqref{eq:solitonsmallL} similar to that in sec.~\ref{general} gives at leading order $\s = 2\Ttt \tent^{-1}$, with $\tent$ for the strip in eq.~\eqref{eq:striptent}. In other words, we find precisely twice the result expected from the FLEE. Given $\Ttt<0$ from eq.~\eqref{eq:casimir}, $\s \propto -L$ at leading order at small $L/z_0$, as shown in fig.~\ref{fig:solitonED}.

As $L/z_0$ increases, the ED decreases until the transition from connected to disconnected minimal surface in the bulk at $L/z_0=0.615$, where a kink (discontinuous first derivative) appears, due to the transition from connected to disconnected minimal surface. As $L/z_0$ increases further, the ED decreases to a global minimum at
\beq
L = 2 (d-1)^{\frac{1}{d-2}} \, \pi^{\frac{d-1}{2(d-2)}}\,\left(\frac{\Gamma\left(\frac{d}{2(d-1)}\right)}{\Gamma\left(\frac{1}{2(d-1)}\right)}\right)^{\frac{d-1}{d-2}} \, z_0,
\eeq
and then increases until eventually $\s \to 0^-$ as $L/z_0 \to \infty$. Indeed, in eq.~\eqref{eq:solitonlargeL} the area term $\propto -1/L$ obviously dominates over the term $\propto 1/L^{d-1}$ at large $L/z_0$, hence $\s \to 0^-$ as $L/z_0 \to \infty$ for all $d$. In other words, $\Delta \a>0$ for all $d$, consistent with the area theorem.

In summary, the AdS soliton is our only example with a mass gap, \textit{i.e.} no massless IR degrees of freedom. The AdS soliton is not Lorentz-invariant, and hence the proofs of the area theorem in refs.~\cite{Casini:2012ei,Casini:2016udt} do not apply. Nevertheless, the AdS soliton has $\Delta \a>0$, consistent with the idea that the area term's coefficient counts degrees of freedom: with zero IR degrees of freedom, $\Delta \a$ should be positive. The question of whether $\Delta \a>0$ in all (holographic) systems with a mass gap we leave for future research.

\section*{Acknowledgements}

We thank Tom\'as Andrade, Jyotirmoy Bhattacharya, Andrej Ficnar, Nabil Iqbal, Keun-Young Kim, Anton Pribytok, Kostas Skenderis, and Marika Taylor for useful conversations and correspondence, and Pablo Bueno, Mohammed Reza Mohammadi Mozaffar, and William Witczak-Krempa for comments on the manuscript. We especially thank Johanna Erdmenger and Nina Miekley for useful conversations and for sharing the results of ref.~\cite{Erdmenger:2017pfh} with us prior to publication. N.~I.~G. and A.~O'B. also thank the University of W\"urzburg for hospitality during the completion of this project. N.~I.~G. and A.~O'B. were partially supported by the Royal Society research grant ``Strange Metals and String Theory'' (RG130401). N.~I.~G. was also supported by the European Research Council under the European Union's Seventh Framework Programme (ERC Grant agreement 307955). A.~O'B. is a Royal Society University Research Fellow. R.~R. acknowledges support from STFC through Consolidated Grant ST/L000296/1.

\bibliographystyle{JHEP}
\bibliography{eedensity_v2}

\end{document}